        \documentstyle[]{l-aa}
        \input psfig.tex
        \newcommand{\Hbeta}{\mbox{${ H_{\beta}}$}}
        \newcommand{\MgFe}{\mbox{${ [MgFe]} $}}

        \def\smallskip{\vskip 6pt}
        \def\littleskip{\vskip 4pt}
        \def\M12{${\rm M_{12}}$}
\begin{document}
	\thesaurus{    }

        \title{A new scenario of galaxy evolution under a universal IMF  }

	\author{ C. Chiosi$^{1,2}$, A. Bressan $^3$, L. Portinari$^2$, 
                 R. Tantalo$^2$  }

	\institute{ 
$^1$ European Southern Observatory, Karl-Schwarzschild-Strasse 2,
            D-85748, Garching bei Muenchen, Germany \\
$^2$ Department of Astronomy, University of Padova, 
                 Vicolo dell'Osservatorio 5, 35122 Padova, Italy \\ 
$^3$ Astronomical Observatory, Vicolo dell'Osservatorio 5, 
                  35122 Padova, Italy    }

	\offprints{C. Chiosi  }
	\date{Received July  1997; accepted }
	\maketitle
	\markboth{IMF: new scenario }{}
%%%%%%%%%%%%%%%%%%%%%%%%%%%%%%%%%%%%%%%%%%%%%%%%%%%%%%%%%%%%%%%%%%%%%%%

\begin{abstract}
In this paper, basic observational properties of elliptical galaxies such as
the integrated spectra, the chemical abundances and the enhancement of
$\alpha$-elements inferred from broad-band colors and line strength indices
$Mg_2$ and $\langle Fe \rangle$ (and their gradients), the color-magnitude
relation, the UV fluxes, and the mass-luminosity ratios, are examined in the
light of current theoretical interpretations, and attention is called on
several points of internal contradiction. Indeed existing models for the
formation and evolution of elliptical galaxies are not able to 
simultaneously account for all
of the above observational features. Specifically, in the
context of standard star formation in the galactic-wind driven models, that
are at the base of present-day understanding of the color-magnitude relation
and UV fluxes, it is difficult to explain the slope of the $M/L_B$ versus
$M_B$ relation (tilt of the Fundamental Plane) and enhancement of the
$\alpha$-elements in the brightest elliptical galaxies. We suggest that the
new initial mass function (IMF) by Padoan et al. (1997), which depends on the
temperature, density, and velocity dispersion of the medium in which stars are
formed, may constitute a break-through in the difficulties in question. Models
of elliptical galaxies incorporating the new IMF (varying with time and
position inside a galaxy) are presented and discussed at some extent. In
brief, in a hot, rarefied medium the new IMF is  more skewed toward the high
mass end than in a cool, dense medium, a situation which is met passing from
low to  high mass galaxies or from the center to the external regions of a
galaxy. As a result of the changing IMF, the enhancement of $\alpha$-elements
and tilt of the Fundamental Plane are easily explained leaving unaltered the
interpretation of the remaining properties under examination. Finally, some
implications concerning the relative proportions of visible stars, collapsed
remnants (baryonic dark matter), and gas left over by the star forming process
are examined.

\end{abstract}

%%%%%%%%%%%%%%%%%%%%%%%%%%%%%%%%%%%%%%%%%%%%%%%%
\section{Introduction}

The conventional picture of star formation in elliptical galaxies is one in
which the galaxies and their stellar content formed early in the universe and
have evolved quiescently ever since. This view is supported by the apparent
uniformity of ellipticals in photometric and chemical appearance (cf.
Matteucci 1997 for a recent review) and the existence of scaling relations,
e.g. the fundamental plane (cf. Bender 1997). In contrast, the close scrutiny
of nearby ellipticals makes evident a large variety of morphological and
kinematic peculiarities and occurrence of star formation in a recent  past
(Schweizer et al. 1990, Schweizer \& Seitzer 1992, Longhetti  et al. 1997a,b).
All this leads to a different picture in which elliptical galaxies are formed
by mergers and/or accretion of smaller units over a time scale comparable to
the Hubble time. Furthermore, strong evolution in the population of early type
galaxies has been reported by Kauffmann et al. (1996) which has been
considered to support the hierarchical galaxy formation model (Kauffmann et
al. 1993, Baugh et al. 1996).

Tracing back the formation mechanism of elliptical galaxies from the bulk of
their chemo-spectro-photometric properties is a cumbersome affair because
studies of stellar populations in integrated light reveal only luminosity
weighted ages and metallicities, and are ultimately unable to distinguish
between episodic (perhaps recurrent) and monolithic histories of star
formation and between star formation histories in isolation or interaction. In
any case, there are a number of primary observational constraints that should
be met by any model of galaxy formation and evolution (they will be briefly
summarized below). As nowadays most properties of elliptical galaxies have
been studied with sophisticated chemo-spectro-photometric models in which the
dynamical process of galaxy formation is reduced to assuming either the closed
box or infall scheme and a suitable law of star formation (Arimoto \& Yoshii
1987, 1989; Bruzual \& Charlot 1993; Bressan et al. 1994; Worthey 1994; Einsel
et al. 1995; Tantalo et al. 1996; Bressan et al. 1996; Gibson 1996a,b; Gibson
1997; Gibson \& Matteucci 1997; Tantalo et al. 1997a, TCBMP97). In contrast, the highly
sophisticated dynamical models of galaxy formation (Davis et al. 1992; Navarro
\& Steinmetz 1966; Haehnelt et al. 1996a,b; Katz 1992;  Katz \& Gunn 1991;
Katz et al. 1996; Steinmetz \& Mueller 1995; Navarro et al. 1996; Steinmetz
1996a,b and references), owing to the complexity of the problem, are still
somewhat unable to make detailed predictions about the
chemo-spectro-photometric properties of the galaxies in question. Attempts to
bridge the two aspects of the same problem are by Theis et al. (1992 and
references) and most recently  Contardo, Steinmetz \& Fritze-von Alvensleben
(1997 in preparation) and Carraro et al. (1997). However, even in the case of
the chemo-spectro-photometric approach, extant models, which ultimately stem
from the classical galactic wind driven model of Larson (1974), often come to
solutions that are mutually discordant, when trying to explain one or another
of the properties of elliptical galaxies. In this study  we would like to
propose a coherent scenario in which a new IMF plays an important role and 
proves to be able to remove some of the points of contradiction.

The plan of the paper is as follows. Section 2 shortly reviews  the basic
observational data and their current interpretation. Section 3 presents the
new IMF proposed by Padoan et al. (1997), which is a function of the
temperature, density, and velocity dispersion of the interstellar medium in
which stars are formed. Section 4  describes our prescriptions for models of
elliptical galaxies, in which the new IMF is an essential ingredient. Section
5 discusses at some extent the heating-cooling balance of the gas that is
required to derive temperature, density, and velocity dispersion of the IMF.
Section 6  presents the reference models and discusses some of their general
properties. Section 7 analyzes in detail the  model results concerning the
history of star formation and chemical enrichment, the fractionary masses of
visible and remnant stars, and the variation of the IMF with age and  galactic
mass. Section 8 presents the broad-band colors and magnitudes and the mass to
light ratios of the models. In particular, it presents our fits of the
color-magnitude relation and fundamental plane. Finally, a summary and some
concluding, critical  remarks are given in Section 9.

%%%%%%%%%%%%%%%%%%%%%%%%%%%%%%%%%%%%%%%%%%%%%%%%%%%
\section{Learning from observations}

\subsection{The G-Dwarf Analog}
As first noticed by Bressan et al. (1994, BCF94), the near UV spectrum (2000
\AA\ to 3500 \AA) of elliptical galaxies shows that the relative percentage of
old metal-poor stars is small, thus indicating that the metallicity partition
function $N(Z)$, i.e. number of stars per metallicity bin, cannot be the one
predicted by the closed-box scheme. In analogy with the G-dwarf problem in the
solar vicinity (cf. Tinsley 1980), whose solution is found in models with
infall (Lynden-Bell 1975), Tantalo et al. (1996, TCBF96) show that also in the
case of elliptical galaxies the infall scheme removes the above discrepancy.
Similar conclusion is reached by Greggio (1996), who however favors the
prompt enrichment alternative.

\subsection{The Chemical Abundances}
Abundances in elliptical galaxies up to now have been deduced either from
integrated colors, spectra, or line-strength indices (Carollo et al. 1993,
Carollo \& Danziger 1994a,b, Davies et al. 1993; Schombert et al. 1993)
together with their spatial gradients.  In general what is measured is a
complicated result of age and metallicity, so that disentangling age from
metallicity effects is the primary task of any study aimed at evaluating the
metallicity of these galaxies (cf. the recent review by Matteucci 1997).
Metallicity indicators such as $ Mg_2$ and   $\langle Fe \rangle$ are the most 
used to
derive information on metallicity. Although passing from the line strength
indices $ Mg_2$ and  $\langle Fe \rangle$  to chemical abundances is not a straight
process (cf. Tantalo et al. 1997b, TBC97), arguments are given to conclude
that the mean  $ [Mg/Fe]$ ratio  exceeds that of the most metal-rich stars
in the solar vicinity by about 0.2-0.3 dex (enhancement of $\alpha$-elements;
O, Mg, Si, etc.. with respect to Fe), and the ratio $ [Mg/Fe]$  increases
with the galactic mass up to this value (cf. Matteucci 1994, 1997). The
enhancement of $\alpha$-elements is particularly demanding as it implies that
a unique source of nucleosynthesis has been contributing to chemical
enrichment (type II SN from massive stars, the main  producers of Mg). It
follows from this that in the standard star formation scenario, i.e. with
constant IMF, the maximum duration of the star forming activity should be
inversely proportional to the galaxy mass ($\Delta t_{SF} \propto M_G^{-1}$).

\subsection {The Color-Magnitude Relation}
Elliptical galaxies obey a mean color-magnitude relation (CMR): colors get
redder at increasing luminosity, cf. Bower et al. (1992) for galaxies in the
Virgo and Coma clusters and Schweizer \& Seitzer (1992) for galaxies in small
groups and field. The cluster CMR is particularly narrow, whereas the field
CMR is more disperse. Long ago Larson (1974) postulated that the present-day
CMR is  the consequence of SN-driven galactic winds.  In the classical
scenario, massive galaxies eject their gaseous content much later and get
higher mean metallicities than less massive ones. This implies that $\Delta
t_{SF} \propto  M_G$, contrary to the trend required by the
$\alpha$-enhancement problem. The tightness of the cluster CMR (in the U-V
color) suggests that most galaxies are nearly coeval with ages ranging from 13
to 15 Gyr (Bower et al. 1992). The CMR for field galaxies is likely compatible
with more recent episodes of star formation, perhaps interactions spread over
several Gyrs (Schweizer \& Seitzer 1992, Longhetti et al. 1997a,b). Finally,
the CMR is a  mass-metallicity sequence and not an age sequence along which
bluer galaxies are significantly younger than red galaxies (cf. BCF94, TCBF96,
TCBMP97, and Kodama \& Arimoto 1996).

\subsection{The Age-Metallicity Dilemma}
The stellar content of a galaxy gets redder  both at increasing age and metallicity
thus giving rise to the well known  age-metallicity dilemma. The \Hbeta\ and
\MgFe\ indices are particularly suited to cast light on their separate
effects, because \Hbeta\ is a measure of the turn-off color and luminosity, 
and age in turn, whereas   \MgFe\ is more sensitive to the RGB color and hence
metallicity. The analysis of existing data (Gonzales 1993) leads to the
following provisional conclusions: (i) most galaxies seem to possess nearly
identical chemical structures, i.e. high mean metallicities and narrow range
of metallicities (Bressan et al. 1996, BCT96; Greggio 1996); (ii) galaxies do
not distribute along the locus expected from the CMR of coeval old objects (BCT96). In
contrast, they seem  to follow a sequence of about constant metallicity and
varying age  so that  recent episodes of star formation (bursts) have
been suggested. The major  drawback  with this idea is that a sort of
synchronization is required. Looking at the  difference in \Hbeta\ and \MgFe\ 
between the nuclear region (i.e. {\it Re}/8 : N) and the whole galaxy (i.e.
{\it Re}/2:  W), and translating   $\Delta$\Hbeta$_{NW}$ and 
$\Delta$\MgFe$_{NW}$ into $\Delta t_{NW}$ and  $\Delta Z_{NW}$,  i.e. in  age
and metallicity difference, respectively, BCT96 pointed out that in most
galaxies the nucleus is  younger and more metal-rich than the external
regions, and suggested that $\Delta t_{NW} \simeq  \Delta t_{SF} \propto 1/
\Sigma \propto 1/ M_G$.

\subsection{The Ultraviolet Excess}
All studied elliptical galaxies have detectable UV flux short-ward of about
2000 \AA\ (Burstein et al. 1988) with large variations from galaxy to galaxy.
The intensity of the UV emission as measured by the (1550-V) color correlates
with the $\rm Mg_2$ index, the central velocity dispersion $\Sigma_c$, and the
luminosity (mass) of the galaxy. Finally, the HUT observations by Ferguson et
al. (1991) and Ferguson \& Davidsen (1993) of the UV excess in the bulge of
M31, in which a drop-off short-ward of about  1000 \AA\ is detected, indicate
that the temperature of the emitting source must be about 25,000 K. Only a
small percentage of the $912 \leq \lambda \leq 1200 $ \AA\ flux can be coming
from stars hotter than 30,000 K and cooler than 20,000 K.

Most likely, the UV excess owes its origin to an old component that gets hot
enough to power the ISED of a galaxy in the far UV regions. Several possible
candidates are envisaged (cf. Greggio \& Renzini 1990, BCF94 and TCBF96). The
appearance of the various types of UV sources is governed by several important
physical factors, each of which is affected by a certain degree of uncertainty
still far from being fully assessed. These are the efficiency of mass loss
during the RGB and AGB phases, the enrichment law $\Delta Y/\Delta Z$ and
finally, for the specific case of P-AGB stars the detailed relation between
the initial and final mass of the stars at the end of the AGB phase. Popular
sources of UV radiation are:

\noindent
(1) the classical post asymptotic giant branch (P-AGB) stars (see Bruzual \&
Charlot 1993, Charlot \& Bruzual 1991), which are always present in the
stellar mix of a galaxy. They cannot, however, be the sole source of UV flux
because of their high mean temperature (about 100,000 K) and lack of
sufficient fuel (cf. Greggio \& Renzini 1990). Furthermore, they hardly
explain the correlation with $Mg_2$ and $\Sigma_c$.

\noindent
(2) Very blue HB (VB-HB) stars of extremely low metallicity (Lee 1994). These
stars have effective temperatures  hotter than about 15,000 K but much cooler
than those of the P-AGB stars. Therefore, depending on their actual $\rm
T_{eff}$, they can generate ISEDs in agreement with the observational data
provided that the age is let vary from galaxy to galaxy (Park \& Lee 1997). It
must be checked, however, whether the percentage  of low metallicity stars is
compatible with the G-Dwarf problem above.

\noindent
(3) The H-HB and AGB-manqu\'e stars of high metallicity (say $Z\!\!>\!\!0.07$) which are expected to be present albeit in small percentages
in the stellar content of bulges and elliptical galaxies in general (Bertelli
et al. 1995). Indeed, these stars have effective temperatures in the right
interval and generate ISEDs whose intensity drops short-ward of  about 1000
\AA\ by the amount indicated by the observational data. With normal mass loss
and  $\Delta Y/ \Delta Z = 2.5$ (Pagel et al. 1992), the first  H-HB and
AGB manqu\'e stars occurs at the age of about 5.6 Gyr. This age is lowered if
$\Delta Y/ \Delta Z$ is higher than 2.5.

\noindent
(4) Finally, the analog of the above H-HB and AGB-manqu\'e stars, but
generated by enhancing the mass loss rate during the RGB phase at increasing
metallicity. These type of stars have been named by Dorman et al. (1993, 1995)
extremely hot HB objects (E-HB). They share nearly the same properties of the
H-HB and AGB-manqu\'e stars. The main difficulty with this option is the
uncertainty concerning the metallicity dependence of the mass loss rate (cf.
Carraro et al. 1996).

The most probable channels, i.e. items (3) and (4) above, require a suitable
metallicity partition $N(Z)$, in which high metallicity bins are populated (a
few percent of the total are fully adequate), provide a simple explanation for
the dependence of the (1550-V) color on $ Mg_2$ (metallicity ?), velocity
dispersion $\Sigma_c$, and luminosity of the parent galaxy, but in turn pose
strong constraints on the past history of star formation and chemical
enrichment so that the right $N(Z)$ is generated.

\subsection{The Iron Discrepancy}
Elliptical galaxies have extended haloes of gas emitting in the X-ray band.
Other indications of the existence of such haloes come from HI kinematics,
planetary nebulae, gravitational lensing, and dynamical studies (cf. Carollo
et al. 1995 and references). Measurements of the iron content in the X-ray
emitting gas (hot inter-stellar medium, H-ISM) from the  ASCA data  indicate a
very low abundance, much lower (some time up a factor of 10) than the iron
content of the stellar populations in these galaxies as derived from
integrated optical spectra (Arimoto et al. 1997). Several causes are
thoroughly examined by Arimoto et al. (1997) going from supernova enrichment,
chemical evolution of galaxies and clusters of galaxies, gas flows, heating of
intra-cluster medium to reliability of the diagnostic used to derive iron
abundances from X-ray observations. The problem is still unsettled.

\subsection{Oxygen over-abundance and iron-luminosity relation of ICM}
X-ray observations of the hot intra-cluster medium (ICM)  of galaxies by
Mushotzsky (1994) show an oxygen overabundance relative to iron of [O/Fe]=0.1
$\div$ 0.7. This result is an important constraint to the past chemical
history of elliptical galaxies, since thought to be the major contributors to
the enrichment of the ICM. In brief, galactic winds are the favored
mechanism to eject metal-rich gas (Matteucci \& Vettolani 1988) provided that
a flatter IMF than the Salpeter case is used to account for the amount of Fe
and the [O/Fe] ratio at the same time (David et al 1991; Matteucci  \& Gibson
1995). Bimodal star formation (Elbaz et al. 1995), while explaining the
abundance ratios, is unable to match the CMR (Gibson 1996b). All these models,
 while recovering the total ICM iron mass, could only  account for part of
the total mass of gas measured in the X-ray band (Arnaud 1994). Following a
suggestion by Trentham (1994) that the precursors of the dwarf galaxies
populating the faint end of the luminosity function may in fact be the source
for 100\% of the ICM gas,  Gibson \& Matteucci (1997) show that the bulk of
the ICM gas cannot originate within the dwarf precursors unless the pre-wind
binding energy was 4-5 times lower than the present day values. Finally, the
ICM iron mass increases as a function of the cluster optical luminosity
(E+S0 galaxies) such that $M_{Fe}^{ICM} \simeq 0.02 L_V$ (Arnaud et al.
1992).

\subsection{The Fundamental Plane}
Elliptical galaxies do not populate uniformly the parameter space with central
velocity dispersion $\Sigma_c$, effective radius $R_e$, and surface brightness
$I_e$ as coordinates. They cluster around a plane called the {\it Fundamental
Plane}. Using the coordinate system defined by Bender et al. (1992) and the
identities

\begin{equation}
         L = c_1 I_e R_e^2
\end{equation}

\begin{equation}
         M = c_2 \Sigma_{c}^2 R_e
\end{equation}

\noindent
with $c_1$ and $c_2$, suitable constants, the physical coordinates are
translated into

\begin{equation}
k_1 =  { 1 \over \sqrt { 2} }  log[{ M \over c_2 }]
\end{equation}

\begin{equation}
  k_2  = { 1 \over \sqrt { 6} } log [{c_1 \over c_2} ({M \over L }) Ie^3] 
\end{equation}

\begin{equation}
k_3  =  {1 \over \sqrt{3} } log[{c_1 \over c_2 } ( {M \over L })].
\end{equation}
  
Of particular relevance is the projection of the FP  onto the $k_1-k_3$ plane,
 where the FP is seen edge on. Limited to the case of the Virgo elliptical
galaxies to avoid distance uncertainties, the relation $k_3=0.15 k_1+0.36$
with $\sigma(k_3)=0.05$ is found to hold (cf. Ciotti et al. 1996 for details).
The ratio $(M/L)$  increases with galaxy mass (tilt of the FP).

Unfortunately, current models of elliptical galaxies that fairly match other
key properties of these systems are still unable to account for the tilt of
the FP. Specifically, under standard assumptions for the IMF and star
formation rate, at any given age the models predict $M/L_B$ ratios that are
nearly constant at increasing luminosity (mass) of the galaxy, and that
increase with the age at fixed galactic mass. See for instance the $M/L_B$
versus $L_B$ relations of the BCT96  and TCBF96 models. An easy way out of the
problem would be that low mass galaxies are truly young objects (in the sense
that dominant star formation started much later than in less massive ones).
For the arguments discussed above, this alternative seems to be unlikely.

Renzini \& Ciotti (1993) investigated two possible origins of the tilt: a
systematic variation of the IMF (of power-law type) and a trend in the
relative proportions and distributions of bright and dark matter. The
conclusion is that in both cases, in order to explain the tilt and tightness
of the FP at the same time, major changes and fine tuning are required. In
brief, the IMF or relative bright/dark matter distribution should change along
the FP, but at every position a small dispersion  in the IMF or relative
bright/dark matter distribution are required to preserve the tightness. In a
subsequent paper along the same vein, Ciotti et al. (1996) looked at various
effects of structural and dynamical nature (such as orbital radial anisotropy,
relative bright/dark matter distributions, shape of the light profiles) under
the assumption of a constant stellar mass to light ratio. While anisotropy
gives a marginal effect, variations in the bright/dark matter distributions
and/or shape of the light profiles can produce the tilt. Also in this case
fine tuning is however required to preserve the tightness of the FP.

\subsection {How to reconcile things ?}
It is evident that up to now no coherent explanation is available for the
pattern of properties of elliptical galaxies. Most striking is that,  while
the gradients in \Hbeta\ and \MgFe\ and trend for the ratio $[Mg/Fe]$  with
galaxy mass (Davies et al. 1993; Matteucci 1994) suggest that the duration of
the star formation activity ought to be shorter in massive galaxies than in
low mass ones, the opposite should occur to account for the slope of the CMR.

In the contest of convectional picture of star formation in elliptical
galaxies (formation and evolution in isolation), we address the question
whether relaxing some of the standard assumptions a viable solution can be
found. As already recalled, among the various suggestions advanced to explain
the FP tilt,  Renzini \& Ciotti (1993) explored the effect of varying the IMF
and concluded that major but suitable change of the IMF are required. Padoan
et al. (1997) have proposed a new type of IMF that may turn out to be able to
reconcile the above points of difficulty.

%%%%%%%%%%%%%%%%%%%%%%%%%%%%%%
\section{A universal IMF}

\begin{figure}
%\picplace{8.0cm}
%\psfig{file=blank.ps,height=8.0truecm,width=8.5truecm}
%\psfig{file=imf.ps,height=8.0truecm,width=8.5truecm}
\psfig{file=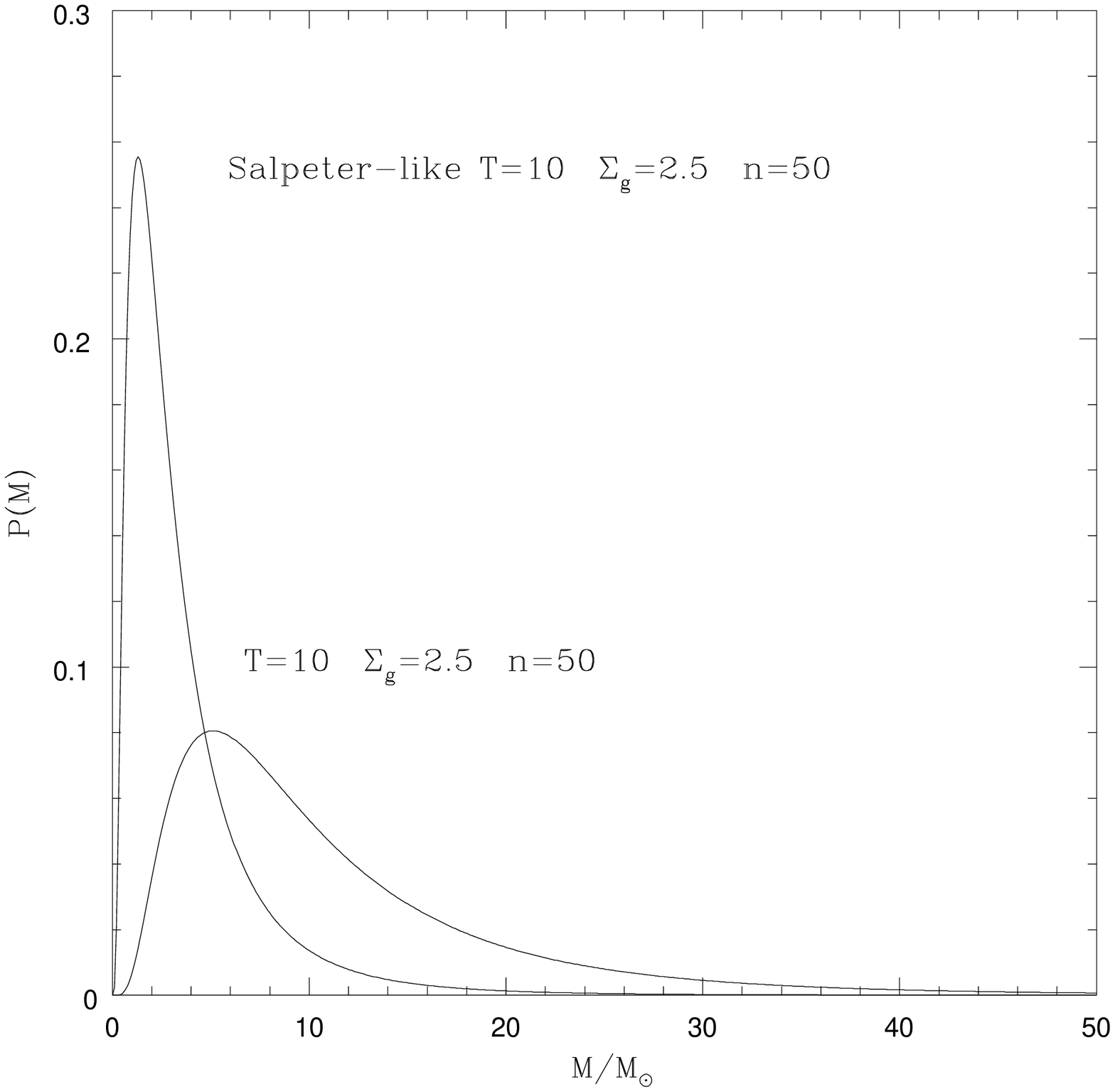,height=8.0truecm,width=8.5truecm}
\caption{The IMF of Padoan et al. (1997) for two different values of the
temperature (10 and 30 K) and constant density ($n=50 ~cm^{-1}$) and velocity
dispersion $\Sigma_g=2.5~km~s^{-1}$. Similar curves are possible at varying
density and velocity dispersion. Note how the IMF may change from a
Salpeter-like law (very small $M_{P}$) to cases very different from the
classical behavior.}
\label{imf}
\end{figure}

Padoan's et al. (1997) IMF stems from a statistical description of the density
field emerging from randomly forced supersonic flows in star forming regions.
Hydrodynamical simulations of supersonic flows indicate that very large
density contrasts can develop with log-normal probability distribution. The
resulting IMF is

\begin{displaymath}
\int_{0}^{\infty}\phi(M)dM = ~~~~~~~~~~~~~~~~~~~~~~~~
\end{displaymath}
\begin{equation}
  \int_{0}^{\infty}
                {2 B^2 \over (2\pi\sigma^{2})^{0.5} }  M^{-3} 
              exp [ -0.5 ({ 2 ln M - A \over \sigma} )^2 ] dM = 1.
\label{eq_1}
\end{equation}
The quantities $A$, $B$ and $\sigma$ are defined by the following relations:

\begin{equation}
A= 2\ln B + 0.5 \sigma^2,  
\end{equation}

\begin{equation}
B =1.2 \times (T/10)^{1.5} \times (n/1000)^{-0.5}, 
\end{equation}

\begin{equation}
\sigma^2= \ln [1 + 0.36 ({\mathcal{M}}^2 - 1)], 
\end{equation}

\noindent
in which $\sigma$ is the standard deviation of  the number density
distribution in the field with respect to the mean. $\mathcal{M}$ is the Mach
number given by  ${\mathcal{M}}^2=(\Sigma_g/v_s)^2$, $T$ is the temperature in
K, $n$ is the number density in $cm^{-3}$, $v_s$ is the sound velocity, and
$\Sigma_g$ (in km/s) is  the velocity dispersion of the star forming gas. This
IMF has a long tail at high masses, an exponential cutoff at the smallest
masses, a characteristic peak mass

\begin{equation}
M_{P} \simeq 0.2 M_{\odot} \times (T/10)^2 \times (n/1000)^{-0.5} 
               \times (\Sigma_g/2.5)^{-1},
\label{eq_2}
\end{equation}
and a slope continuously varying with the  mass. The IMF gets flatter and
$M_{P}$  gets higher at increasing  temperature or decreasing  density and
velocity dispersion. The opposite occurs reversing the variations of the three
physical parameters. The response of the IMF to variations in temperature  is
shown in Fig.~\ref{imf} for values of density and velocity dispersion
$\Sigma_g$ typical of molecular clouds, i.e. $n=50$  particles per cm$^3$, and
$\Sigma_g=2.5$ km~s$^{-1}$. See Padoan et al. (1997) for all other details.

The IMF is analytically integrable. Performing the change of variable $M=e^t$ 
and writing

\begin{displaymath}
\alpha   = {2 B^2 \over (2\pi)^{0.5} \sigma };~~~~~~~~
\beta    = 2;~~~~~~~~ 
\gamma   = {2 \over {\sigma}^2 };~~~~~~~~
\delta   = {A \over 2}
\end{displaymath}

\noindent
one gets

\begin{equation}
\int_{o}^{\infty} \phi(M) dM = \int_{-\infty}^{\infty}
         \alpha e^{-\beta t} e^{-[-\gamma (t-\delta)^2]} dt
\label{eq_3}
\end{equation}
\noindent
whose general solution is

\begin{equation}
\int_{-\infty}^{t^{*}}\phi(t)dt = 0.5 [1. +  Erf(x)]
\label{eq_4}
\end{equation}

\noindent
with 
\begin{equation}
  x= {\beta -2 \gamma \delta + 2 \gamma t^{*} \over 2 (\gamma)^{0.5} }
\label{eq_5}
\end{equation}
\noindent
and
\begin{equation}
Erf(x) = {2 \over \sqrt{\pi} } \int_{0}^{x} exp(-z^2)dz.
\label{erf}
\end{equation}

%%%%%%%%%%%%%%%%Figure 1 %%%%%%%%%%%%%%%%%%%%%%%%%%%%%%%%%%%%%%%%%%%%
%%%%%%%%%%%%%%%%%%%%%%%%%%%%%%%%%%%%%%%%%%%%%%%%%%%%%%%%%%%%%%%%%%%%

%%%%%%%%%%%%%%%%%%%%%%%%%%%%%%%%%%%%%%%%%%
\section{Modelling elliptical galaxies}
The new IMF has been implemented in a simplified version of the 
TCBMP97 models of elliptical galaxies in which spatial gradients of 
mass density (baryons and dark matter) and star formation are taken into
account. For the sake of simplicity and better understanding of the role
played by the sole IMF, we adopt  here the closed-box formulation (no infall
of primordial gas).

\subsection{Sketch of the models}
Elliptical galaxies are modelled as a spherical distribution of baryonic
material with total mass $M_{L,T}$  (originally in form of gas) embedded in a
large halo of dark matter with  total mass $M_{D,T}$. Both components  have
suitable radial distributions. For the purposes of this study, we will limit
ourselves to consider only the region of the galaxy inside the effective
radius $R_e$, treated as single density-averaged zone. As far as dark matter
is concerned this is included assuming that the ratios of dark to luminous
mass  and radius are $ M_{D,T}/M_{L,T} = \zeta = 5$ and $ R_D/R_L =
\zeta = 5$ (cf. Bertin et al. 1992; Saglia et al. 1992). Other values of
$\zeta$ are possible, whose net effect is to change the gravitational
potential.

\subsection{The basic notation}

Let $\rho_{L}(r)$ be the density distribution of luminous matter, and
${\overline \rho}_{L}(r)$ its mean value within the sphere of radius $r$.
Originally this material is in form of gas and it is gradually turned into
stars so that at each time ${\overline \rho}_{L}(r,t)={\overline
\rho}_{s}(r,t) + {\overline \rho}_{g}(r,t)$, with obvious meaning of the
symbols. The masses of luminous material, gas and stars within the sphere of 
radius $r$ are simply given by

\begin{displaymath}
 ~~~~~~~~~~M_L(r) =  {\overline \rho}_{L}(r) \times V(r) 
\end{displaymath}

\begin{displaymath}
 ~~~~~~~~~~M_g(r,t) =  {\overline \rho}_{g}(r,t) \times V(r) 
\end{displaymath}

\begin{displaymath}
 ~~~~~~~~~~M_s(r,t) =  {\overline \rho}_{s}(r,t) \times V(r) 
\end{displaymath}
\noindent
where $V(r)$ is the volume of the sphere, supposedly constant in time.

\noindent
In the following galaxies are identified by their total luminous mass
 $M_{L,T}$, whereas the luminous mass within  the $R_e$-sphere is named 
$M_{L,R_e}$. All masses are in units of $10^{12}\times M_{\odot}$.

\noindent
Limited to the $R_e$-sphere, let us now define the a-dimensional variables 

\begin{displaymath}
G_g(r,t) = {M_g(r,t) \over M_{L,R_e} } =  
   { { \overline \rho}_{g}(r,t)  \over  {\overline \rho}_{L,R_e}} 
\end{displaymath}

\begin{displaymath}
G_s(r,t) = {M_s(r,t) \over M_{L,R_e} } =  
   { { \overline \rho}_{s}(r,t)  \over  {\overline \rho}_{L,R_e}} 
\end{displaymath}

\noindent
where ${\overline \rho}_{L,R_e}$ is the mean density of luminous mass within
it.

\noindent
Finally, we introduce the gas components $G_{g,i}(r,t)= G_g(r,t)\times
X_i(r,t)$ where $X_i(r,t)$ are the abundances by mass of the elemental species
$i$. Summation of  $X_i(r,t)$ over all components is equal to unity.

\subsection{The spatial distribution of $M_{L,T}$ and $M_{D,T}$}

{\it Luminous matter}. 
The spatial distribution of the luminous matter is supposed to follow the Young
(1976) density profile based on the $r^{1/4}$-law. Given the effective radius
of a galaxy, one can immediately derive from the tabulations of Young (1976)
the density $\rho_{L}(r)$, the mean density ${\overline \rho}_{L}(r)$ within
the sphere of radius $r$, and the  gravitational potential $\varphi_L(r)$.
\littleskip

{\it Dark matter}.
The mass distribution and gravitational potential of the dark matter are
derived from the density profiles by Bertin et al. (1992) and Saglia et al.
(1992) adapted to the Young formalism for the sake of internal consistency.
The details of this procedure are given in TCBMP97. In brief we start from the
density law

\begin{equation}
\rho_{D} (r) = \frac{\rho_{D,0} \times  
                 r_{D,0}^{4}}{(r_{D,0}^{2} + r^{2})^{2}}
\end{equation}
\noindent
where $r_{D,0}$ and $\rho_{D,0}$ are two scale factors of the distribution.

We assume $r_{D,0}= \frac{1}{2} R_{D,e}$, where $R_{D,e}$ is the effective
radius of dark matter. Furthermore, for this latter we assume that it scales
according to the same law as for the total radii of dark and luminous matter,
i.e. $R_{D,e} = \frac{\zeta}{2} R_{e} = \frac {5}{2} R_e$.

The density scale factor $\rho_{D,0}$ is derived from imposing that the total
mass of dark-matter $M_{D,T} = \zeta M_{L,T}$, where $M_{D,T}$ is given by:

\begin{equation}
M_{D,T} = 4\pi \int_{0}^{\infty} r^{2}\rho(r) dr = 
4\pi \rho_{D,0} r_{D,0}^{3} m(\infty)
\label{ddm}
\end{equation}

\noindent
with 
\begin{equation}
m(\infty) = \int_{0}^{\infty}
\frac{r^{2}}{r_{D,0}^{3}\left(1+\left(\frac{r}{r_{D,0}}\right)^{2}\right)^{2}}dr
\label{int}
\end{equation}
\noindent
for which a numerical integration is required. Finally, the density profile of
dark-matter is

\begin{equation}
\rho_{D}(r) = \frac{M_{D,T}}{m(\infty)}\frac{1}{4\pi r_{D,0}^{3}}
\frac{1}{\left(1+\left(\frac{r}{r_{D,0}}\right)^{2}\right)^{2}} .
\label{dend}
\end{equation}

The radial dependence of the dark-matter gravitational potential is

\begin{equation}
\varphi_{D}(r) = - G \int_{0}^{r} \frac{M_{D}(r)}{r^{2}} dr
\label{poten}
\end{equation}

\noindent
which upon integration becomes

\begin{equation}
\varphi_{D}(r) = - 4\pi G \rho_{D,0} r_{D,0}^{2}
\widetilde{\varphi_{D}}\left(\frac{r}{r_{D,0}}\right)
\label{poten1}
\end{equation}

\noindent
where $\widetilde{\varphi_{D}}\left(\frac{r}{r_{D,0}}\right)$ is given by

\begin{equation}
\int_{0}^{r/r_{D,0}} \frac{m(r/r_{D,0})}{r_{D,0}\left(\frac{r}{r_{D,0}}\right)^{2}}dr
\label{poten2}
\end{equation}
which must be integrated numerically. Finally, the total binding energy of the
gas is

\begin{equation}
\Omega_{g}(r,t)= 
         \overline{\rho}_{g}(r,t) V(r) \varphi_{L}(r) +
         \overline{\rho}_{g}(r,t) V(r) \varphi_{D}(r) .
\label{gas_pot}
\end{equation}

\subsection{The mass-effective radius relation}
To proceed further we need to adopt a suitable relationship between the
effective radius and the total mass of the luminous material. For the purposes
of this study, we derive  from the data of Carollo \& Danziger (1994a,b),
and Carollo et al. (1993, 1995) the following relation

\begin{equation}
       R_e = 17.13 \times M_{L,T}^{0.557} 
\label{re_mass}
\end{equation}
\noindent
with $R_e$ in kpc and $M_{L,T}$ in units of $10 ^{12} M_{\odot}$.

\noindent
Another useful relation is between the total radius and  total mass of the luminous
material for which using the same sources of data we get

\begin{equation}
       R_L = 39.10 \times M_{L,T}^{0.402} 
\end{equation}

\noindent
in the same units as above.

\subsection{The chemical equations}
The chemical evolution of elemental species is governed by the same set of
equations as in TCBF96 and TCBMP97 however adapted to the density formalism and improved
as far as the ejecta and the contribution from the Type Ia and Type II
supernovae are concerned (cf. Portinari et al. 1997).

Specifically, we follow in detail the evolution of the abundance of thirteen
chemical elements ( $H$, $^{4}He$, $^{12}C$, $^{13}C$, $^{14}N$, $^{16}O$,
$^{20}Ne$, $^{24}Mg$, $^{28}Si$, $^{32}S$, $^{40}Ca$, $^{56}Fe$, and the
isotopic neutron-rich elements $nr$ obtained by $\alpha$-capture on $^{14}N$,
specifically $^{18}O$, $^{22}Ne$, $^{25}Mg$). Furthermore, the stellar yields
in usage here take  into account the effects of different initial chemical
compositions (cf. Portinari et al. 1997).

The equations governing the time variation of the $G_{i}(r,t)$ and hence
$X_{i}(r,t)$ are:

\begin{displaymath}
\frac{dG_{i}(r,t)}{dt}= - X_{i}(r,t) \Psi(r,t) + 
\end{displaymath}

\begin{displaymath}
 \int_{M_{min}}^{M_{B_m}}\Psi(t-\tau_M)Q_{M,i}(t-\tau_M)\phi(M)dM +
\end{displaymath}

\begin{displaymath}
A \int_{M_{B_m}}^{M_{B_M} } \phi(M) [\int_{\mu_{min}}^{0.5} f(\mu) 
            \Psi(t-\tau_{M_2})Q_{M,i}(t-\tau_{M_2})d\mu ] +
\end{displaymath}

\begin{displaymath}
(1-A) \int_{M_{B_m} }^{B_M} \Psi(t-\tau_M) Q_{M,i}(t-\tau_M) 
         \phi(M) dM +
\end{displaymath}

\begin{equation}
\int_{M_{B_M} }^{M_{max}} \Psi(t-\tau_M) Q_{M,i}(t-\tau_M) \phi(M) dM
\label{degas_i}
\end{equation}
\noindent
where all the symbols have their usual meaning. Specifically, $\Psi(t)$ is the
normalized rate of star formation, $Q_{M,i}(t)$ are the restitution fractions
of the elements $i$ from stars of mass $M$ (cf. Talbot \& Arnett 1973),
$\phi(M)$ is the IMF, whose lower and upper mass limits are $M_{min}$ and
$M_{max}$ (see below), $\tau_M $ is the lifetime of a star of mass $M$.
Contrary to many other models in literature (cf. Matteucci 1997) we take into
account the important variation of $\tau_M$ with the initial chemical
composition using the tabulations by Bertelli et al. (1994). The various
integrals appearing in eq.(\ref{degas_i}) represent the separated
contributions of Type II and Type Ia supernovae as introduced by Matteucci \&
Greggio (1986). In particular, the second integral stands for all binary
systems having the right properties to become Type Ia supernovae. $M_B$$_{m}$
and $M_B$$_{M}$ are the lower and upper mass limit for the total mass of the
binary system, $f(\mu)$ is the distribution function of their mass ratios and
$\mu_{min}$ is the minimum value of this, finally $A$ is the fraction of such
systems with respect to the total. We adopt $B_{m}=1.5M_{\odot}$,
$B_{M}=12M_{\odot}$, and $A=0.2$. The stellar ejecta are from Marigo et al.
(1996, 1997),  Portinari (1995), and Portinari et al. (1997) to whom we refer
for all details.

\subsection{The star formation rate}
The rate of star formation (SFR) is assumed to depend on the gas density
according to the Schmidt (1959) law (see also Larson 1991):

\begin{equation}
 {d {\overline\rho}_g(r,t) \over dt }=\nu(r,t) 
                             {\overline\rho}_g(r,t)^{\kappa}
\label{eq_6}
\end{equation}
\noindent
where the specific efficiency of star formation 
$\nu(r,t)$ is a suitable function to be specified below. 
All the models we are going to describe are for $\kappa=1$. 

\noindent
Upon normalization, the star formation rate becomes

\begin{equation}
\Psi(r,t)= \nu(r,t) [{ {\overline\rho}_{L,R_e} }]^{k-1} G_g(r,t)^{k} .
\label{sfr1}
\end{equation}

\subsection{Mass limits on the IMF}
Although the IMF of Padoan et al. (1997) is self-normalizing to unity over 
the whole range of masses from zero to infinity (cf. equation \ref{eq_1}), 
we prefer to limit  the mass range of validity
from $M_{min} = 0.07 M_{\odot}$ to $M_{max} = 200 M_{\odot}$, values that 
likely are more suited to real stars, and re-normalize the IMF by imposing
that
 
\begin{equation}
           \int_{M_{min}}^{M_{max}} C \phi(M)dM = 1
\label{eq_7}
\end{equation}
\noindent
where 

\begin{equation}
C = {1 \over 1 - \int_{0}^{M_{min}} \phi(M)dM - \int_{M_{max}}^{\infty}\phi(M)dM  } .
\label{eq_8}
\end{equation}
\noindent

\subsection{The basic energy equation}
In order to calculate the IMF we need to derive the density, temperature and
velocity dispersion $\Sigma_g$ of the collapsing clouds. These quantities are obtained by solving the energy equation

\begin{equation}
 {dE(T,\rho_g,t) \over dt} =  H_{R}(T,\rho_g,t) + H_{M} + H_{C}  
                  - {\Lambda(T,\rho_g,t) \over \rho_g }
\label{eq_9}
\end{equation}

\noindent
where $E(T,\rho_g,t)$ is the energy input per unit mass, $H_{R}$, $H_{M}$ and
$H_{C}$ are the heating rates (per unit mass) of radiative, mechanical, and
collapse origin, respectively, and $\Lambda$ is  the cooling rate (in units of
$\rm ergs~ cm^{-3}~ s^{-1}$).

All physical quantities are  functions of time and position. Therefore, the
IMF continuously varies with time, radial distance, and galaxy mass.  The
coupling of energy equation with those governing the chemical model and the
IMF, in particular, is described in  Section 5 below.

\subsection{Galactic winds}
Depending on the competition between heating and cooling, the energy stored
into the interstellar medium continuously grows thus making gas hotter and
hotter. The case can be met in which either the thermal energy of the gas
eventually exceeds its gravitational binding energy, or equivalently the gas
temperature exceeds the virial temperature, or the amount of gas ionized by th
UV radiation emitted by massive stars (see below) equals the total gas mass.
If any of these conditions is met, star formation is halted. The conditions in
usage are

\begin{equation}
 E_{th}(t) \geq \Omega_g(t)
\label{cond_therm}
\end{equation}

\begin{equation} 
 T_g(t) \geq T_{virial} 
\label{cond_temp}
\end{equation}

\begin{equation}M_{g,ion} \geq M_g
\label{cond_strom}
\end{equation}

\noindent
where $E_{th}(t)$ and $\Omega_g(t)$ are the current total thermal energy and
gravitational binding energy of the gas, respectively, $T_g(t)$ is the current
gas temperature, and $M_{g,ion}$ and $M_g$ are the masses of the ionized gas
and total gas,  respectively.

%%%%%%%%%%%%%%%%%%%%%%%%%%%%%%%%%%%%%%%%%%%%%%%%%%%%%%%%%%%%%%%
\section{The energy equation: details on heating and cooling}

\subsection{Heating}
Heating of the interstellar medium is caused by many processes among which we
consider the thermalization of the energy deposit by supernova explosions
(both Type I and II) and stellar winds from massive stars, the ultraviolet
flux from massive stars,  and finally two sources of mechanical nature.
\littleskip

{\bf Supernovae}. 
The rate of supernova explosions $R_{SNI,II}(t)$ over the time interval
$\Delta t$ is calculated according to standard prescriptions (cf. BCF94,
TCBF96, TCBMP97,  and Greggio \& Renzini 1983 for Type I SN in particular).
Knowing the amount of energy released by each SN explosion, the total  energy
injection over the time interval $\Delta t$ centered at age $t$ is the sum of
the two terms $E_{SNI}$ and  $E_{SNII}$ of type

\begin{equation}
E_{SNI,II} = \int_{\Delta t} \epsilon_{SNI,II}(t-t') R_{SNI,II}(t') dt'
\label{eq_10}
\end{equation}

\noindent
where $R_{SNI,II}(t)$ is the number of supernovae per unit galactic mass and
time, and $\epsilon_{SNI,II}(t)$ is the thermalization law of the energy
released by a SN explosion, in which the effects of cooling are suitably taken
into account, cf. BCF94, TCBF96, TCBMP97, and Gibson (1995).
\littleskip

{\bf Stellar winds}. 
The rate of energy injection by stellar winds is 

\begin{equation}
E_{W} = \int_{\Delta t} \epsilon_{W}(t-t') R_{W}(t') dt'
\label{eq_11}
\end{equation}

\noindent
where $R_{W}$ is the number of stars per unit galactic mass and time expelling
their envelopes during the time interval $\Delta t$ and, in analogy with the
SN remnants, $\epsilon_{W}(t)$ is the thermalization law of the kinetic energy
of stellar winds. A losing mass star is expected to deposit  into the
interstellar medium the energy

\begin{equation}
   \epsilon _{W0} = \eta_W \times 
      { M_{ej}(M) \over 2 } ({ Z \over Z_{\odot} })^{0.75} \times v(M) ^2 
\label{eq_12}
\end{equation}
\noindent
where $M_{ej}(M)$ is the amount of mass ejected by each star of mass $M$,
$v(M)$ is the velocity of the ejected material,  and $\eta_W$  is an
efficiency factor of the order of 0.3 (Gibson 1994 and references therein).
For the velocity $v(M)$ we take the maximum between the terminal velocity of
the wind and the galaxy velocity dispersion $\Sigma$. For the cooling law
$\epsilon_{W}(t)$ one could adopt the same formalism as in BCF94.
\littleskip

{\bf Ultraviolet flux from massive stars.}
The rate of energy injection from the ultraviolet flux emitted by massive
stars, whose mass is the range $10~\div~200~ M_{\odot}$, is

\begin{equation}E_{UV} = \int_{\Delta t} \epsilon_{UV}(t-t') R_{UV}(t') dt'
\label{eq_UV}
\end{equation}
\noindent
where $R_{UV}$ is the number of massive stars per unit galactic mass and time
and  $\epsilon_{UV}$ is the amount of ultraviolet energy emitted by such
stars. To derive a reasonable estimate of the UV flux  we adopt the following
procedure. The problem is simplified considering that significant fluxes of UV
radiation will be emitted only by evolutionary stages at high effective
temperature. Therefore we limit ourselves to stars along the main sequence for
which the relations between  mass, luminosity,  and effective temperature are
known from stellar model calculations (cf. Bertelli et al. 1994). For each
value of the star  mass we integrate the spectral energy distribution up to
1000 \AA\ (our ultraviolet region) and calculate  its ratio to the total
bolometric luminosity, shortly indicated by $F_{UV}$. The UV flux emitted by
each star is therefore

\begin{equation}
 \epsilon_{UV} = \eta_{UV} \times  F_{UV} \times 
                       \langle{ L \over L_{\odot} } \rangle  L_{\odot}
\label{UV}
\end{equation}
 \noindent
where $\langle L/L_{\odot} \rangle$ is the mean luminosity of a massive star
in solar units (the mean is performed over the main sequence phase), and  
$\eta_{UV}$ is an efficiency parameter. The fraction $F_{UV}$ increases from
about 0.1 to 0.7 as the stellar mass increases from 10 to $200M_{\odot}$. The
effects of different chemical composition are neglected here. Furthermore, the
relationship between mass and lifetime of the star is used to calculated the
integral of eq. (\ref{eq_UV}). The major problem with the UV radiation is
whether or not it must be included in the energy balance equation. In presence
of dust, nearly all UV radiation is absorbed and re-emitted in
the far infrared so that it can freely escape from the galaxy; indeed, the
recent models of population synthesis from the far UV to the far infrared by
Granato et al. (1997) show that about 99\% of the UV radiation is re-processed
in presence of even a modest amount of dust. On the basis of the above
considerations, only 0.01\% of $\epsilon_{UV}$ above is injected into the energy
equation and, accordingly, $\eta_{UV}=0.01$. Perhaps the most important use of
the UV radiation is to set a new condition for the occurrence of galactic
winds. Knowing the amount of UV flux we calculate the radius of the Stroemgren
sphere and associated gas mass, in which 99\% of the UV flux is absorbed and
re-emitted in the infrared. When the mass of the Stroemgren sphere equals the 
total gas mass, no further absorption of the UV radiation is possible, so that
the whole UV energy becomes suddenly available to the gas. Many numerical
tests proved that, soon after, the gas gets so hot to exceed the virial
temperature, so that condition (\ref{cond_strom}) applies.
\littleskip

{\bf Remarks on $\epsilon_{SN}$ and $\epsilon_W$}.
According to standard prescriptions for  $\epsilon_{SN}(t)$, the
thermalization time scale of SN energy is short (a few $10^6$ yr) compared to
the typical time scale of chemical and photometric evolution (a few $10^7$ to
$10^8$ yr).  This  means that only supernovae exploding in the recent past
will basically contribute to the current energy injection. This allows us to
simplify the calculation of the energy input  and to treat SN cooling by means
of equation (\ref{eq_9}). The same considerations apply to stellar winds, for
which the main source of energy are massive stars with short lifetimes and
terminal velocities up to 2000 - 3000 km/s. Therefore, we drop the time
dependence both in $\epsilon_{SNI,II}$ and $\epsilon_{W}$. They are set equal
to the initial energy of explosion or wind respectively. Sufficiently small
time intervals $\Delta t$ secure the desired accuracy and the recovery of
results very close to those obtained with the standard treatment.
\littleskip

{\bf Total radiative heating.} 
The total heating rate $H_{R}(T,\rho,t)$ of equation (\ref{eq_9}) is 

\begin{equation}
  H_{R}(T,\rho_g,t)= { E_{SNI} + E_{SNII} + E_{W} + E_{UV} \over \Delta t }  .
\label{eq_13}
\end{equation}
\littleskip

{\bf Mechanical sources}. In addition to this, we include two other sources of
heating whose origin is of mechanical nature (i.e $H_{C}$ and $H_{M}$ to be
defined below).
\littleskip
 
{\it Initial flame}. The first term, $H_{C}$, is related to the very initial 
period of galaxy formation, when gas is collapsing with the infall time
scale to form the galaxy. To evaluate this term in a simple fashion we 
follow Binney \& Tremaine (1987). Let $R_h$ be the initial radius of the 
protogalaxy 
(say $R_h=100 \times R_e$) and $ \tau_{ff}$ the corresponding free-fall 
time scale 

\begin{equation}
\tau_{ff} = \pi  \sqrt { { R_h^3 \over 2 G M  } } \label{tau_ff}
\end{equation}

\noindent
where $M$ is the total mass of the galaxy, i.e. 
$M_{L,T} + M_{D,T}=(1+\zeta)\times M_{L,T}$. 
In absence of energy sources of any type, a proto-galactic cloud at the 
virial temperature would have cooled on a time scale

\begin{equation}\tau_{cool} = 6.3 \times 10^5 (R_h/10 kpc)^2   ~~~~~~~{\rm yr}.
\label{tau_cool}
\end{equation}

\noindent
Expressing $R_h=\lambda R_e$ and using the relation between the effective
radius and the total luminous mass, equation (\ref{re_mass}) above, we get

\begin{equation} {\tau_{ff} \over \tau_{cool}} = 120.7\times    
          (1+\zeta)^{-0.5}\times \lambda^{-0.5}\times M_{L,T}^{-0.78}
\end{equation}
\noindent
with $\lambda \simeq 100$, and $\zeta=5$ (for this particular set of models).
We suppose that at the end of this phase the system has the energy and 
velocity dispersion $\Sigma$ given by

\begin{equation}
E_{vir} = E_{vir,0} (1 - {\tau_{cool} \over \tau_{ff}} )
\end{equation}

\begin{equation}
\Sigma = (1 - {\tau_{cool} \over \tau_{ff}} )^{0.5} \Sigma_0
\end{equation}
\noindent
where $E_{vir,0}$ and $\Sigma_0$ are the initial virial energy and velocity
dispersion, respectively. The  energy left over at the end of this phase is
then converted into heat at the rate (per unit galactic mass and time)

\begin{equation}
H_{C} = { E_{vir} \over \tau_{ff} }  .
\label{heat_vir}
\end{equation}
\noindent
 This source of energy
is supposed to be active as long as the galaxy age is smaller than
$\tau_{ff}$, thus providing the initial gas temperature. The effect of this
energy source on the final result is however of marginal importance, which
somehow justifies the crudeness of the present approach.
\littleskip

{\it Cloud-cloud collisions}.
The second term $H_{M}$ owes its origin to mechanical interactions of
different nature among gas clouds, such as shock waves, friction, cosmic rays,
etc. In order to get an estimate of this source of energy from simple
arguments, we have looked at the cloud-cloud collisions. Limiting to the case
of head-on collisions, the kinetic energy per unit mass of the relative motion
is $\Sigma^2$ ( galaxy velocity dispersion), whereas the number of such
collisions per unit time can be simply expressed by $1/t_{ff}$, where $t_{ff}$ is the current free-fall time scale. Therefore the
rate of  energy injection is cast as follows

\begin{equation}
  H_{M} = \eta_M \times   { \Sigma^2 \over t_{ff}  }
\end{equation}

\noindent
where $\eta_M$ is an efficiency parameter masking the real complexity of the
energy injection by mechanical interactions. Preliminary test calculations
show that the parameter $\eta_M$ ought to be of the order of $10^{-6}$. This
source of energy actually plays  the dominant role in the models we are going
to describe because  it will turn out to drive the minimum temperature
attainable by the collapsing gas clouds (see below).
\littleskip

{\bf The CBR limit}.
Although not explicitly included in equation (\ref{eq_9}), there is another
source of heating to be considered, i.e. the cosmic background radiation
(CBR). It is used to limit the lowest temperature attainable by gas as a
consequence of cooling. Specifically, we assume that at each time the gas
temperature cannot be cooler than the limit $T_{CBR}$ set  by

\begin{equation}
         T_{CBR}[z(t)] = {273 \over 100} \times [1 + z(t)]
\end{equation}

\noindent
where $z(t)$ is the red-shift corresponding to the  galactic age $t$. The
red-shift $z(t)$ depends on the choice of the cosmological model of the
Universe, i.e. $H_0$,  $q_0$ (Hubble constant and deceleration parameter,
respectively, in a Friedmann model), and red-shift of galaxy formation
$z_{for}$. All of the results below are for the  standard case of $H_0 = 50 $
$ \rm km~s^{-1}~Mpc^{-1}$, $q_0 = 0$, and $z_{for} = 5$.

%%%%%%%%%%%%%%%%Figure 2   %%%%%%%%%%%%%%%%%%%%%%%%%%%%%%%%%%%%%%%%%%%% 
\begin{figure}
%\picplace{8.0cm}
%\psfig{file=blank.ps,height=8.0truecm,width=8.5truecm}
%\psfig{file=cool_rate.ps,height=8.0truecm,width=8.5truecm}
\psfig{file=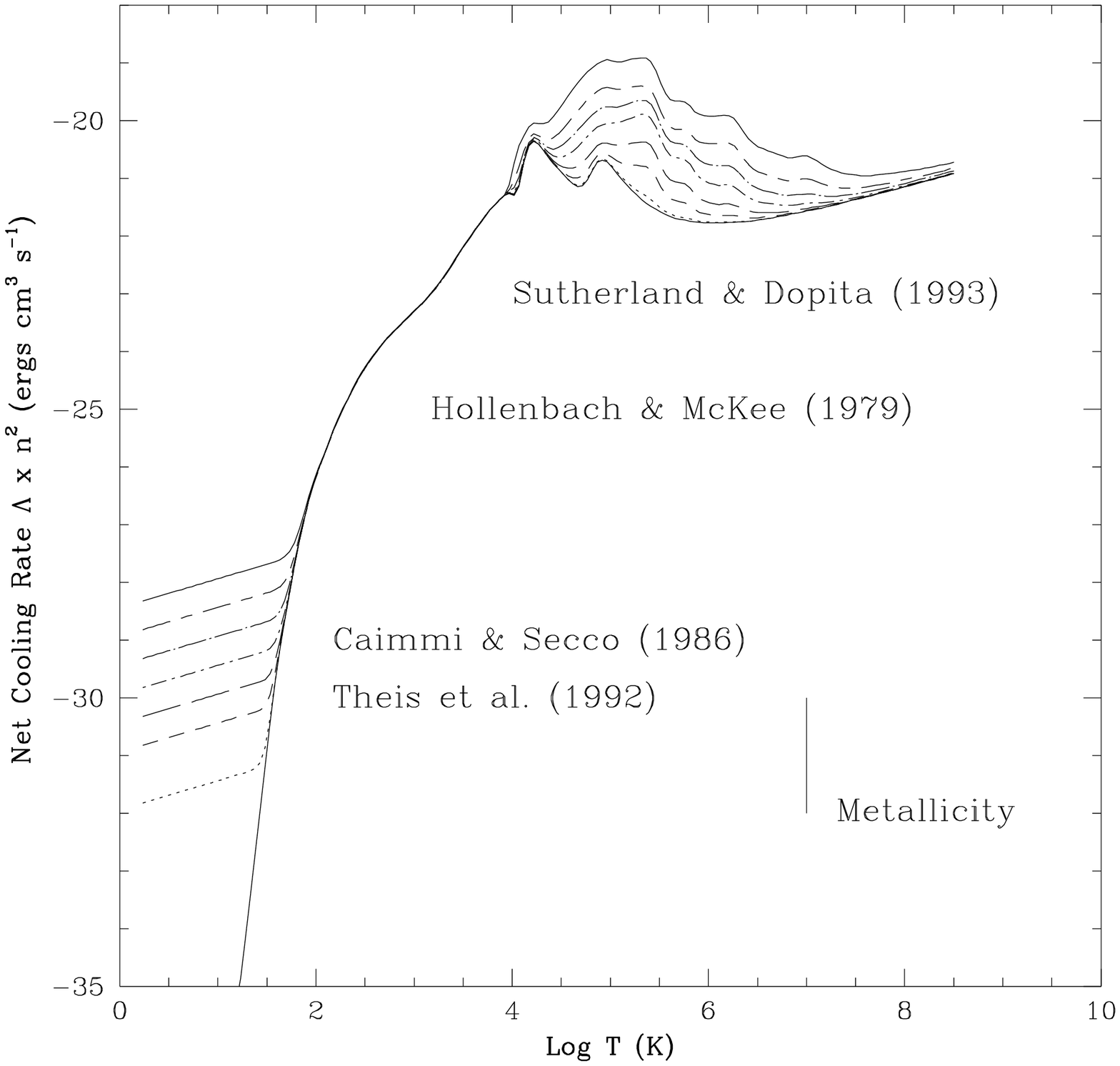,height=8.0truecm,width=8.5truecm}
\caption{ The net cooling rate $\Lambda\times n_j n_k$ in $\rm ergs~cm^{3}~
s^{-1}$ as a function of temperature: from Sutherland \& Dopita (1993) for $T
\geq 10^4$, Hollenbach \& McKee (1979) and Tegmark et al. (1996) for $100 \leq
T \leq 10^4$, and Caimmi \& Secco (1986) and Theis et al. (1992) for $T <
10^2$ K. The product $n_j \times n_k$ contains  the number densities of
particles. The meaning of $n_j$ and $n_k$ changes according to the portion of
the curve under consideration: in Sutherland \& Dopita (1993) $n_j \times n_k
=n_e \times n_i$, where $n_e$ and $n_i$ are the number densities of electrons
and sum of the ion number densities, respectively. In all the others, the
product $n_j \times n_k  = n^{2}$, with $n$ the number density of all
particles lumped together for the sake of simplicity. The number density in
usage is $n \simeq 10$. From  bottom to top, each curve corresponds to an
increasing value of the metallicity $[Fe/H]$. See the text for more details.}
\label{cool_rate}
\end{figure}
%%%%%%%%%%%%%%%%%%%%%%%%%%%%%%%%%%%%%%%%%%%%%%%%%%%%%%%%%%%%%%%%%%

\subsection{Cooling}
The cooling term $\Lambda(T,\rho_g,t)$ of equation (\ref{eq_9}) is derived
from literature data,  and it includes several radiative processes. For
temperatures greater than $10^4$ K we lean on the Sutherland \& Dopita (1993)
tabulations for a plasma under equilibrium conditions and  metal abundances 
$[Fe/H]$=-10 (no metals), -3, -2, -1.5, -1, -0.5, 0 (solar), and 0.5. For
temperatures in the range $100 \leq T \leq 10^4 $ the dominant source of
cooling is the $H_2$ molecule getting rotationally or vibrationally excited
through a collision with an $H$ atom or another $H_2$ molecule and decaying
through radiative emission. The data in use have been derived from the
analytical expressions of Hollenbach \& McKee (1979) and Tegmark et al.
(1996). Finally, for temperatures lower than 100 K, starting from the relation
of Theis et al. (1992) and Caimmi \& Secco (1986), we have implemented the
results of Hollenbach \& McKee (1979), and Hollenbach (1988) for CO  as the
dominant coolant. The following analytical relation in which the mean
fractionary abundance of CO is given as a function of $[Fe/H]$, is found to
fairly represent the normalized cooling rate (i.e. $\Lambda_{CO}/n^2$ with $n$
the number density of particles)

\begin{equation}
 { \Lambda_{CO} \over n^2 }  =  1.6 \times 10^{-29} 10^{([Fe/H] -1.699) }  
                T^{0.5}~~~~\rm erg~cm^{-3}~s^{-1} 
\label{eq_14}
\end{equation}

The net cooling rate ($\Lambda \times n^2$) over the whole temperature range
is shown in Fig.~\ref{cool_rate}, in which the effects of metallicity are
clearly visible. The number density in usage for the purposes of
Fig.~\ref{cool_rate} is $n \simeq 10$. It is worth mentioning that no
re-scaling of the cooling rate from the various sources has been applied to
get the smooth curves shown in Fig.~\ref{cool_rate}.

%%%%%%%%%%%%%%%%Figure 3   %%%%%%%%%%%%%%%%%%%%%%%%%%%%%%%%%%%%%%%%%%%
\begin{figure}
%\picplace{8.0cm}
%\psfig{file=blank.ps,height=8.0truecm,width=8.5truecm}
%\psfig{file=ene.ps,height=8.0truecm,width=8.5truecm}
\psfig{file=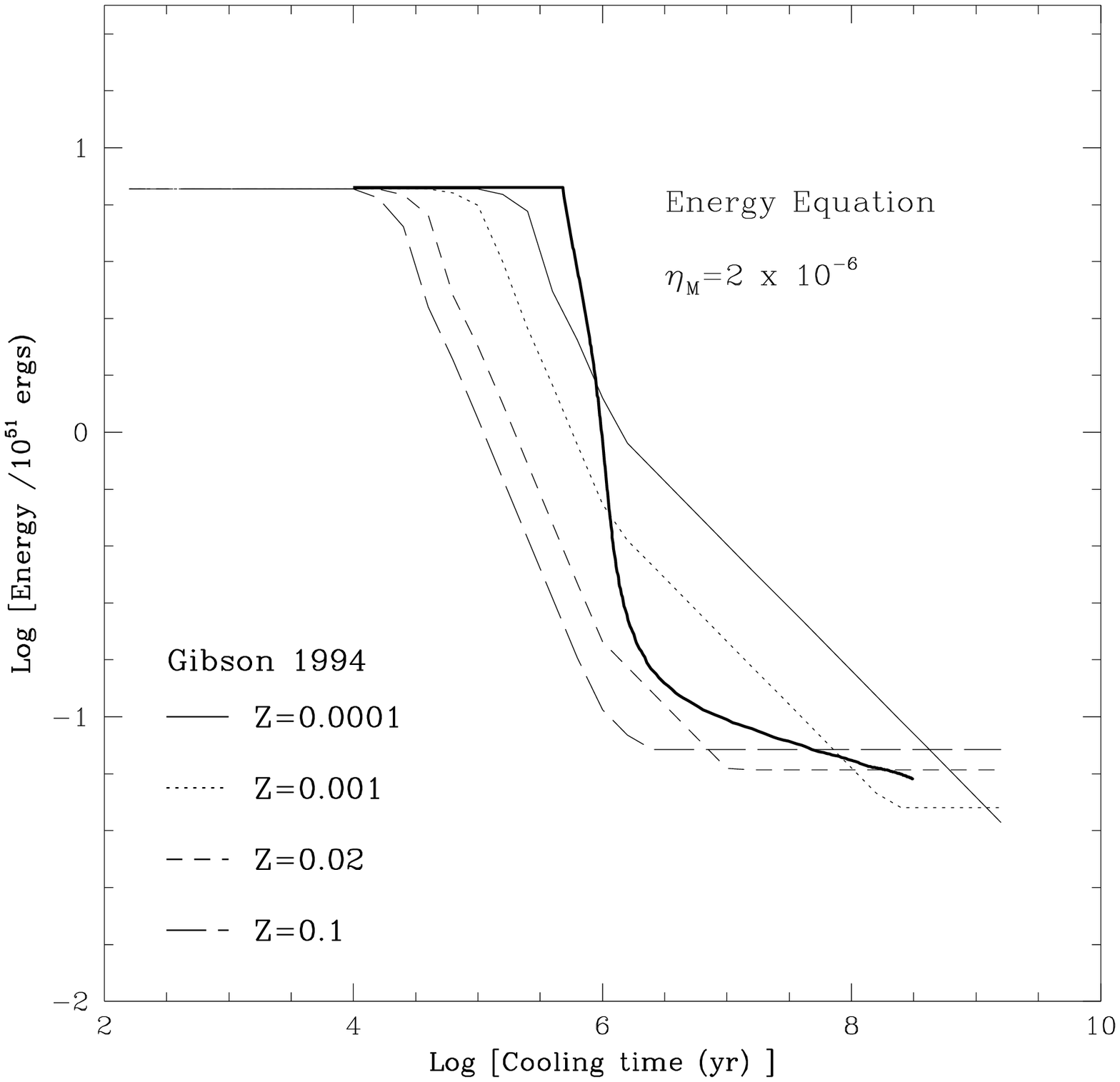,height=8.0truecm,width=8.5truecm}
\caption{ Cooling law of supernova remnants $\epsilon_{SN}(t)$ from Gibson
(1994). Each curve is for a different metallicity of the interstellar medium
as indicated. Superposed is the cooling law of eq.(30) for the 3 $M_{L,T}$
model with age of 0.024 Gyr and metallicity $Z=0.010$ (thick line) described
in the text. Re-normalization of the input energies is applied.}
\label{ene}
\end{figure}
%%%%%%%%%%%%%%%%%%%%%%%%%%%%%%%%%%%%%%%%%%%%%%%%%%%%%%%%%%%%%%%%%%%%%%%

How does equation (\ref{eq_9}) compare with respect to the classical cooling
laws for supernova ejecta ? To answer this question we display in
Fig.~\ref{ene} the path  in the energy-cooling time plane followed by
supernova ejecta according to the most recent discussion of the subject by
Gibson (1995) and  one of the models we are going to present below, i.e. the 3
$M_{L,T}$ galaxy during its stages of stellar activity. The cooling laws of
supernova ejecta displayed in Fig.\ref{ene} are for different metallicities as
indicated,  typical density of the medium of about $10^{-22}$ g/cm$^3$, and
injected   energy  of  $10^{51}$ erg. In the model galaxy  the metallicity is
about $Z=0.010$, the parameter of $H_{M}$ is $\eta_M=2 \times 10^{-6}$, the
density is about $10^{-23}$ g/cm$^3$, and the energy injected is about
$10^{11}$ erg per unit mass of gas. Re-normalization of the energies is
applied. Although the general trend is similar to that of supernova cooling
laws, at least as far as the drop-off is concerned, the cooling time is much
shorter.

\subsection{Coupling energy equation and IMF}
\begin{figure}
%\picplace{{8,2}.0cm}
%\psfig{file=blank.ps,height=8.0truecm,width=8.5truecm}
%\psfig{file=cool_gas.ps,height=8.0truecm,width=8.5truecm}
\psfig{file=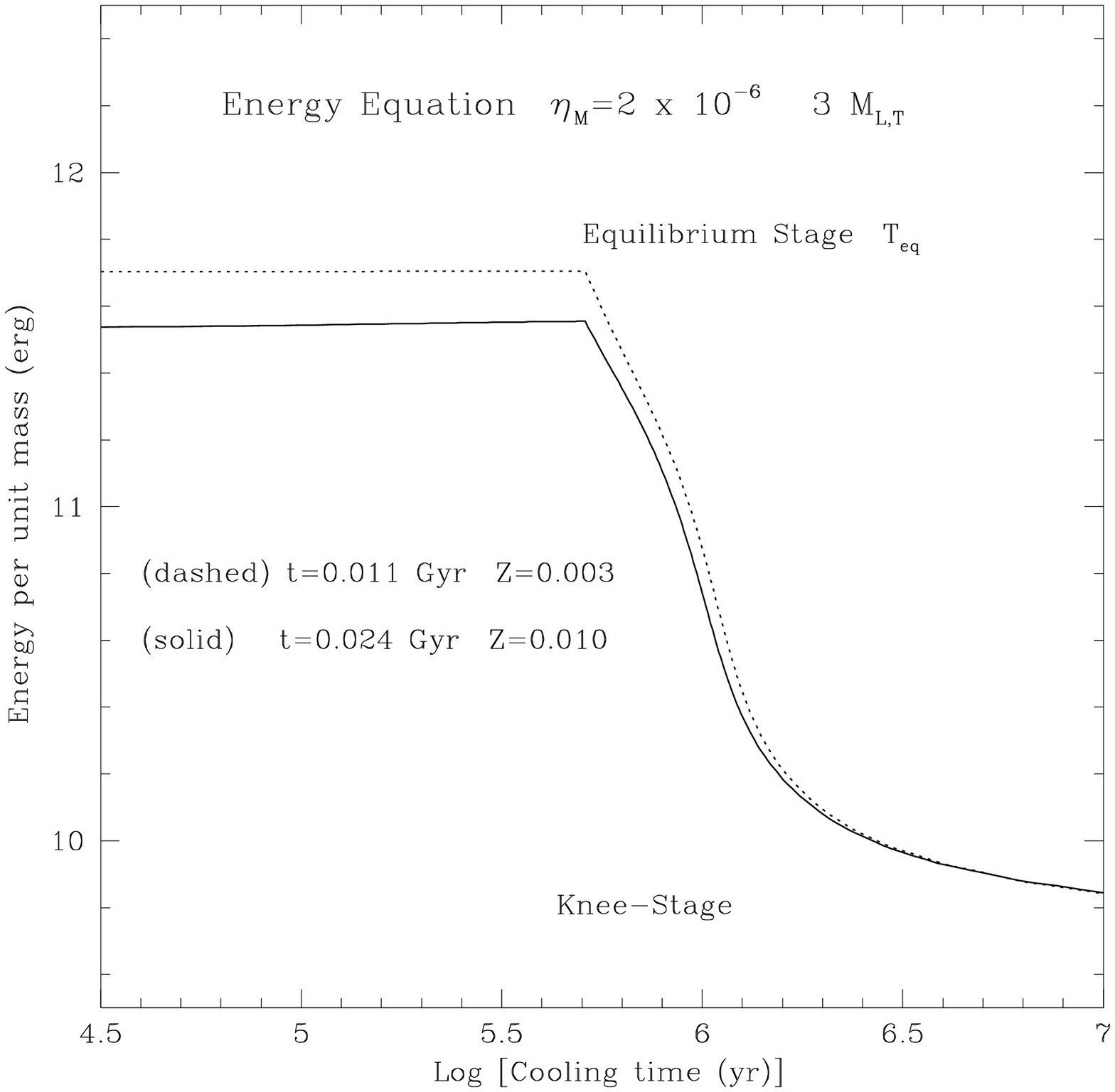,height=8.0truecm,width=8.5truecm}
\caption{The cooling law the  3 $ M_{L,T}$ model with $\nu=10$. Each curve
corresponds to a different age and metallicity. The galactic age (in Gyr) is
annotated along each curve. The particular stages at which the equilibrium
temperature ($T_{eq}$) is reached and  rapid contraction (collapse) begins,
and the final temperature ($T_{knee}$) at which fragmentation is supposed to
occur are indicated. See the text for more details.}
\label{cool_gas}
\end{figure}

The energy equation (\ref{eq_9}) regulates  heating and cooling of gas
according to the following scheme:

\begin{itemize}
\item{At each time step $\Delta t$, the energy input from the various sources
heats up and expands the gas, whereas cooling acts in the opposite sense. }

\item{Equation (\ref{eq_9}) is integrated in time assuming that
heating-expansion and cooling-contraction of gas follow the constant-pressure
law ($P= {K \over \mu H} \rho T = const)$. }

\item{ An equilibrium stage with temperature $T_{eq}$ and density $\rho_{g,eq}$ 
is  soon
reached that yields the Jeans mass $M_{J,eq}$ above which elemental gas clouds
are gravitationally unstable and prone to collapse. }

\item{In principle we can suppose that of the existing clouds only the
fraction whose Jeans mass is larger than $M_{J,eq}$ can form stars.  
In order to determine
the temperature, density and velocity dispersion at the start of the star
forming process we need to follow the thermo-dynamical fate of the gas 
component prone to instability. 
This is conceived as
thermally decoupled from the surrounding medium, so that no further energy
input of radiative nature is possible, but still able to mechanically interact with it. Therefore, the radiative term $H_{R}(T,\rho_g,t)$
of equation (\ref{eq_9}) is switched off, whereas the mechanical term $H_{M}$
is retained. }

\item{The fraction of gas with Jeans mass in excess of $M_{J,eq}$ is let cool
down and  collapse until a new equilibrium stage is reached characterized by
certain values of temperature and density, $T_{knee}$ and $\rho_{knee}$,
respectively. The existence of the {\it knee-stage }is secured by the energy
input of mechanical nature. In any case, $T_{knee}$ is not let be lower than
$T_{CBR}[z(t)]$. In such a case the integration of the energy equation is stopped,
and $T_{knee}$ and $\rho_{knee}$ are the last computed values.}

\item{ During all these stages the cooling and free-fall times scales turn out
to be nearly  identical. The temporal evolution of the gas energy (per unit mass 
of gas) and the
temperatures $T_{eq}$ and $T_{knee}$ are shown in Fig.~\protect{\ref{cool_gas}
} for the $3 M_{L,T}$ galaxy at different evolutionary stages as indicated.
  The efficiency of star
formation characterizing this model will be defined below. }

\item{We identify the fraction of gas that underwent the cooling process
described above and was able to reach $T_{knee}$ and $\rho_{knee}$ as the
molecular component in which fragmentation according to the Padoan et al.
(1997) scheme can occur. }

\item{ It is reasonable to assume that these clouds acquire their  kinetic
energy from the surrounding medium under the condition of energy
equipartition. Accordingly their velocity dispersion scales as

\begin{equation}
\Sigma_g(T_{eq}) = \Sigma_g(T) \times \sqrt {M_J(T) \over M_J(T_{eq})} 
\end{equation}

with obvious meaning of the symbols. The starting value in the recursive
process to determine the velocity dispersion of collapsing clouds is the
galaxy velocity dispersion $\Sigma$.}

\item{Finally, the temperature and density $T_{knee}$ and $\rho_{knee}$, and
the velocity dispersion $\Sigma_g(T_{eq})$  are the values to be plugged into
the IMF. Each cloud prone to star formation will fragment into stars according
to this IMF. }

\item{The case may arise, in which  for the values of $T_{knee}$, $\rho_{knee}$, and velocity dispersion $\Sigma_g(T_{eq})$ derived from the energy equation the
peak mass of the IMF is comparable or even lower than the minimum mass
$M_{min}=0.07 M_{\odot}$ for the existence of real stars (i.e. burning
hydrogen in their cores). In such a case the peak mass is not let fall below 
$M_l=0.1 M_{\odot}$, and the IMF is calculated with the temperature obtained
from inverting equation (\ref{eq_2}) in which $\rho_{knee}$ and
$\Sigma(T_{eq})$ are used. Even if there is a marginal inconsistency, this can
be neglected for any practical purpose. }
\end{itemize}

%%%%%%%%%%%%%%%%Figure 4  %%%%%%%%%%%%%%%%%%%%%%%%%%%%%%%%%%%%%%%%%%%% 
%%%%%%%%%%%%%%%%%%%%%%%%%%%%%%%%%%%%%%%%%%%%%%%%%%%%%%%%%%%%%%%%%%%%%%

\subsection{Deriving the  specific efficiency of star formation} 
The discussion above has been referred to the fraction of gas with Jeans mass
in excess of $M_{J,eq}$ and ready to collapse and fragment into stars. This fraction is, however, still unknown. Indeed to be determined a law governing
the mass or number distribution of such clouds is needed.
Similarity arguments would suggest that the same IMF governing star formation should also
regulate the cloud mass (number) distribution. The major difference is in the
temperature,  density, and velocity dispersion to be adopted. It goes without
saying that $T_{eq}$ and $\rho_{eq}$ must be used as they represent the stage
at which elemental clouds become thermo-dynamically distinct units out of the
surrounding medium. In such a case, the mass fraction of unstable clouds  would
be

\begin{equation}
  f_{jeans} =  \int_{M_{J,eq}}^{\infty} C \phi(M, T_{eq}, 
                 \rho_{g,eq}, \Sigma_g(T_{eq})) dM 
\end{equation}

In principle, with aid of this quantity one could derive $\nu(r,t)$. 
Recalling that  cooling  occurs on a certain time scale $\tau_{cool}$, known
from solving equation (\ref{eq_9}), we expect that the rate of star formation
will be  inversely proportional to $\tau_{cool}$. On the base of the above
arguments, we could write the specific efficiency $\nu(r,t)$ of the star formation
rate as

\begin{equation}
\nu(r,t) = \nu_0 \times { f_{jeans}  \over \tau_{cool} } 
\end{equation}
\noindent
where $\nu_0$ is a suitable scale factor of the order of unity.

With this formulation for $\nu(r,t)$, there would  no longer be free
parameters in the rate of star formation, which would be tightly coupled to the
energy storage of the medium in which stars are born. Star formation
 drives the energy input, and this latter in turn affects  star
formation: a self-regulating mechanism is soon established. 

The 
application of the above scheme rises, however, some difficulties both of conceptual and  numerical
nature. First it is not granted that the number (mass) distribution of
unstable clouds is governed by the same law determining the mass distribution of stars at each formation event, second   both $f_{jeans}$ and $\tau_{cool}$ fluctuate from time step to time
step, thus rendering the associated value of $\nu(r,t)$ highly uncertain. The preliminary analysis of the problem has shown that the quantity
$f_{jeans}/\tau_{coll}$ is of the order of $5\div 50$. Therefore, we
considered the specific efficiency $\nu(r,t)$ as an adjustable  parameter to
be eventually fixed by comparing model results with observational data (e.g
CMR, chemical abundances, etc..) and did not make use of the above scheme.

%%%%%%%%%%%%%%%%%%%%%%%%%%%%%%%%%%%%
\section{The reference models}
In this section we present our reference models for elliptical galaxies of
different total luminous mass $M_{L,T}$ and luminous mass $M_{L,Re}$ within the
effective radius.

All the models are calculated assuming  $\eta_M=2\times
10^{-6}$ (the parameter governing the efficiency of mechanical energy sources)
and $\nu=10$ (the parameter driving the efficiency of the star formation
rate), but two slightly different conditions for the onset of the galactic
wind. The main reason for it resides in the fact that the energy equation is
solved separately from those  governing gas consumption, star formation, and
chemical enrichment. Therefore, there is an implicit ambiguity concerning the
mode and time at which the conditions for galactic winds are met. In short,
what is uncertain is whether the energy liberated by the various sources going
into the thermal content of the gas ought to be tested against wind conditions
before of after cooling has taken place. In order to check the sensitivity of
model results to this kind of uncertainty, at each time step we examine two
possible schemes:

\begin{itemize}
\item{ At the beginning of the time step, all the energy generated by
supernova explosions, stellar winds,  etc. is  tested against the condition

\begin{equation}
E_{th}(t) \geq \Omega_g(t).
\end{equation}

In the case of a negative answer (no galactic wind yet), the total energy
budget is given to the energy equation and all remaining conditions for the
onset of galactic wind are examined only at the end of the time step. These
models are named Type A.}

\item{Alternatively, at the beginning of the time step the total energy budget
is given to the energy equation and the whole set of conditions for the onset
of galactic winds is tested only at the end of the time step. These models are
named Type B.}
\end{itemize}

The two sets of models are presented in Table~1, which lists the total
luminous mass $M_{L,T}$, the total luminous mass  $M_{L,R_e}$ inside the
effective radius, the mean mass density within $R_e$, the galaxy velocity
dispersion $\Sigma$, the maximum value for the  gas temperature $T_g$ (in K)
and peak mass $M_P$  (in $M_{\odot}$) of the IMF reached in the course of
galactic evolution, the age $t_w$ (in Gyr) at the onset of the galactic wind,
the fractionary masses in gas, $M_g/M_{L,R_e}$, and stars $M_s/M_{T,R_e}$
inside the $R_e$-sphere at the onset of the wind, the current ($Z$) and mean
metallicity ($\langle Z \rangle$), and finally the index $Mg_{2}$ at the same
epoch.  The whole discussion below will refer to models with 
$\eta_M=2\times 10^{-6}$. However, when examining the CMR and FP (section 8 below), we will also
consider models of Type A but with $\eta_M=3\times 10^{-6}$ which are much
similar to those described here.

%%%%%%%%%%%%%%%Table 1 %%%%%%%%%%%%%%%%
\begin{table*}[htb]
\begin{center}
\caption{Basic data of the models. $M_{L,T}$ and $M_{L,R_e}$ are the total
luminous mass and the mass within the effective radius, respectively. They are
in units of $10^{12}\times M_{\odot}$. $\langle \rho \rangle$ is the mean mass
density (in $gr~cm^{-3}$) within the $R_e$-sphere. $\Sigma$ is the galaxy
velocity dispersion. $T_g$ and $M_P$ are the maximum values of the gas
temperature (in K) and peak mass (in $M_{\odot}$) of the IMF in the course of
galactic evolution. $t_w$ is the age (in Gyr) of the galaxy at the onset of
galactic wind. $M_g/M_{L,R_e}$ and $Ms/M_{L,R_e}$ are the fractionary masses
of gas and stars, respectively, at the galactic wind stage. $Z$ and $\langle Z
\rangle$ are the current and mean metallicity reached by the galaxy at the
wind stage. All the models are calculated with $\eta_M=2\times 10^{-6}$.
Finally, two groups of models are listed: models of case A have early galactic
wind; models of case B eject their wind at ages that progressively increase at
decreasing galactic mass. Those with $t_w=16.45$ continue to form stars albeit
at minimal levels up to the present epoch.}
\begin{tabular*}{145mm}{l ccc ccc ccc cc }
\hline
\hline
  &   &    &    &    &    &    &     &    &    &    &         \\
$M_{L,T}$& $M_{L,R_e}$ &$\rho $ &$\Sigma$ & $T_g$ & $M_P$ &$t_w$& 
  $M_g \over M_{L,R_e}$ & $M_s \over M_{L,R_e}$&  Z    & $<Z>$    & $Mg_2$   \\
       &         &    &    &      &       &     &      &      &       &       &\\
\hline
       &         &    &    &      &       &     &      &      &       &       &\\
\multicolumn{11}{c}{Type A~~~~~~$\eta_M=2\times 10^{-6}$}\\
       &         &    &    &      &       &     &      &      &       &       &\\
\hline
       &         &    &    &      &       &     &      &      &       &       & \\
3      & 1.2420  &1.33(-23) &354 & 43.1 & 1.010 &0.22 & 0.281& 0.646& 0.0782& 0.0443& 0.3353\\  
1      & 0.4140  &2.79(-23) &278 & 42.6 & 0.801 &0.39 & 0.123& 0.770& 0.1031& 0.0511& 0.3437\\  
0.1    & 0.0410  &1.31(-22) &167 & 38.0 & 0.549 &0.52 & 0.053& 0.858& 0.0755& 0.0295& 0.3114\\  
0.01   & 0.0041  &6.16(-22) &100 & 33.9 & 0.417 &0.74 & 0.019& 0.907& 0.0561& 0.0155& 0.2738\\  
0.001  & 0.0004  &2.89(-21) &60 & 29.5 & 0.300 &0.82 & 0.011& 0.942& 0.0418& 0.0059& 0.2168\\  
       &         &    &    &      &       &     &      &      &       &       &       \\
\hline
       &         &    &    &      &       &     &      &      &       &       &       \\
\multicolumn{11}{c}{Type B~~~~~~$\eta_M=2\times 10^{-6}$}\\
       &         &    &    &      &       &     &      &      &       &       &       \\
\hline
       &         &    &    &      &       &     &      &      &       &       &        \\
3      &1.2420   &1.33(-23) &354 & 43.1 & 1.010 &0.22 & 0.281& 0.646& 0.0782& 0.0443& 0.3353\\  
1      &0.4140   &2.79(-23) &278 & 42.6 & 0.801 &0.49 & 0.089& 0.790& 0.1169& 0.0552& 0.3481\\  
0.1    &0.0410   &1.31(-22) &167 & 38.0 & 0.549 &4.99 & 0.003& 0.735& 0.1897& 0.0530& 0.3457\\  
0.01   &0.0041   &6.16(-22) &100 & 33.9 & 0.417 &16.45& 0.000& 0.709& 0.3818& 0.0405& 0.3300\\  
0.001  &0.0004   &2.89(-21) &60 & 29.5 & 0.300 &16.45& 0.000& 0.774& 0.3274& 0.0202& 0.2893\\  
       &         &    &    &      &       &     &      &      &       &       &       \\
\hline
\hline
\end{tabular*}
\end{center}
\label{tab_mod}
\end{table*}

%%%%%%%%%%%%%%%%Figure 5   %%%%%%%%%%%%%%%%%%%%%%%%%%%%%%%%%%%%%%%%%%%% 
\begin{figure}
%\picplace{8.0cm}
%\psfig{file=blank.ps,height=8.0truecm,width=8.5truecm}
%\psfig{file=te_age.ps,height=8.0truecm,width=8.5truecm}
\psfig{file=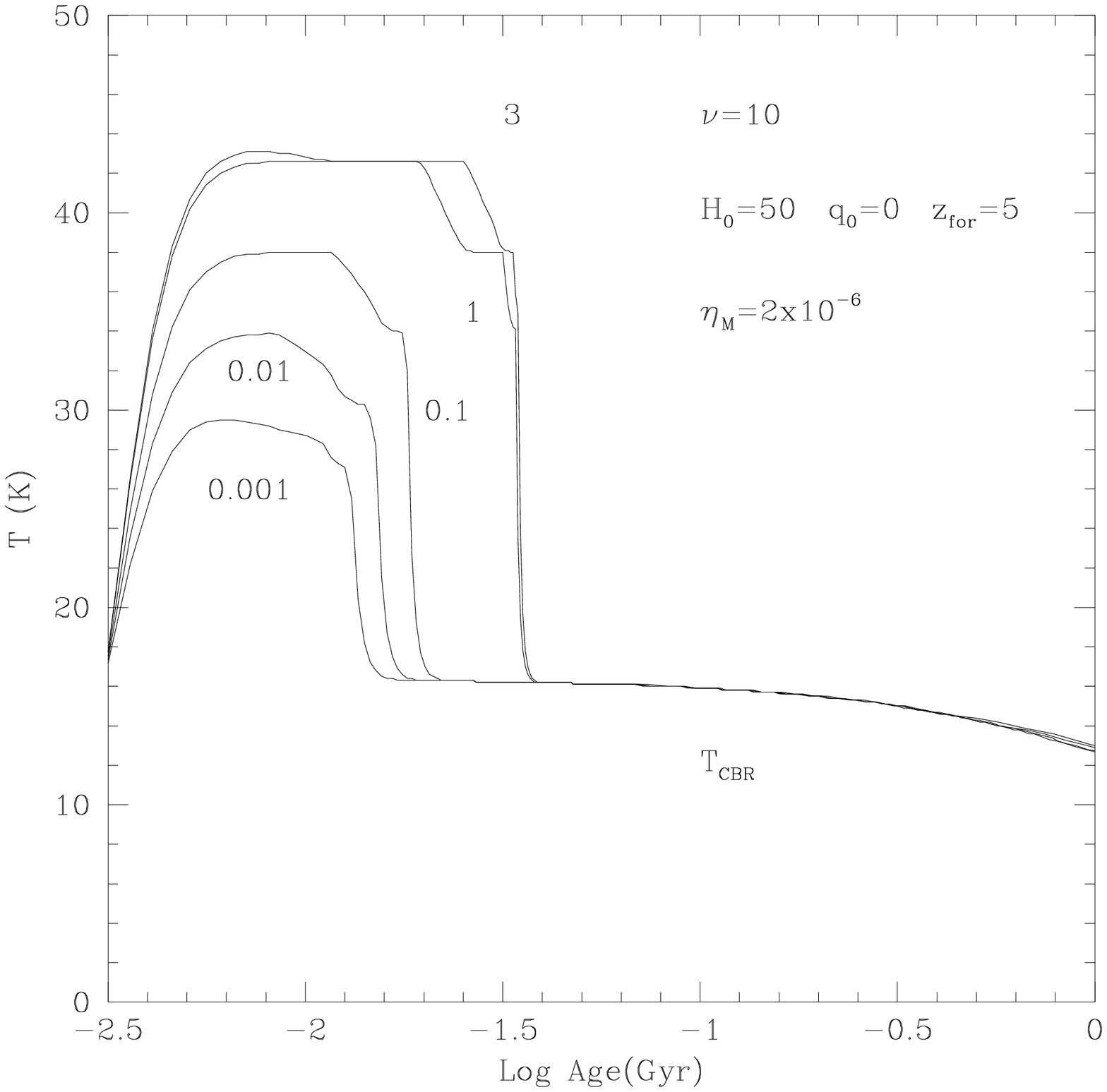,height=8.0truecm,width=8.5truecm}
\caption{ The gas temperature (in K) as a function of the age (in Gyr) for the
models  galaxies with $M_{L,T}$ equal to 0.001, 0.01, 0.1, 1 and 3 as
indicated. The same curves apply  both to Type A and B models.}
\label{te_age}
\end{figure}
%%%%%%%%%%%%%%%%%%%%%%%%%%%%%%%%%%%%%%%%%%%%%%%%%%%%%%%%%%%%%%%%%%%%%%%

%%%%%%%%%%%%%%%%Figure 6   %%%%%%%%%%%%%%%%%%%%%%%%%%%%%%%%%%%%%%%%%%%% 
\begin{figure}
%\picplace{8.0cm}
%\psfig{file=blank.ps,height=8.0truecm,width=8.5truecm}
%\psfig{file=mp_age.ps,height=8.0truecm,width=8.5truecm}
\psfig{file=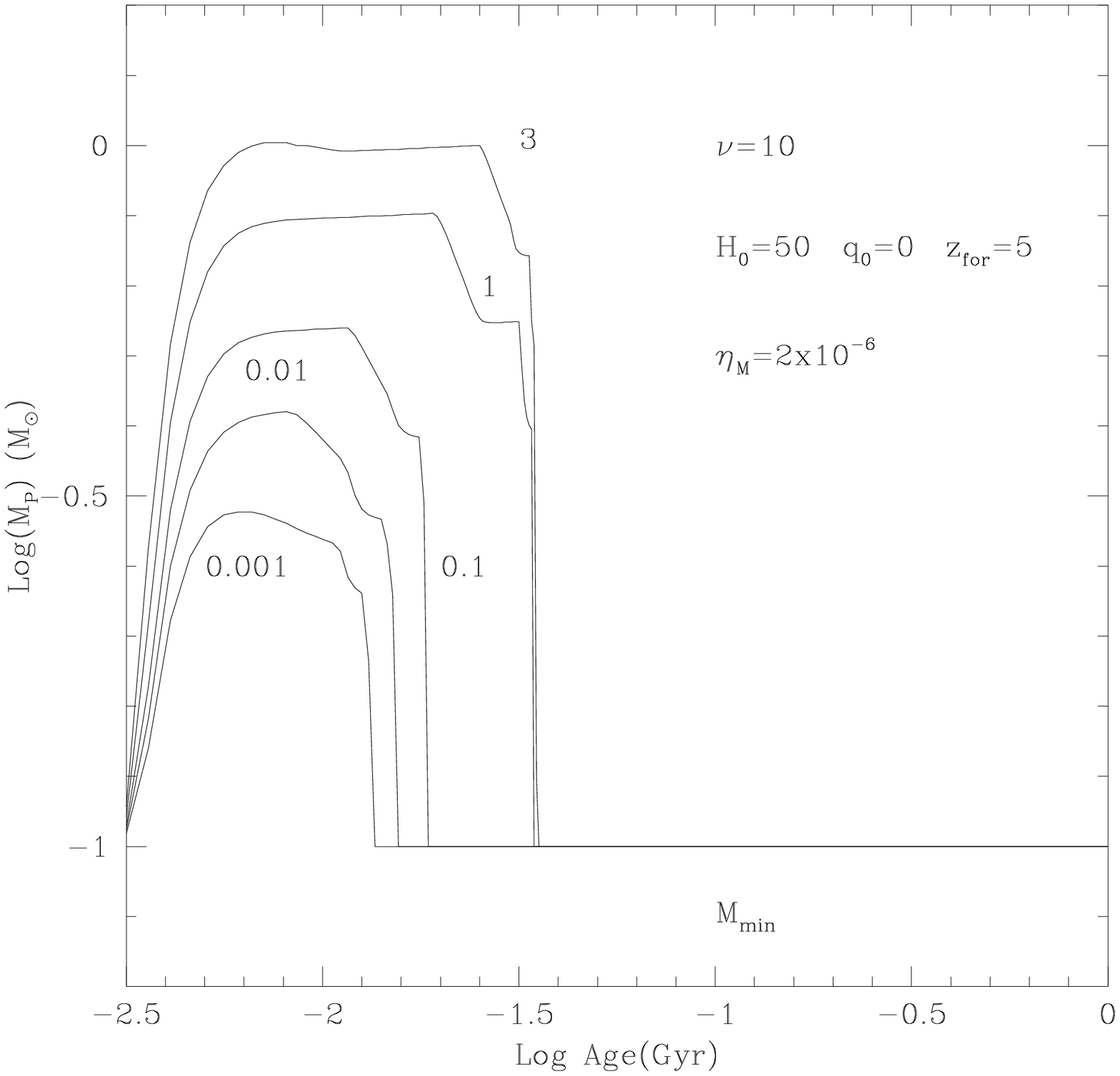,height=8.0truecm,width=8.5truecm}
\caption{The peak mass $M_P$ (in solar units) of the IMF as a function of the
age (in Gyr) for the models galaxies with $M_{L,T}$ equal to 0.001, 0.01, 0.1,
1 and 3 as indicated. The same curves apply to both Type A and B models.}
\label{mp_age}
\end{figure}
%%%%%%%%%%%%%%%%%%%%%%%%%%%%%%%%%%%%%%%%%%%%%%%%%%%%%%%%%%%%%%%%%%%%%%%

%%%%%%%%%%%%%%%%Figure 7  %%%%%%%%%%%%%%%%%%%%%%%%%%%%%%%%%%%%%%%%%%%% 
\begin{figure}
%\picplace{8.0cm}
%\psfig{file=blank.ps,height=8.0truecm,width=8.5truecm}
%\psfig{file=mp_zeta.ps,height=8.0truecm,width=8.5truecm}
\psfig{file=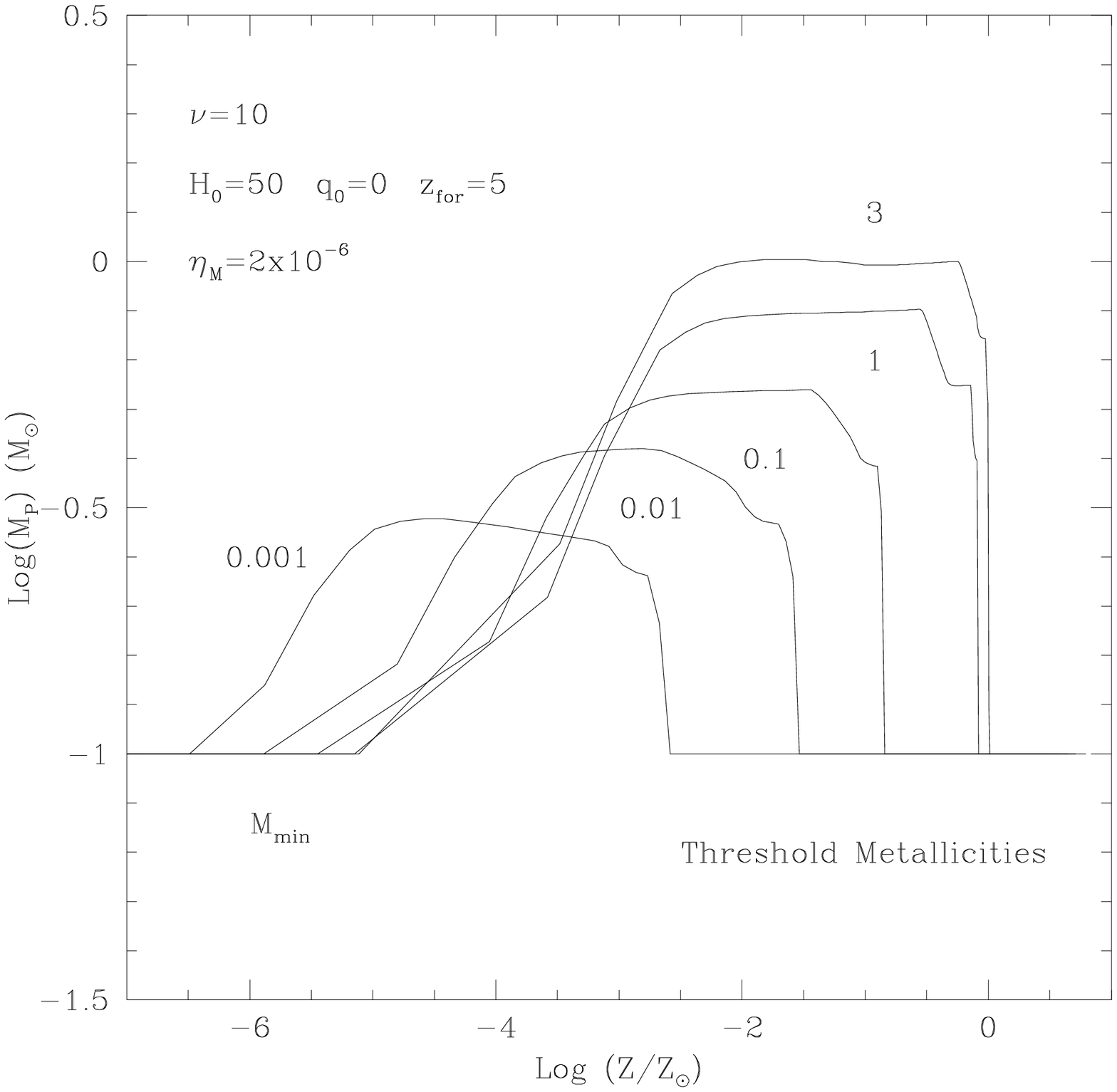,height=8.0truecm,width=8.5truecm}
\caption{ The variation of the peak mass $M_P$ as a function of the gas
metallicity for all models under consideration.}
\label{mp_zeta}
\end{figure}
%%%%%%%%%%%%%%%%%%%%%%%%%%%%%%%%%%%%%%%%%%%%%%%%%%%%%%%%%%%%%%%%%%%%%%

\subsection{General properties}
All the models obey some general properties that are determined by the
interplay between the physical conditions of the star forming gas and the IMF.
We will sketch the problem in several steps for the sake of an easier
understanding:

\begin{itemize}
\item{At increasing galaxy mass the mean  gas density gets lower, the velocity
dispersion gets higher together with the initial gas temperature (the one
determined by the very first energy source $H_{C}$). }

\item{As a consequence of it, the IMF tends to skew towards the high mass
end ($M_P$ moves to higher masses and the slope above it gets slightly flatter
at increasing galaxy mass). This means that more energy from supernova
explosions and stellar winds is injected into the interstellar gas making it
hotter and hotter till a sort of equilibrium stage is reached, during which
both the gas temperature and $M_P$ remain nearly constant.}

\item {However, at proceeding star formation, the gas gets richer in metals
thus increasing the cooling efficiency. A situation is eventually met in which
cooling is so efficient that the gas temperature and peak mass of the IMF in
turn  start decreasing. This implies that less energy is injected into the
interstellar medium with consequent catastrophic fall-off of both temperature
and peak mass.}
\end{itemize}

The temporal evolution of the gas temperature, and peak mass is shown in
Fig.~\ref{te_age} and Fig.~\ref{mp_age}, respectively, whereas the
relationship between $M_P$ and metallicity is presented in  Fig.~\ref{mp_zeta}.
Notable features of the three diagrams are:  (i) the dependence of the maximum
value attainable by the gas temperature and $M_P$  on the galaxy mass (both
increase with this); (ii) the existence of a threshold value of the
metallicity, above which cooling gets so efficient that bot temperature and
$M_P$ dramatically drop to their minimum permitted values, i.e.
$T_{CBR}[z(t)]$ and $0.07 M_{\odot}$, respectively; (iii) the dependence of
the threshold metallicity on the galaxy mass (it gets higher at increasing
mass).

\subsection{The onset of galactic winds}
In all Type A models the galactic wind occurs at early epochs, i.e. within the
first Gyr, even though a systematic trend with the galaxy mass can be noticed.
High mass galaxies undergo their wind ejection earlier than low mass galaxies
(0.22 Gyr for the $3 M_{L,T}$ galaxy and 0.82 Gyr for the $0.001 M_{L,T}$
object). In contrast,  galactic winds in Type B models  span a much wider
range of ages that go from 0.22 Gyr for the $3 M_{L,T }$ galaxy to 4.99 Gyr
for the $0.1 M_{L,T}$ object, to even no ejection at all within the maximum
age range permitted by the assumed cosmological parameters $H_0$, $q_0$ and
$z_{for}$ as in the case of the two lowest mass galaxies. In this latter case,
star formation continues all over the whole galactic lifetime albeit at  a
minimal level of activity.

\begin{figure}
%\picplace{8.0cm}
%\psfig{file=blank.ps,height=8.0truecm,width=8.5truecm}
%\psfig{file=sfr.ps,height=8.0truecm,width=8.5truecm}
\psfig{file=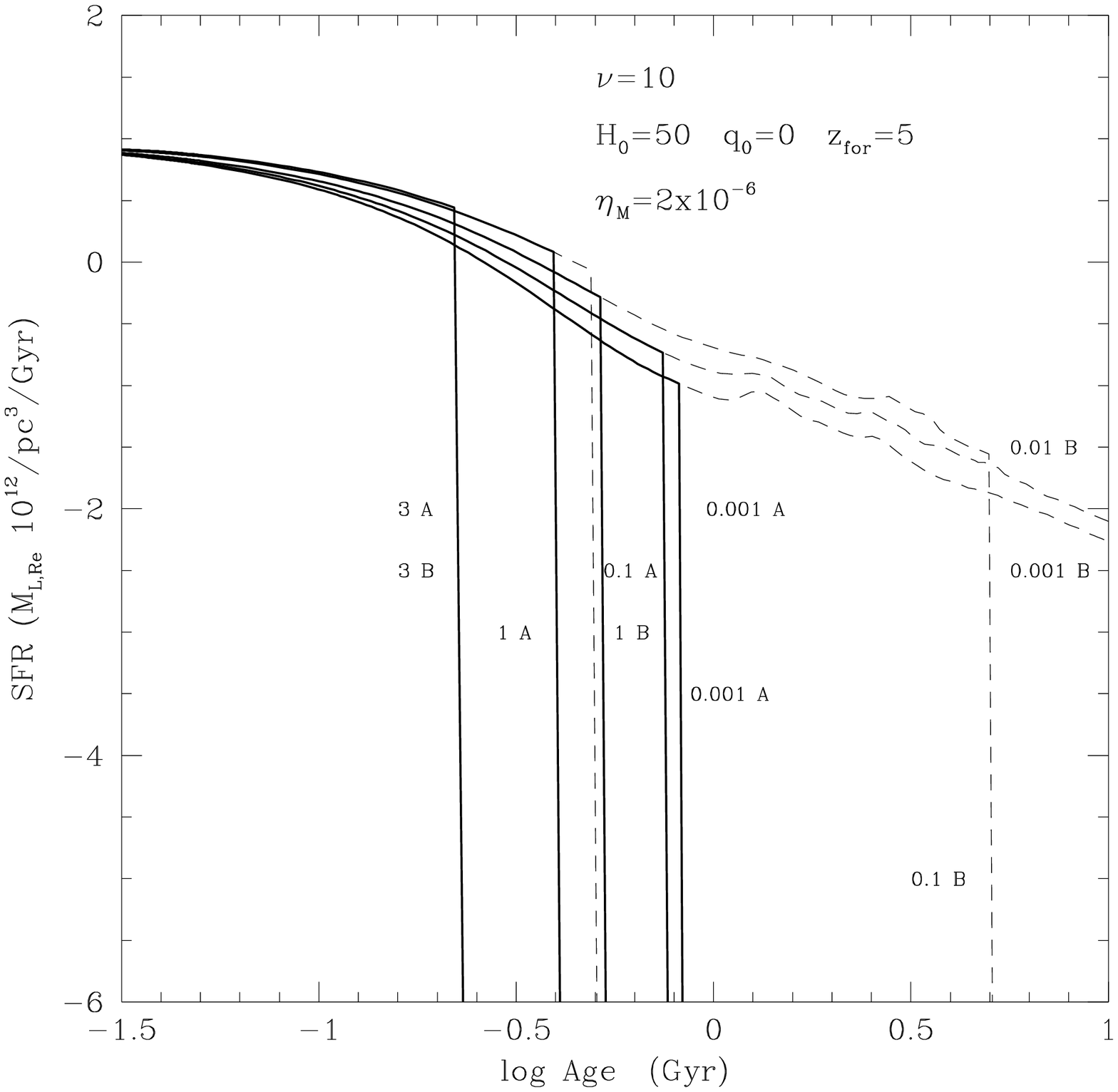,height=8.0truecm,width=8.5truecm}
\caption{ The normalized rate of star formation SFR as a function of the age
for Type A (thick solid lines) and B (thin dashed lines) models. The  SFR is
expressed in units of $ M_{L,Re} \times 10^{12}\times M_{\odot}/pc^3/Gyr$.
For the sake of clarity the age is limited to the first 10 Gyr.}
\label{sfr}
\end{figure}

The star formation history of  our model galaxies is shown in Fig.~\ref{sfr},
whereby the solid and dotted lines are for Type A and B models, respectively.
The rate of star formation, in units of $ M_{L,Re} \times 10^{12} \times
M_{\odot}/ pc^3/Gyr$, is plotted against age (in Gyr). For the sake of
clarity, we limit the age to the first 10 Gyr. With the efficiency of star
formation we have assumed ($\nu=10$) most of the gas is turned into stars
within the first few $10^8$ years with consequent fall-off of the gas content
and star formation rate.

The time variation of the gas content is shown in Fig.~\ref{gas_cont} where
one can notice that even in models with ever lasting star formation (the 
0.001 and 0.01 $M_{L,T}$ galaxies of case B), the amount of gas left over
after the first 0.5 Gyr is less than 1\% or so. Therefore, in these galaxies
the real star formation process reduces to very low levels of activity. After
the minimum in each curve, which corresponds to the galactic wind stage, the
gas content rises again because continuously ejected (into the interstellar
medium) by dying stars. However, as star formation does no longer occur in our
models, this gas is supposed to freely escape into the intergalactic medium,
because it is rapidly heated up to high temperatures corresponding to the
velocity dispersion of the galaxy (cf. Fabbiano 1986).
 
What is the reason for the different behavior between models A and B ? and how
much do the models of the two groups differ each other ? Close inspection of
their structure shows that there is no significant difference among them, the
only reason for the difference in $t_w$ being the wind criterion adopted in
Type A models (weaker condition) with respect to Type models B (stronger
condition). It is worth recalling that in Type A  models the thermal and
gravitational energies of the gas are compared to each other {\it before
cooling}, whereas in the Type B ones they are compared {\it after cooling}.

It goes without saying that, in Type B models, the prolonged activity of star
formation will eventually induce somewhat high metallicities in the  residual
gas content. This however is less of a  problem because only  a very small
fraction of the galaxy mass is stored in stars with such high metallicities
(see below).

Although the above uncertainty on $t_w$ constitutes a point of weakness in our
modelling of the interplay between gas thermo-dynamics and galactic winds, both
the results we are going to presents  and the very low gas content in those
galaxies that do not suffer from galactic winds, hint that  this phenomenon is
a marginal event over the whole evolutionary history of  a galaxy  not 
affecting its global structure.

{\it It is worth emphasizing that in both cases the relationship between
galaxy mass and age of the galactic wind has the opposite trend as compared to
the standard supernova driven galactic wind model (cf. section 1) in which low
mass galaxies eject their wind much earlier than the high mass ones}. We will
come back later to this point.

\subsection{What wedge to the energy equation? The role of $\eta_M$}
As already anticipated, the term $H_{M}$ of the energy equation (together with
the CBR limit) has an important role on determining the properties of the
models. Why and how are easy to understand. They indeed provide a lower limit
to the temperature attainable by the gas during the cooling process. While the
effect of $T_{CBR}[z(t)]$ is straightforward, the one of $H_M$ is
by far more indirect and difficult to understand. This term is constant for  a
given galaxy because  so are $\Sigma$ and $t_{ff}$ are, but it varies from
galaxy to galaxy. According to its definition, it represents a threshold value
for the energy input below which gas clouds prone to collapse cannot go. Were
this term neglected, owing to the efficiency of cooling (in presence of
metals, in particular), gas clouds would fall down to extremely low temperatures on a very short time scale. Since the peak mass of the IMF scales with
temperature, $H_{M}$ has a direct effect on $M_P$ and mean slope of the IMF
above it. The higher $H_{M}$, the higher is $M_P$ and the flatter is the IMF
in the mass range above the peak. The quantitative  effects depend of course
on $\eta_M$. If $H_{M}$ is efficient, and $M_P$  in turn is shifted toward the
high mass end (say above $1 M_{\odot}$), of the generations of stars born
under these conditions, a large fraction of these will soon evolve, become
white dwarf or collapsed remnants in virtue of their short lifetime as
compared to the evolutionary time scale of a galaxy. More gas is refuelled to
the interstellar medium, more metals are created, and star formation will tend
to last longer. The extreme case can be met in which, depending on the
competition between gas restitution and heating, no galactic wind occurs in
all model galaxies. Therefore $H_{M}$ eventually drives the mean metal
content, the relative fraction of living stars and remnants (white dwarfs,
neutron stars, black holes as appropriate), and many other properties of the
models,  e.g. the colors (see below). All this hints that the parameter
$\eta_M$ can perhaps be constrained by looking at global properties either of
a galaxy, such as the mean and maximum metallicity, the colors at the present
epoch etc., or of   a galaxy manifold, such as the CMR. Many preliminary
models calculated to this aim have indicated that $\eta_M=2 \div 3\times
10^{-6}$ is a plausible choice leading to acceptable  models. Other values
cannot of course be excluded.

%%%%%%%%%%%%%%%%%%%%%%%%%%%%%%%%%%%%%%%%%%%%%
\section{Discussion of the main results}

\subsection{Star formation history }
As a result of the complex game between IMF and  gas thermo-dynamics, the
standard supernova driven wind model  is largely invalidated. Recalling that
this latter was suggested to explain the CMR as a mass-metallicity sequence
(the mean metallicity grows with the galactic mass), from the entries of
Table~1 we see that the same trend is now possible even for models in which
the overall duration of the star forming activity goes inversely proportional
to the galaxy mass {\it i.e.}  $\Delta t_{SF} \propto M_G^{-1}$ {\it as
indicated by the chemical abundances}. Although the age range is uncertain
because of the above arguments, yet the possibility exists that, following the
dominant initial activity of rather short duration, traces of star formation
might occur over much longer periods of time. Remarkably, this is easier to
get in low mass galaxies than in high mass ones. In this context, an elegant
way out can be found to the difficulties encountered by standard models as far
as  the distribution of nearby galaxies in the \Hbeta\ versus \MgFe\ plane,
the trend in the suspected enhancement of $\alpha$-elements, and the gradients
in \Hbeta\ and \MgFe\ across a galaxy are concerned (see  Section 2 and the
discussion below). We will see  that the standard interpretation of the CMR is
maintained as also these models yield the right mass-metallicity sequence.
Finally, there is another consequence of the density dependence of the IMF:
since within a galaxy the central regions are naturally  denser than the
external ones, we expect over there the same trend as passing from a low mass
(high density) to a high mass (low density) galaxy. Therefore, the central
regions of a galaxy should form stars for periods of time longer than  their
outskirts. If so, galaxies should be  built up by an out-in process.

%%%%%%%%%%%%%%%%Figure 8  %%%%%%%%%%%%%%%%%%%%%%%%%%%%%%%%%%%%%%%%%%%
%%%%%%%%%%%%%%%%%%%%%%%%%%%%%%%%%%%%%%%%%%%%%%%%%%%%%%%%%%%%%%%%%%%%%%
 
%%%%%%%%%%%%%%%%Figure 9   %%%%%%%%%%%%%%%%%%%%%%%%%%%%%%%%%%%%%%%%%%% 
\begin{figure}
%\picplace{8.0cm}
%\psfig{file=blank.ps,height=8.0truecm,width=8.5truecm}
%\psfig{file=gas_cont.ps,height=8.0truecm,width=8.5truecm}
\psfig{file=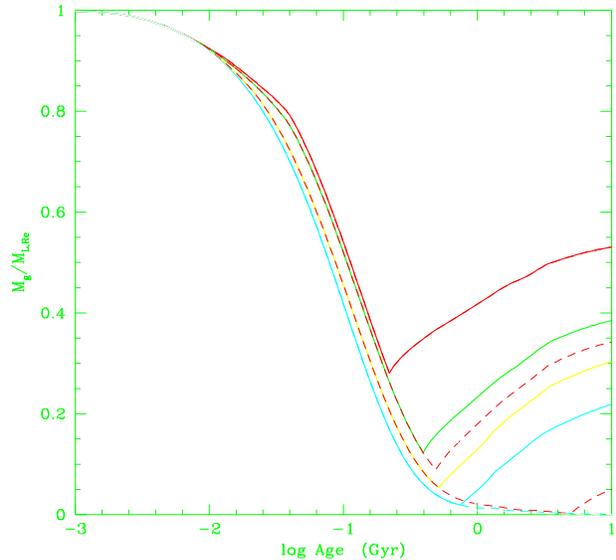,height=8.0truecm,width=8.5truecm}
\caption{ The fractionary mass of gas as a function of the age for Type A
(thick solid lines)  and Type B (thin dashed lines) models. For the sake of
clarity the age is limited to the first 10 Gyr.}
\label{gas_cont}
\end{figure}
%%%%%%%%%%%%%%%%%%%%%%%%%%%%%%%%%%%%%%%%%%%%%%%%%%%%%%%%%%%%%%%%%%%%%%

%%%%%%%%%%%%%%%%%Figure 10 %%%%%%%%%%%%%%%%%%%%%%%%%%%%%%%%
\begin{figure}
%\picplace{8.0cm}
%\psfig{file=blank.ps,height=8.0truecm,width=8.5truecm}
%\psfig{file=star_rem.ps,height=8.0truecm,width=8.5truecm}
\psfig{file=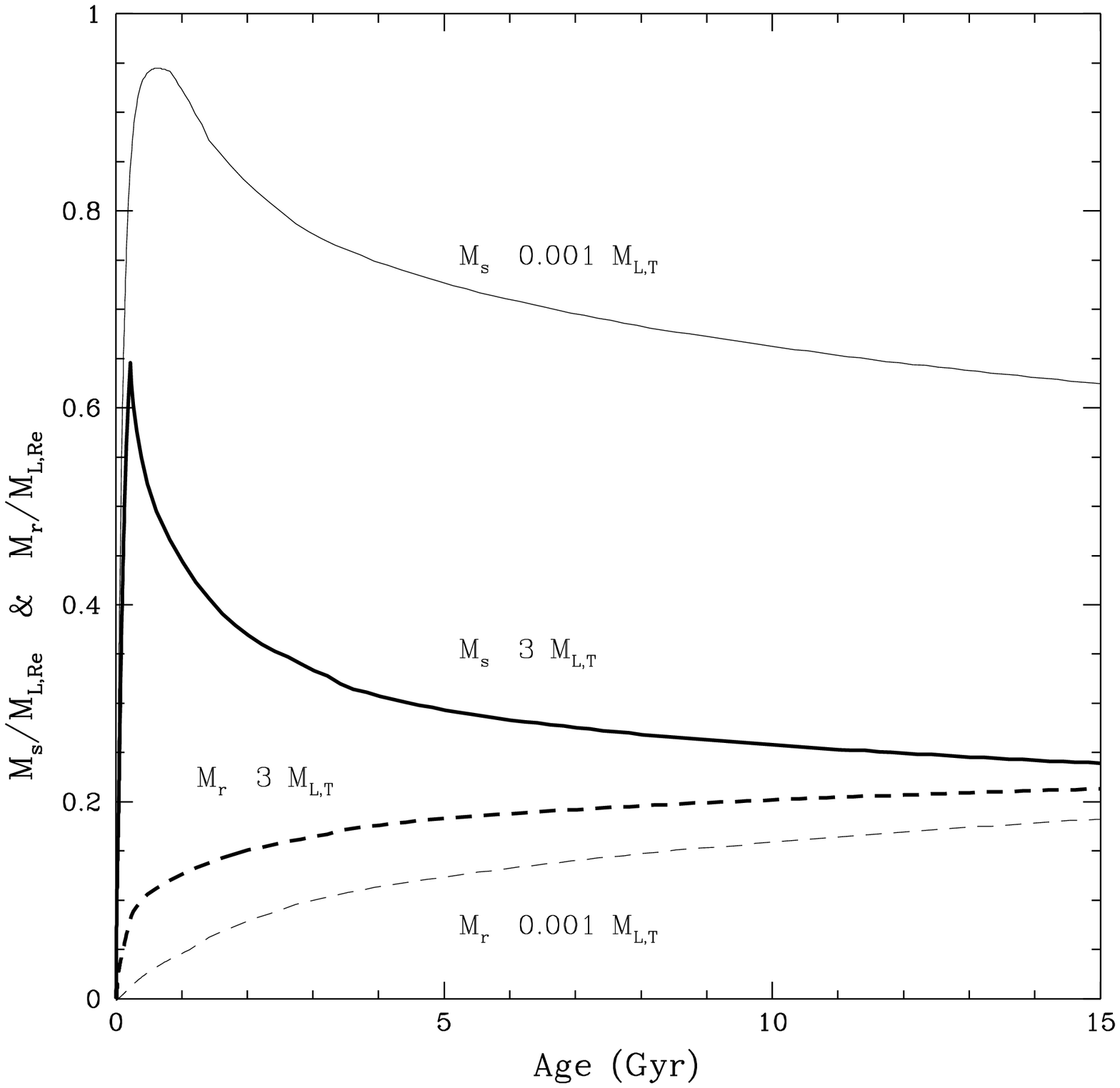,height=8.0truecm,width=8.5truecm}
\caption{ Fractionary masses in stars (solid lines), and remnants (dashed
lines)  as a function of the age for the  0.001  $ M_{L,T}$ (thin lines) and
the 3 $M_{L,T}$  (thick lines) galaxies.}
\label{star_rem}
\end{figure}
%%%%%%%%%%%%%%%%%%%%%%%%%%%%%%%%%%%%%%%%%%%%%%%%%%%%%%%%%%%%%%%%%%%%%%

%%%%%%%%%%%%%%%%Figure 11   %%%%%%%%%%%%%%%%%%%%%%%%%%%%%%%%%%%%%%%%%%% 
\begin{figure}
%\picplace{8.0cm}
%\psfig{file=blank.ps,height=8.0truecm,width=8.5truecm}
%\psfig{file=o_fe_001.ps,height=8.0truecm,width=8.5truecm}
\psfig{file=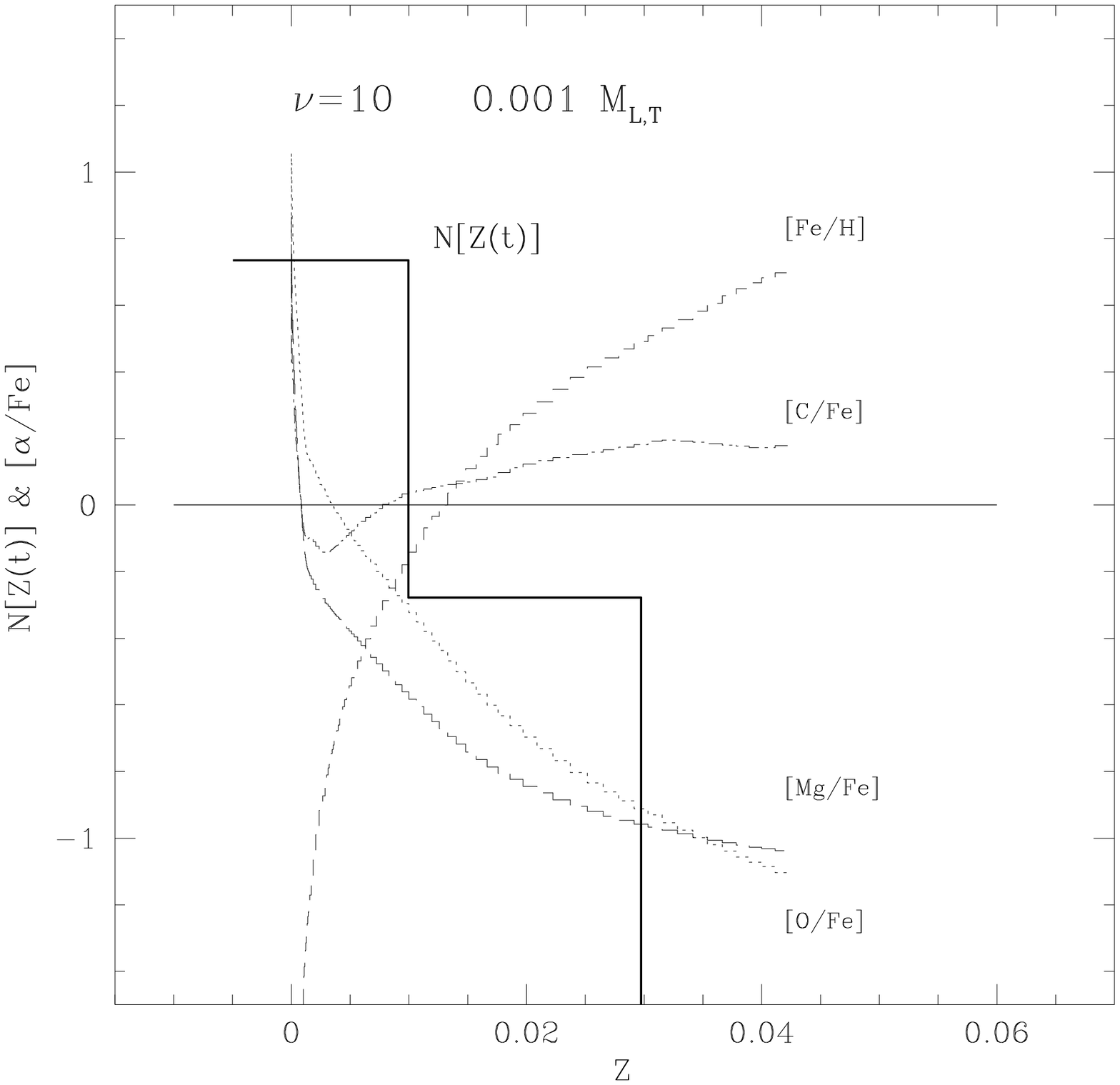,height=8.0truecm,width=8.5truecm}
\caption{ The relative number distribution of living stars $N[Z(t)]$ per
metallicity bin (solid line). $N[Z(t)]$ is on logarithmic scale and in
arbitrary units. The relationship between $[Fe/H]$, $[C/Fe]$, $[O/Fe]$, $[Mg/Fe]$, and
the total metallicity $Z$. The galaxy on display is the 0.001 $M_{L,T}$ model
of Type A at the age of 15 Gyr. The largest fraction of stars in this galaxy
have abundance ratios well below solar, this in spite of the short duration of
the star forming period (0.82 Gyr). This model is the prototype of a galaxy
without enhancement of the $\alpha$-elements.}
\label{o_fe_001}
\end{figure}
%%%%%%%%%%%%%%%%%%%%%%%%%%%%%%%%%%%%%%%%%%%%%%%%%%%%%%%%%%%%%%%%%%%%%%

%%%%%%%%%%%%%%%%Figure 12   %%%%%%%%%%%%%%%%%%%%%%%%%%%%%%%%%%%%%%%%%%% 
\begin{figure}
%\picplace{8.0cm}
%\psfig{file=blank.ps,height=8.0truecm,width=8.5truecm}
%\psfig{file=o_fe_3.ps,height=8.0truecm,width=8.5truecm}
\psfig{file=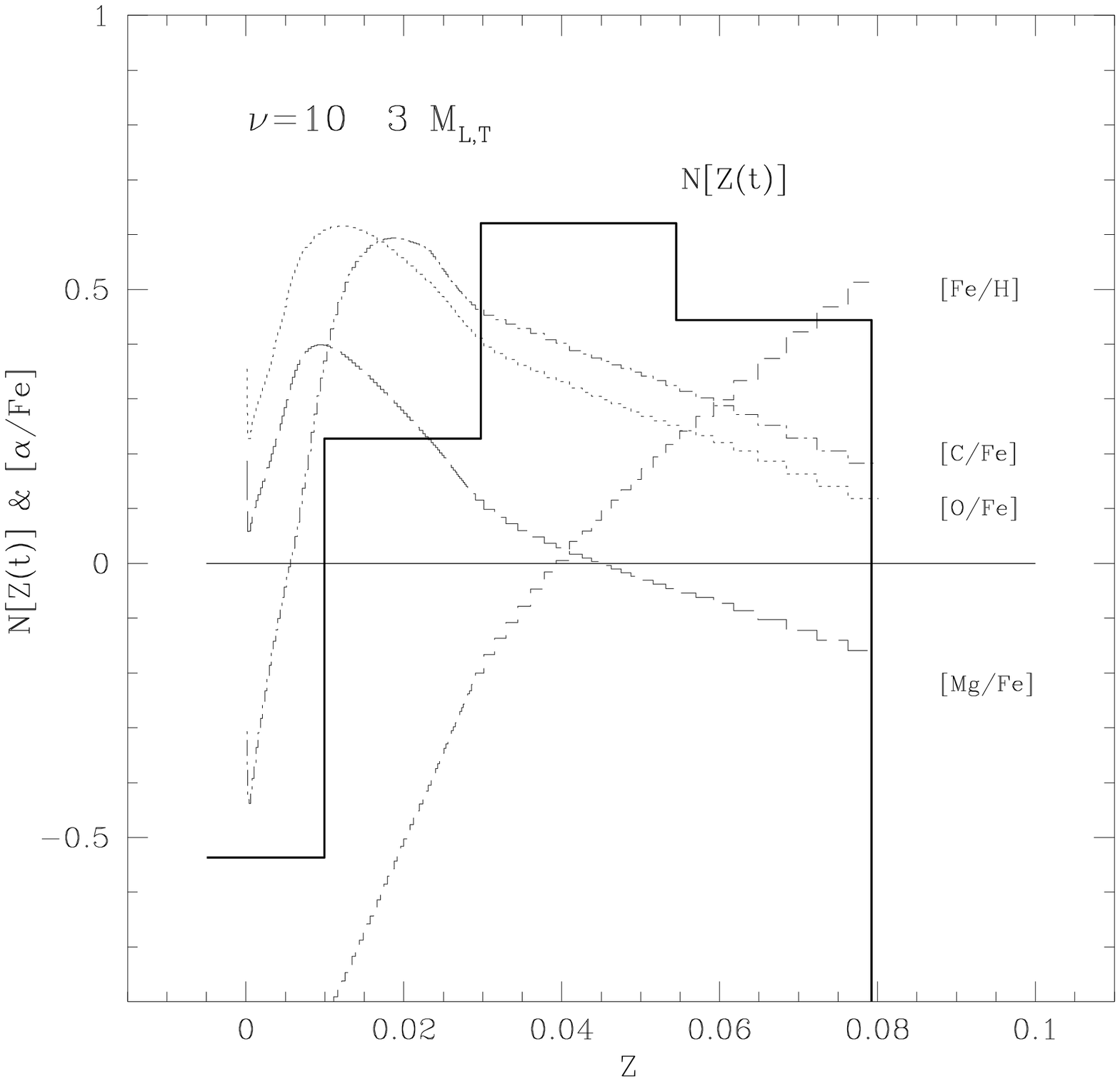,height=8.0truecm,width=8.5truecm}
\caption{ The relative number distribution of living stars $N[Z(t)]$ per
metallicity bin (solid line). $N[Z(t)]$ is on logarithmic scale and in
arbitrary units. The relationship between $[Fe/H]$,  $[C/Fe]$, $[O/Fe]$, $[Mg/Fe]$,
and the total metallicity $Z$. The galaxy on display is the 3 $M_{L,Re}$ model
of Type A at the age of 15 Gyr. The largest fraction of stars in this galaxy
have abundance ratios well above solar. This model is the prototype of a
galaxy with enhancement of the $\alpha$-elements.}
\label{o_fe_3}
\end{figure}
%%%%%%%%%%%%%%%%%%%%%%%%%%%%%%%%%%%%%%%%%%%%%%%%%%%%%%%%%%%%%%%%%%%%%%

%%%%%%%%%%%%%%%%Figure 13   %%%%%%%%%%%%%%%%%%%%%%%%%%%%%%%%%%%%%%%%%%% 
\begin{figure}
%\picplace{8.0cm}
%\psfig{file=blank.ps,height=8.0truecm,width=8.5truecm}
%\psfig{file=zeta_gas.ps,height=8.0truecm,width=8.5truecm}
\psfig{file=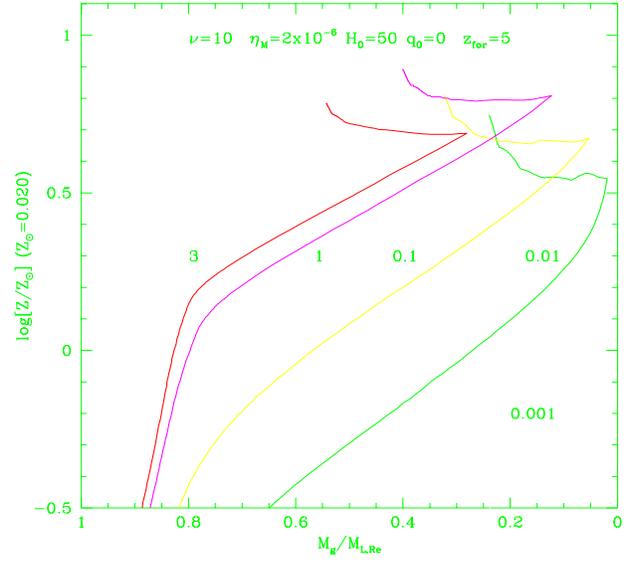,height=8.0truecm,width=8.5truecm}
\caption{ The relation between the current gas fraction $M_g/M_{L,Re}$ and the
current metallicity $Z(t)/Z_{\odot}$ ($Z_{\odot}=0.020$) for all  Type A models
with $\eta_M=2\times 10^{-6}$. The minimum value in $M_g/M_{L,Re}$ corresponds
to the galactic wind stage. The increase in $M_g/M_{L,Re}$ afterwards is due
to dying stars of the previous generations.}
\label{zeta_gas}
\end{figure}
%%%%%%%%%%%%%%%%%%%%%%%%%%%%%%%%%%%%%%%%%%%%%%%%%%%%%%%%%%%%%%%%%%%%%%

%%%%%%%%%%%%%%%%Figure 14   %%%%%%%%%%%%%%%%%%%%%%%%%%%%%%%%%%%%%%%%%%%%%%%% 
\begin{figure}
%\picplace{8.0cm}
%\psfig{file=blank.ps,height=8.0truecm,width=8.5truecm}
%\psfig{file=imf_time.ps,height=8.0truecm,width=8.5truecm}
\psfig{file=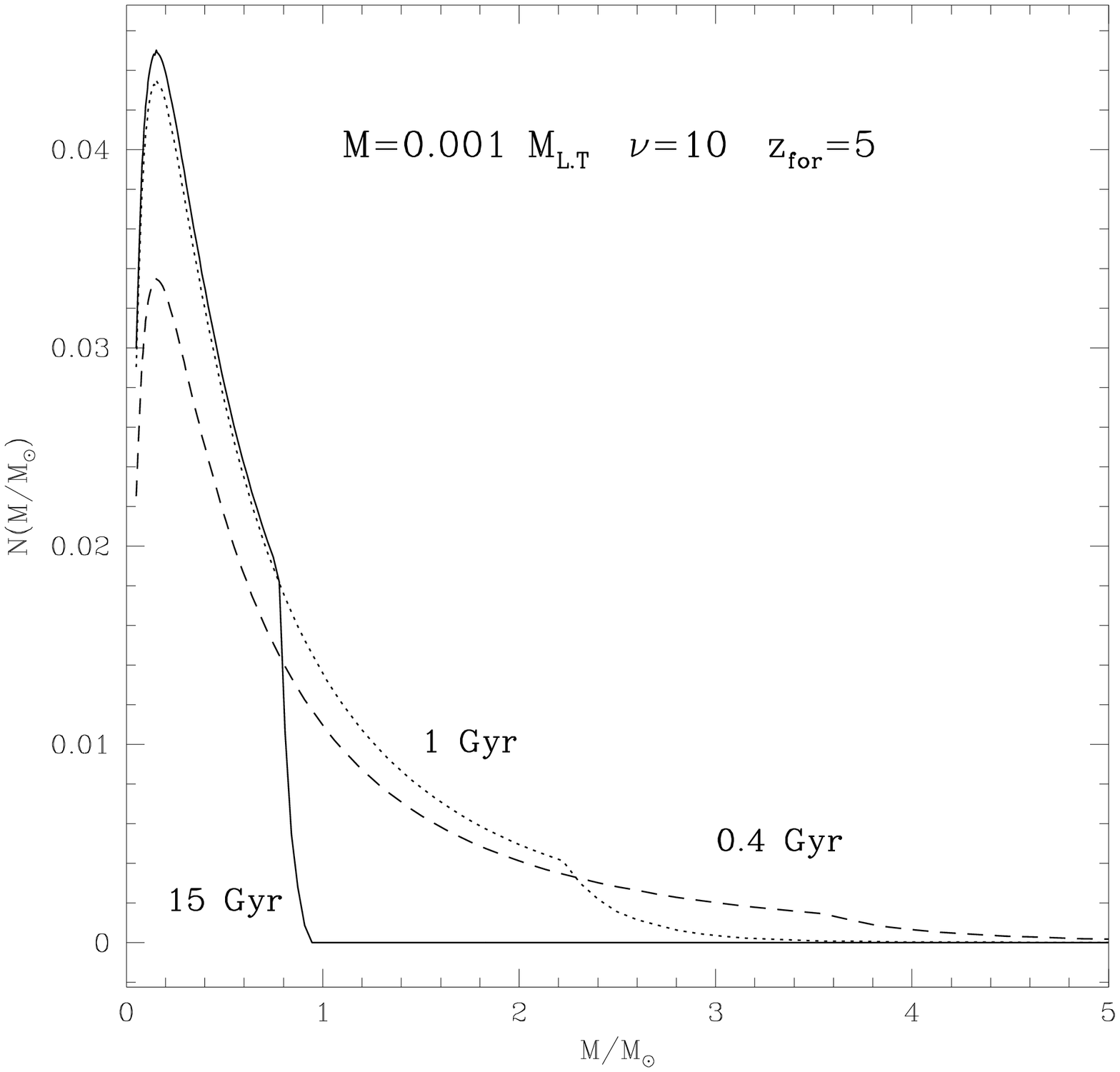,height=8.0truecm,width=8.5truecm}
\caption{The theoretical $N(M)$ for the 0.001 $ M_{L,T}$ model at three
different ages: 0.4 Gyr (solid), 1 Gyr (dotted), and 15 Gyr (dashed).}
\label{imf_time}
\end{figure}
%%%%%%%%%%%%%%%%%%%%%%%%%%%%%%%%%%%%%%%%%%%%%%%%%%%%%%%%%%%%%%%%%%%%%%%
     
%%%%%%%%%%%%%%%Table 2 %%%%%%%%%%%
\begin{table*}[t]
\begin{center}
\caption{Amounts of mass in form of oxygen and iron ejected into the
intra-cluster medium by galaxies of different mass $M_{L,T}$ over their whole
life. The data refer to Type A models with $\eta_M=2\times 10^{-6}$. The symbol $gw$ indicates the amount of material ejected at the stage of
galactic wind. The symbol $sw$ indicates the total amount of material emitted
by dying stars from the wind epoch to the present time. $Fe_T $ and $ O_T$ are
the totals. $M_g^{ej}$ is the total amount of gas thrown into the
intra-cluster medium.  All masses are in $M_{\odot}$. $y_{Fe}$ and  $y_O$ are
crude estimates of the galactic yields of the two elements, i.e.
$M_{j}/M_{L,R_e}$. Finally, $O_T/Fe_T$ is a crude estimate of the degree of
$\alpha$-enhacement in the ejected material.}
\begin{tabular*}{160mm}{ l  ccc ccc ccc c  }
\hline
\hline
   &        & &         &        &         &        &         &      &      &\\ 
$M_{L,T}$ & $M_g^{ej}$& $Fe^{gw}$ & $O^{gw}$ & $Fe^{sw}$& $O^{sw}$  & $Fe_T $ & $ O_T$ & $y_{Fe}$ & $y_O$& $O_T/Fe_T$ \\   
   &        & &         &        &         &        &         &      &      &\\
\hline
   &        & &         &        &         &        &         &      &      &\\  
3     & 6.94(11)& 1.86(9)& 1.24(10)& 6.02(9)& 1.07(10)& 7.88(9)& 2.31(10)& 0.005& 0.019& 2.9 \\
1     & 1.66(11)& 5.84(8)& 2.10(9) & 3.24(9)& 3.52(9) & 3.82(9)& 5.60(10)& 0.009& 0.013& 1.5 \\
0.1   & 1.35(10)& 2.73(7)& 6.15(7) & 3.21(8)& 2.34(8) & 3.48(8)& 2.91(8) & 0.008& 0.010& 0.8 \\
0.01  & 1.00(9) & 1.14(7)& 8.03(5) & 2.56(7)& 9.25(6) & 3.66(7)& 1.00(7) & 0.008& 0.009& 0.3 \\
0.001 & 7.60(7) & 3.95(4)& 1.65(4) & 2.10(6)& 4.35(5) & 2.14(6)& 2.15(6) & 0.005& 0.0005&0.1 \\
   &        & &         &        &         &        &         &      &      &\\ 
\hline
\hline
\end{tabular*}
\end{center}
\label{tab_o_fe}
\end{table*}

%%%%%%%%%%%%%%%Table 3 %%%%%%%%%%%
\begin{table*}[t]
\begin{center}
\caption{Normalized integrated magnitudes $m_i$ of Type A model galaxies with
different ages. Real magnitudes are given by $M_i = - 2.5\log (M_{L,R_e} \times
10^{12}) + m_i$. Magnitudes are for the Johnson photometric system. $M_{L,T}$
and $M_{L,R_e}$ are the mass of the galaxy and the mass inside the effective
radius, both are in units of $10^{12} M_{\odot}$; $M_s$  and   $ M_r$ are the
fractionary masses in living stars and collapsed remnants, respectively. The
normalization mass is $M_{L,R_e}$.}
\scriptsize
\begin{tabular*}{145mm}{ l cc ccc ccc ccc c c  }
\hline
\hline
     &    &      &      &      &      &      &      &      &      &      &      &      &     \\
$M_{L,T}$ &$M_{L,Re}$ &    Age&  $M_s$  &  $ M_r$ & Mb   &  U   &   B  &  V   &   R  &  I   &   J  &    H &  K \\
     &    &      &      &      &      &      &      &      &      &      &      &      &     \\
\hline
     &    &      &      &      &      &      &      &      &      &      &      &      &     \\
\multicolumn{13}{c}{ $\eta_M=2\times 10^{-6}$~~~~   $\nu=10$~~~~   $M_l=0.07 M_{\odot}$ }\\
     &    &      &      &      &      &      &      &      &      &      &      &      &     \\
\hline
     &    &      &      &      &      &      &      &      &      &      &      &      &     \\
3& 1.241&  15. &0.239 &0.213 &6.939 &9.146 &8.637 &7.610 &6.827 &6.159 &5.405 &4.695 &4.500\\  
3& 1.241&   8. &0.270 &0.195 &6.305 &8.355 &7.929 &6.944 &6.187 &5.536 &4.795 &4.087 &3.893\\  
3& 1.241&   5. &0.296 &0.182 &5.831 &7.744 &7.380 &6.435 &5.700 &5.066 &4.344 &3.641 &3.453\\  
     &    &      &      &      &      &      &      &      &      &      &      &      &     \\
1& 0.414&  15. &0.333 &0.271 &6.490 &8.831 &8.254 &7.199 &6.404 &5.717 &4.934 &4.222 &4.021\\
1& 0.414&   8. &0.375 &0.246 &5.869 &8.055 &7.567 &6.555 &5.784 &5.110 &4.331 &3.617 &3.416\\
1& 0.414&   5. &0.411 &0.227 &5.401 &7.446 &7.023 &6.050 &5.301 &4.644 &3.887 &3.175 &2.982\\
     &    &      &      &      &      &      &      &      &      &      &      &      &     \\
0.1& 0.041&  15. &0.435 &0.245 &6.262 &8.323 &7.882 &6.885 &6.117 &5.471 &4.754 &4.053 &3.864\\
0.1& 0.041&   8. &0.484 &0.217 &5.666 &7.576 &7.208 &6.251 &5.509 &4.883 &4.187 &3.491 &3.304\\
0.1& 0.041&   5. &0.523 &0.196 &5.230 &7.002 &6.695 &5.781 &5.062 &4.453 &3.774 &3.085 &2.902\\
     &    &      &      &      &      &      &      &      &      &      &      &      &     \\
0.01& 0.004&  15. &0.591 &0.186 &5.891 &7.762 &7.401 &6.441 &5.694 &5.080 &4.425 &3.744 &3.563\\
0.01& 0.004&   8. &0.657 &0.163 &5.320 &7.062 &6.760 &5.836 &5.113 &4.519 &3.882 &3.209 &3.030\\
0.01& 0.004&   5. &0.635 &0.163 &4.926 &6.524 &6.284 &5.409 &4.711 &4.136 &3.514 &2.850 &2.674\\
     &    &      &      &      &      &      &      &      &      &      &      &      &     \\
0.001& 0.0004& 15. &0.625 &0.182 &5.836 &7.466 &7.195 &6.281 &5.561 &4.992 &4.437 &3.796 &3.631\\
0.001& 0.0004&  8. &0.685 &0.146 &5.313 &6.847 &6.621 &5.738 &5.039 &4.486 &3.936 &3.302 &3.135\\
0.001& 0.0004&  5. &0.724 &0.125 &5.008 &6.376 &6.217 &5.393 &4.726 &4.198 &3.663 &3.046 &2.882\\
     &    &      &      &      &      &      &      &      &      &      &      &      &     \\
\hline
     &    &      &      &      &      &      &      &      &      &      &      &      &     \\
\multicolumn{13}{c}{$\eta_M=3\times^{-6}$~~~~  $\nu=10$~~~~      $M_l=0.07 M_{\odot}$   }\\
     &    &      &      &      &      &      &      &      &      &      &      &      &     \\
\hline
     &    &      &      &      &      &      &      &      &      &      &      &      &     \\
3& 1.241&  15. &0.220 &0.213 &7.476 &9.994 &9.324 &8.237 &7.425 &6.716 &5.896 &5.172 &4.967\\
3& 1.241&   8. &0.247 &0.196 &6.824 &9.137 &8.587 &7.551 &6.768 &6.076 &5.268 &4.542 &4.339\\
3& 1.241&   5. &0.271 &0.184 &6.316 &8.472 &7.996 &7.000 &6.239 &5.566 &4.786 &4.065 &3.870\\
     &    &      &      &      &      &      &      &      &      &      &      &      &     \\
1& 0.414&  15. &0.328 &0.278 &6.486 &8.931 &8.382 &7.372 &6.557 &5.817 &4.937 &4.216 &3.991\\
1& 0.414&   8. &0.368 &0.253 &5.834 &7.976 &7.564 &6.649 &5.871 &5.152 &4.296 &3.577 &3.362\\ 
1& 0.414&   5. &0.403 &0.235 &5.356 &7.594 &7.108 &6.162 &5.382 &4.643 &3.775 &3.045 &2.834\\ 
     &    &      &      &      &      &      &      &      &      &      &      &      &     \\
0.1& 0.041&  15. &0.426 &0.248 &6.276 &8.368 &7.914 &6.910 &6.139 &5.488 &4.763 &4.059 &3.868\\
0.1& 0.041&   8. &0.474 &0.220 &5.681 &7.617 &7.238 &6.276 &5.531 &4.900 &4.196 &3.497 &3.309\\
0.1& 0.041&   5. &0.513 &0.200 &5.241 &7.039 &6.722 &5.802 &5.080 &4.467 &3.779 &3.087 &2.904\\
     &    &      &      &      &      &      &      &      &      &      &      &      &     \\
0.05& 0.021&  15. &0.468 &0.242 &6.141 &8.174 &7.746 &6.754 &5.989 &5.347 &4.639 &3.940 &3.752\\
0.05& 0.021&   8. &0.519 &0.213 &5.553 &7.437 &7.079 &6.128 &5.388 &4.767 &4.080 &3.387 &3.201\\
0.05& 0.021&   5. &0.560 &0.192 &5.125 &6.870 &6.574 &5.666 &4.950 &4.346 &3.675 &2.990 &2.808\\
     &    &      &      &      &      &      &      &      &      &      &      &      &     \\
0.001& 0.0004& 15. &0.621 &0.184 &5.833 &7.467 &7.194 &6.280 &5.559 &4.990 &4.432 &3.790 &3.625\\
0.001& 0.0004&  8. &0.679 &0.149 &5.310 &6.846 &6.619 &5.736 &5.036 &4.483 &3.932 &3.297 &3.130\\
0.001& 0.0004&  5. &0.722 &0.126 &5.003 &6.374 &6.214 &5.389 &4.723 &4.194 &3.658 &3.040 &2.876\\
     &    &      &      &      &      &      &      &      &      &      &      &      &     \\
\hline
\hline
\end{tabular*}
\end{center}
\label{tab_magnit}
\normalsize
\end{table*}

%%%%%%%%%%%%%%%%Figure 15   %%%%%%%%%%%%%%%%%%%%%%%%%%%%%%%%%%%%%%%%%%%%%%%% 
\begin{figure}
%\picplace{8.0cm}
%\psfig{file=blank.ps,height=8.0truecm,width=8.5truecm}
%\psfig{file=imf_model_1e9.ps,height=8.0truecm,width=8.5truecm}
\psfig{file=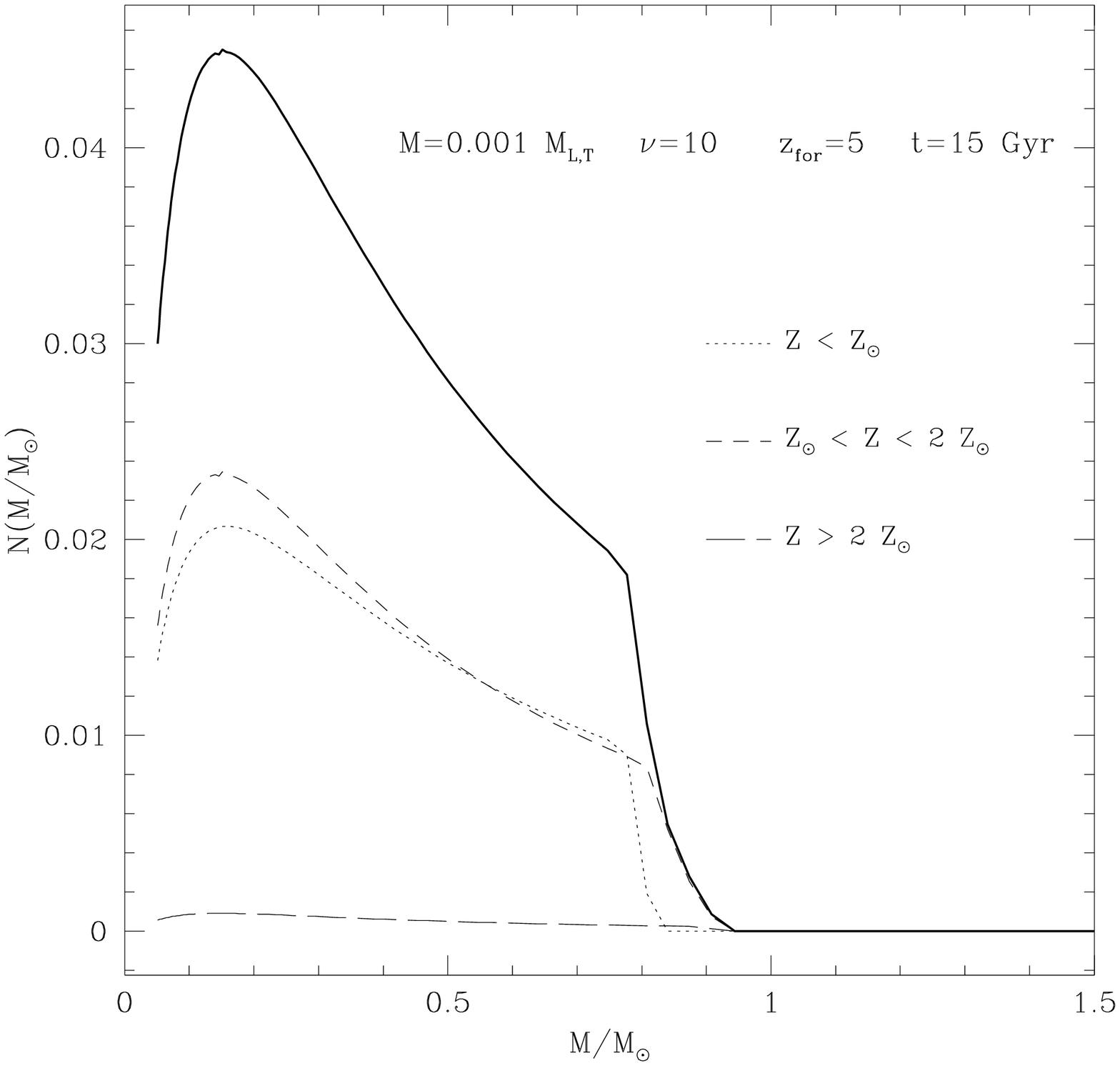,height=8.0truecm,width=8.5truecm}
\caption{ The  theoretical $N(M)$ for the 0.001  $M_{L,T}$ model at the age of
15 Gyr. The top solid line is the total $N(M)$, whereas the other lines are
the partial contributions by   different metallicity intervals:  $Z \leq
Z_{\odot}$ (dotted), $Z_{\odot} < Z \leq 2\times Z_{\odot}$ (dashed), and $Z >
2\times Z_{\odot}$ (long-dashed).}
\label{imf_model_1e9}
\end{figure}
%%%%%%%%%%%%%%%%%%%%%%%%%%%%%%%%%%%%%%%%%%%%%%%%%%%%%%%%%%%%%%%%%%%%%%%

%%%%%%%%%%%%%%%%Figure 16  %%%%%%%%%%%%%%%%%%%%%%%%%%%%%%%%%%%%%%%%%%%%%%%% 
\begin{figure}
%\picplace{8.0cm}
%\psfig{file=blank.ps,height=8.0truecm,width=8.5truecm}
%\psfig{file=imf_model_3e12.ps,height=8.0truecm,width=8.5truecm}
\psfig{file=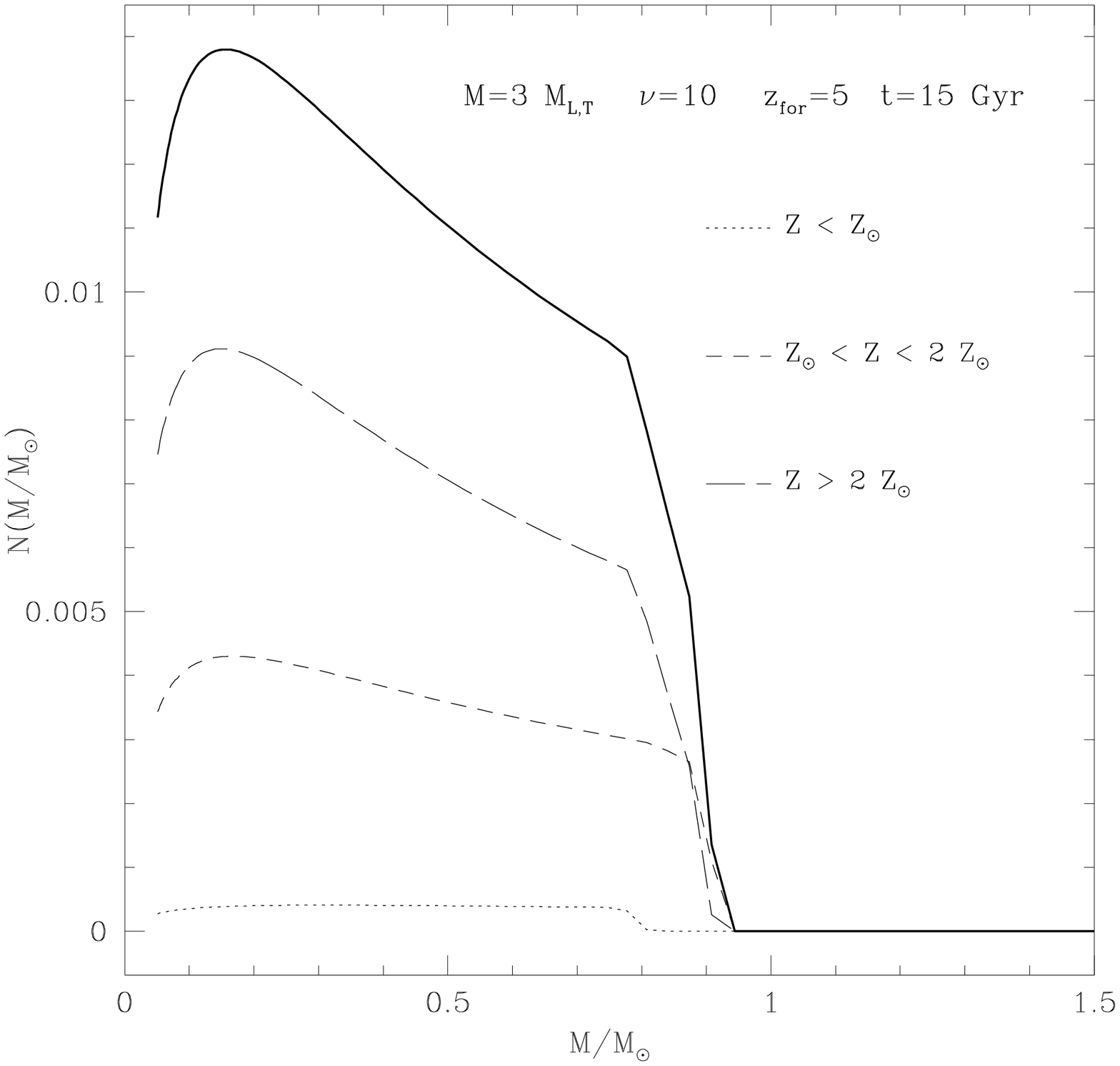,height=8.0truecm,width=8.5truecm}
\caption{The theoretical $N(M)$ for the  3 $ M_{L,T}$ model at the age of 15
Gyr. The top solid line is the total $N(M)$, whereas the other lines are the
partial contributions by different metallicity intervals: $Z \leq Z_{\odot}$
(dotted), $Z_{\odot} < Z \leq 2\times Z_{\odot}$ (dashed), and $Z > 2\times
Z_{\odot}$ (long-dashed).}
\label{imf_model_3e12}
\end{figure}
%%%%%%%%%%%%%%%%%%%%%%%%%%%%%%%%%%%%%%%%%%%%%%%%%%%%%%%%%%%%%%%%%%%%%%%

\subsection{Luminous versus dark material}
Since in the course of evolution, the peak mass of the IMF first grows to a
maximum and then declines, and the maximum value can be as high as $1
M_{\odot}$, we expect different proportions of living stars and collapsed
remnants as a function of time (the present age in particular). In brief,
of the many  stellar generations formed during the activity period $\Delta
t_{SFR}$, those with lifetime shorter than $T_G - \Delta t_{SFR}$, where $T_G$
is the current galactic age, are now in collapsed remnants (black holes,
neutron stars, white dwarfs as appropriate to the initial star mass). They
constitute the "baryonic dark matter" scarcely contributing, if nothing at
all, to the total light. It goes without saying that the remaining stellar
content is in form of visible stars almost fully contributing to the total
light. How much mass is present in "baryonic dark matter" ? How does the
fractionary value of this mass vary with the galaxy mass (local density and
temperature in turn) ? To this aim, in Fig.~\ref{star_rem} we plot the
fractions of visible stars (solid line) and collapsed remnants (dashed lines)
as a function of the age. The corresponding gas fraction was already shown in
Fig.~\ref{gas_cont}. For the sake of clarity, we limit ourselves to display
only the 0.001 and 3 $M_{L,T}$ galaxies.

It is worth noticing  that while the fractionary masses of living stars and
gas are calculated with high precision from the set of equations governing
chemical evolution, the fractionary mass of remnants is  calculated separately
with a precision of about 5\%. This explains why the sum of the three
fractions may sometimes be slightly greater than one ($1 \div 1.05$ in
general).
 
Interestingly enough, in the $R_e$-sphere of a low mass galaxy (0.001
$M_{L,T}$) about 80\% of the original mass $M_{L,Re}$ is in form of visible
stars, the rest is in remnants, and there are almost no traces of gas.
Conversely, in the same region of a massive galaxy (3 $M_{L,T}$)  only 25\% of
 the original mass $M_{L,R_e}$ is in now form of living stars, about 20\%  of
it in form of collapsed remnants, and 55\% of it has remained in form of gas
partly expelled at the stage of galactic winds and partly continuously ejected
by dying stars and never re-used of form stars. For the same arguments advocated
 to predict the gradient in the duration of the star forming activity across a
galaxy, we expect that the fraction of living stars ought to decrease with
galactocentric distance. It goes without saying that as a consequence of this,
one expects the mass-luminosity ratio to increase with the galaxy mass and the
radial distance.

The time dependence of the gas content deserves some final remarks. At the
onset of galactic winds, the gas content has reached its minimum value because
of the past star formation activity. However, generous amounts of gas are
later lost by dying stars. The fractionary mass of the newly emitted  gas is
significantly larger than what customarily possible with standard,
Salpeter-like IMF. Since this gas is likely soon heated up above the virial
temperature and  lost by the galaxy, it will significantly contribute to
enrich and fuel the intra-cluster medium (see below.

\subsection{Metallicity and enhancement of $\alpha$-elements} 
How do the
metallicity and enhancement of $\alpha$-elements vary as a function of time
and galactic mass under this continuously changing IMF ? The answer is
provided by Figs.~\ref{o_fe_001} and \ref{o_fe_3}, in which we plot the
relative distribution of stars, $N[Z(t)]$, per metallicity bin  together with the
relationships between $[Fe/H]$, $[C/Fe]$, $[O/Fe]$, and $[Mg/Fe]$, and the total
metallicity $Z$. Fig.~\ref{o_fe_001} refers to  the 0.001 $M_{L,T}$ galaxy,
whereas Fig.~\ref{o_fe_3} refers to the 3 $M_{L,T}$ object. Both galaxies
belong to group Type A and are displayed at the age of 15 Gyr. Similar
considerations and conclusions apply to Type B models as well. There are two
important  results  out of the data shown  in Figs.~\ref{o_fe_001} and
\ref{o_fe_3}:

\begin{itemize}
\item{ In the 3$M_{L,T}$ galaxy there are very few stars stored in low
metallicity bins say up to $Z=0.01$. This implies that for this galaxy the
so-called G-Dwarf analog does not occur. In the opposite extreme, i.e. the
0.001$M_{L,T}$ object, about 50\% of the stars are in the bins up to $Z=0.01$,
but $[Fe/H]$ increases rapidly to a tenth of solar, so that
very few stars of very low $Z$ are found. Also in this case the G-Dwarf analog
is avoided. {\it The kind of IMF we are using naturally secures   a lower metallicity cut-off in the distribution of living stars, similar to what required by BCF94 to remove low metallicity stars, and a sort of prompt enrichment in the closed-box model. The need of infall-schemes to avoid the G-Dwarf problem is somehow ruled out by the new IMF.}}
\littleskip

\item{ The totality  of stars in the 3$M_{L,T}$ galaxy have abundance ratios
$[C/Fe]$ and $[O/Fe]$ well above solar. As far as the ratio $[Mg/Fe]$ is concerned,
it seems that about 50\% of stars (those with $Z  \leq 0.04$) have the ratio
$[Mg/Fe]$ above solar whereas the remaining stars have $[Mg/Fe]$  below solar.
Before drawing any conclusion out of this, it should be kept in mind
that the stellar yield of Mg used by Portinari et al. (1997), TCBMP97, and in
this paper, underestimates  the production of Mg (see Portinari et al. 1997
for all details), so that the Mg abundance cannot be safely used as an
indicator of $\alpha$-enhancement. No difficulty  exists with the yield of  O,
whereas there is a marginal discrepancy in the yield of C. Therefore, adopting
$[O/Fe]$ as enhancement indicator, we can
conclude that this model can be taken as the {\it prototype of an
$\alpha$-enhanced system}. The opposite occurs with the 0.001 $M_{L,T}$ galaxy
in which almost the totality of stars have $[O/Fe]$ and $[Mg/Fe]$ below solar,
whereas a large fraction of these (the ones with $Z \geq 0.01$) have $[C/Fe]$
above solar. This model can be taken as the {\it prototype of a galaxy with no
enhancement in $\alpha$-elements}. Remarkably, this is possible in a system in
which the past history of stellar activity although shorter than in the high
mass case, still was confined within  0.82 Gyr, too short a time for a
significant contamination by Type I supernovae with the classical Salpeter
law. The new IMF makes it possible because at the densities typical of the
0.001 $M_{L,T}$ galaxy, the IMF tends to skew toward the low mass end, thus
favoring the formation of Type I supernovae.   }
\end{itemize}

These results are particularly interesting because two of the most intriguing
and at some extend demanding constraints imposed by the observational data,
i.e. the G-Dwarf analog and enhancement in $\alpha$-elements as a function of
the galaxy luminosity (mass in turn) are ruled out by the same token.

Finally, in view of the discussion below, we present the relation between the
current gas fraction and current metallicity displayed  in Fig.~\ref{zeta_gas}
(the metallicity is in solar units). The reversal in the trend shown by each
curve beyond the minimum value in the gas fraction reached at the wind stage
corresponds to the further supply by dying stars. Two  notable features can be
pointed out: (i) the metallicity is either ever increasing or has a local
minimum (effect of the gas restitution by stars of the previous generations);
(ii) the fractionary total amount of gas ejected by dying stars is nearly
independent of the galaxy mass, going from 0.25  for the 3 $M_{L,T}$ galaxy to
0.18  for the 0.001 $M_{L,T}$ object (mean value 0.22). Moving from this
diagram and the entries of Table~1, one can estimate the total amount of gas
and metals made available to the intra-cluster medium by each galaxy. This
estimate is a lower limit because we refer only to the $R_e$-sphere. The
amounts of mass in form of oxygen and iron ejected by the galaxies (into the
intra-cluster medium)  are given in Table~2, whereby the meaning of the
symbols is obvious. The data refer only  to Type A models with $\eta_M=2\times
10^{-6}$ for the sake of simplicity. The contributions at the stage of
galactic wind and all over the subsequent history are shown separately.
Remarkably,  the contribution from the post wind phase parallels or even
overwhelms that from the wind stage. The galactic yield ($y_j=M_{j}/M_{L,Re}$,
$j$ stands for the generic element), of $Fe$ varies from  $y_{Fe} \simeq 0.005
$ to 0.008, depending on the galaxy mass, whereas that for oxygen  increases
from $y_O \simeq 0.0005$ for the 0.001 $M_{L,T}$ galaxy to 0.018 for the 3
$M_{L,T}$ object. How do these results  compare with those by Gibson \&
Matteucci (1997), who presented similar calculations? Looking at the two model
galaxies in common, namely the 1 and 0.001 $M_{L,T}$ objects, and keeping in
mind that our models refer to the $R_e$-sphere (therefore the amounts of
ejecta are under-estimated) whereas those by Gibson \& Matteucci (1997) are for 
the  whole galaxy, we get the  following results (GM and CBPT stand for Gibson \&
Matteucci and the present paper, respectively):

\begin{itemize}
\item{1 $M_{L,T}$: $\rm (GM/CBPT)_g=2.2$,  $\rm (GM/CBPT)_O=0.3$  and $\rm
(GM/CBPT)_{Fe}=0.4$ compared to their {\it standard case}, and $\rm
(GM/CBPT)_g=3.8$, $\rm (GM/CBPT)_O=0.3$ and $\rm (GM/CBPT)_{Fe}=0.4$ compared
to their {\it reduced binding energy model}.}

\item{0.001 $M_{L,T}$: $\rm (GM/CBPT)_g=10$,  $\rm (GM/CBPT)_O=10$ and $\rm
(GM/CBPT)_{Fe}=1$ compared to their {\it standard case}, and $\rm
(GM/CBPT)_g=13$, $\rm (GM/CBPT)_O=0.9$ and $\rm (GM/CBPT)_{Fe}=0.1$ compared
to their {\it reduced binding energy model}.}
\end{itemize}

\noindent
We will come back to  this topic in the final discussion.    

%%%%%%%%%%%%%%%Table 4 %%%%%%%%%%%
\begin{table*}[t]
\begin{center}
\caption{Integrated magnitudes, colors, and mass to light ratios.}
\scriptsize
\begin{tabular*}{140mm}{l cc ccc ccc cc }
\hline
\hline
      &     &        &          &          &         &        &       &      &       &       \\
$M_{L,T}$& Age&  $logM_{L,re}$ &$M_s+M_r \over M_{L,Re}$ & $ M_{bol}$ &
                 (B-V) &  (V-K) & $M_B$ & $M_K$ & $log(M/L_B)$ &$log(M/L_K)$\\
      &     &        &          &          &         &        &       &      &       &       \\
\hline
      &     &        &          &          &         &        &       &      &       &       \\
\multicolumn{10}{c} { $\eta_M=2\times 10^{-6}$~~ $\nu=10$~~ $H_0=50$~~ 
                          $q_0=0$~~ $z_{for}=5  $}\\
      &     &        &          &          &         &        &       &      &       &       \\
\hline
      &     &        &          &          &         &        &       &      &       &       \\
3     &  15 & 12.09  &    0.452 &   -23.29 &   1.027 &  3.110 &-21.60 &-25.73& 0.9179& 0.1231\\
1     &  15 & 11.62  &    0.604 &   -22.55 &   1.055 &  3.178 &-20.79 &-25.02& 0.8906& 0.0574\\
0.1   &  15 & 10.62  &    0.680 &   -20.30 &   0.997 &  3.021 &-18.66 &-22.68& 0.7933& 0.0461\\
0.01  &  15 &  9.61  &    0.777 &   -18.11 &   0.960 &  2.878 &-16.63 &-20.47& 0.6588&-0.0163\\
0.001 &  15 &  8.61  &    0.807 &   -15.67 &   0.914 &  2.650 &-14.31 &-17.87& 0.5929& 0.0272\\
      &     &        &          &          &         &        &       &      &       &       \\
3     &  8  & 12.09  &    0.465 &   -23.92 &   0.985 &  3.051 &-22.31 &-26.34& 0.6471&-0.1073\\
1     &  8  & 11.62  &    0.621 &   -23.17 &   1.012 &  3.139 &-21.48 &-25.63& 0.6279&-0.1727\\
0.1   &  8  & 10.62  &    0.701 &   -20.86 &   0.957 &  2.947 &-19.33 &-23.24& 0.5369&-0.1647\\
0.01  &  8  &  9.61  &    0.820 &   -18.68 &   0.924 &  2.806 &-17.27 &-21.00& 0.4258&-0.2062\\
0.001 &  8  &  8.61  &    0.831 &   -16.19 &   0.883 &  2.603 &-14.88 &-18.37& 0.3760&-0.1584\\
      &     &        &          &          &         &        &       &      &       &       \\
3     &  5  & 12.09  &    0.478 &   -24.40 &   0.945 &  2.982 &-22.85 &-26.78& 0.4394&-0.2714\\
1     &  5  & 11.62  &    0.638 &   -23.64 &   0.973 &  3.068 &-22.02 &-26.06& 0.4220&-0.3344\\
0.1   &  5  & 10.62  &    0.719 &   -21.30 &   0.914 &  2.879 &-19.85 &-23.64& 0.3427&-0.3145\\
0.01  &  5  &  9.61  &    0.798 &   -19.07 &   0.875 &  2.735 &-17.75 &-21.36& 0.2236&-0.3606\\
0.001 &  5  &  8.61  &    0.849 &   -16.50 &   0.824 &  2.511 &-15.29 &-18.62& 0.2237&-0.2503\\
      &     &        &          &          &         &        &       &      &       &       \\
\hline
      &     &        &          &          &         &        &       &      &       &       \\
\multicolumn{10}{c} { $\eta_M=3\times 10^{-6}$~~ $\nu=10$~~ $H_0=50$~~
                         $q_0=0$~~ $z_{for}=5  $}\\
      &     &        &          &          &         &        &       &      &       &       \\
\hline
      &     &        &          &          &         &        &       &      &       &       \\
3     & 15  & 12.09  &    0.433 &   -22.75 &   1.087 &  3.270 &-20.91 &-25.27& 1.1740& 0.2913\\
1     & 15  & 11.62  &    0.606 &   -22.55 &   1.010 &  3.381 &-20.66 &-25.05& 0.9433& 0.0468\\
0.1   & 15  & 10.62  &    0.674 &   -20.25 &   1.004 &  3.042 &-18.63 &-22.67& 0.8023& 0.0438\\
0.05  & 15  & 10.32  &    0.710 &   -17.86 &   0.992 &  3.002 &-18.04 &-22.04& 0.7577& 0.0200\\
0.001 & 15  &  8.62  &    0.805 &   -15.67 &   0.914 &  2.655 &-14.31 &-17.88& 0.5914& 0.0238\\
      &     &        &          &          &         &        &       &      &       &       \\
3     & 8   & 12.09  &    0.443 &   -23.41 &   1.036 &  3.212 &-21.65 &-25.90& 0.8892& 0.0500\\
1     & 8   & 11.62  &    0.621 &   -23.20 &   0.915 &  3.287 &-21.48 &-25.68& 0.6267&-0.1941\\
0.1   & 8   & 10.62  &    0.694 &   -20.85 &   0.962 &  2.967 &-19.30 &-23.23& 0.5446&-0.1670\\
0.05  & 8   & 10.32  &    0.732 &   -18.45 &   0.951 &  2.927 &-18.71 &-22.59& 0.5041&-0.1871\\
0.001 & 8   &  8.62  &    0.828 &   -16.19 &   0.883 &  2.606 &-14.89 &-18.38& 0.3736&-0.1620\\
      &     &        &          &          &         &        &       &      &       &       \\
3     & 5   & 12.09  &    0.455 &   -23.91 &   0.996 &  3.130 &-22.24 &-26.36& 0.6644&-0.1260\\
1     & 5   & 11.62  &    0.638 &   -23.08 &   0.946 &  3.328 &-21.93 &-26.21& 0.4560&-0.3936\\
0.1   & 5   & 10.62  &    0.713 &   -21.00 &   0.920 &  2.898 &-19.82 &-23.64& 0.3499&-0.3173\\
0.05  & 5   & 10.32  &    0.752 &   -18.87 &   0.908 &  2.858 &-19.22 &-22.98& 0.3138&-0.3326\\
0.001 & 5   &  8.62  &    0.848 &   -16.50 &   0.825 &  2.513 &-15.29 &-18.63& 0.2220&-0.2532\\
      &     &        &          &          &         &        &       &      &       &       \\
\hline
\hline
\end{tabular*}
\end{center}
\label{tab_masslight}
\normalsize
\end{table*}

\subsection{How does this IMF vary from galaxy to galaxy ?}
Owing to the systematic change of the IMF with the physical conditions of the
interstellar medium in which stars are formed, we expect the distribution of
stars per mass interval $N(M)$ to vary with time and hence metallicity and
from galaxy to galaxy. Figure~\ref{imf_time} shows $N(M)$ at three different
values of the age, namely 0.2, 0.6, and 15.0 Gyr, for the $0.001  M_{L,T}$
galaxy. The drop-off along the three curves at certain values of the mass
corresponds to the current value of the turn-off, i.e. 0.98 $M_{\odot}$ at 15
Gyr, 2.25 $M_{\odot}$ at 0.6 Gyr, and 3.55 $M_{\odot}$ at 0.2 Gyr. Notice how
at increasing  age  the part of the  $N(M)$  below the turn-off  gradually
steepens.  At the age of 15 Gyr the IMF much resembles the Salpeter law but
for the cut-off below 0.15 $M_{\odot}$. Figures~\ref{imf_model_1e9} and
\ref{imf_model_3e12} shows the inner structure of the $N(M)$ as a function of
the metallicity for the $0.001 M_{L,T}$ and $3 M_{L,T}$ models, respectively,
at the constant age of 15 Gyr. The solid line is the total $N(M)$, whereas the
dotted, dashed, and long-dashed lines are the partial $N(M)$ in the
metallicity intervals $Z \leq Z_{\odot}$, $Z_{\odot} < Z \leq 2\times
Z_{\odot}$, and $Z > 2\times Z_{\odot}$, respectively. Note how the trend is
reversed passing from the $0.001 M_{L,T}$ galaxy to the $3 M_{L,T}$ one. In
the former, nearly the totality of stars  have metallicities smaller than
$2\times Z_{\odot}$, in the latter the opposite occurs with the majority of
stars in the range $Z >  Z_{\odot}$.

%%%%%%%%%%%%%%%%%%%%%%%%%%%%%%%%%%%%%%%%%%%%%%%
\section{Photometric synthesis: CMR and FP}
Since the IMF is no longer constant with respect to time the standard
SSP-technique cannot be used to calculate the integrated flux, spectrum,
magnitudes, and colors of a  galaxy (BCF94, TCBF96). They must be derived from
directly calculating  the integrated monochromatic flux generated by the 
stellar content of a galaxy of age $T$

\begin{equation}
F_{\lambda}(T) = \int_0^T \int_{M_L}^{M_U} \Psi(t) \phi(t,M) f_{\lambda}
(M,\tau',Z)  dt  dM
\label{flux}
\end{equation}
where $f_{\lambda}(M,\tau',Z)$ is the monochromatic flux of a star of mass $M$,
metallicity $Z(t)$, and age $\tau'=T-t$, and $\Psi(t)$ is the rate of star
formation. This poses severe problems of numerical nature and computer
storage, of which no details are given here. Finally, the spectral
library, the relationship between effective temperature, metallicity,  and
spectral type of the stars in the various evolutionary stages, in usage
here are as in BCF94
and TCBF96 to whom the reader should refer.

The photometric data (magnitudes and colors) in several pass-bands of the
Johnson system  are listed in Table~3 for three values of the age (15, 8 and 5
Gyr) of Type A models. In view of the discussion below, two sets of models
with $\eta_M=2$ and $\eta_M=3$ are presented. All the magnitudes listed in
Table~3 are normalized to the mass $M_{L,Re}$. Real absolute magnitudes  are
therefore given by

\begin{equation}
   M_i = -2.5 log(M_{L,Re} \times 10^{12}) + m_i
\end{equation}
\noindent

%%@@ To convert our magnitudes M_i in total magnitudes we notice that M_i
%%@@ refers to the effective radius R_e. We thus apply DM_i=-0.75 irrespective
%%@@ of the presence of eventual color gradients.

Since our magnitudes  $M_i$ refer  to the $R_e$-sphere, they should be converted to total magnitudes before comparing them with 
the observational data. To a first approximation, one could apply the maximum correction   
$\Delta M_i = -0.75 $ irrespective of the possible of color gradients.

\subsection{ The color-magnitude relation } 
\begin{figure}
%\picplace{8.0cm}
%\psfig{file=blank.ps,height=8.0truecm,width=8.5truecm}
%\psfig{file=cmr.ps,height=8.0truecm,width=8.5truecm}
\psfig{file=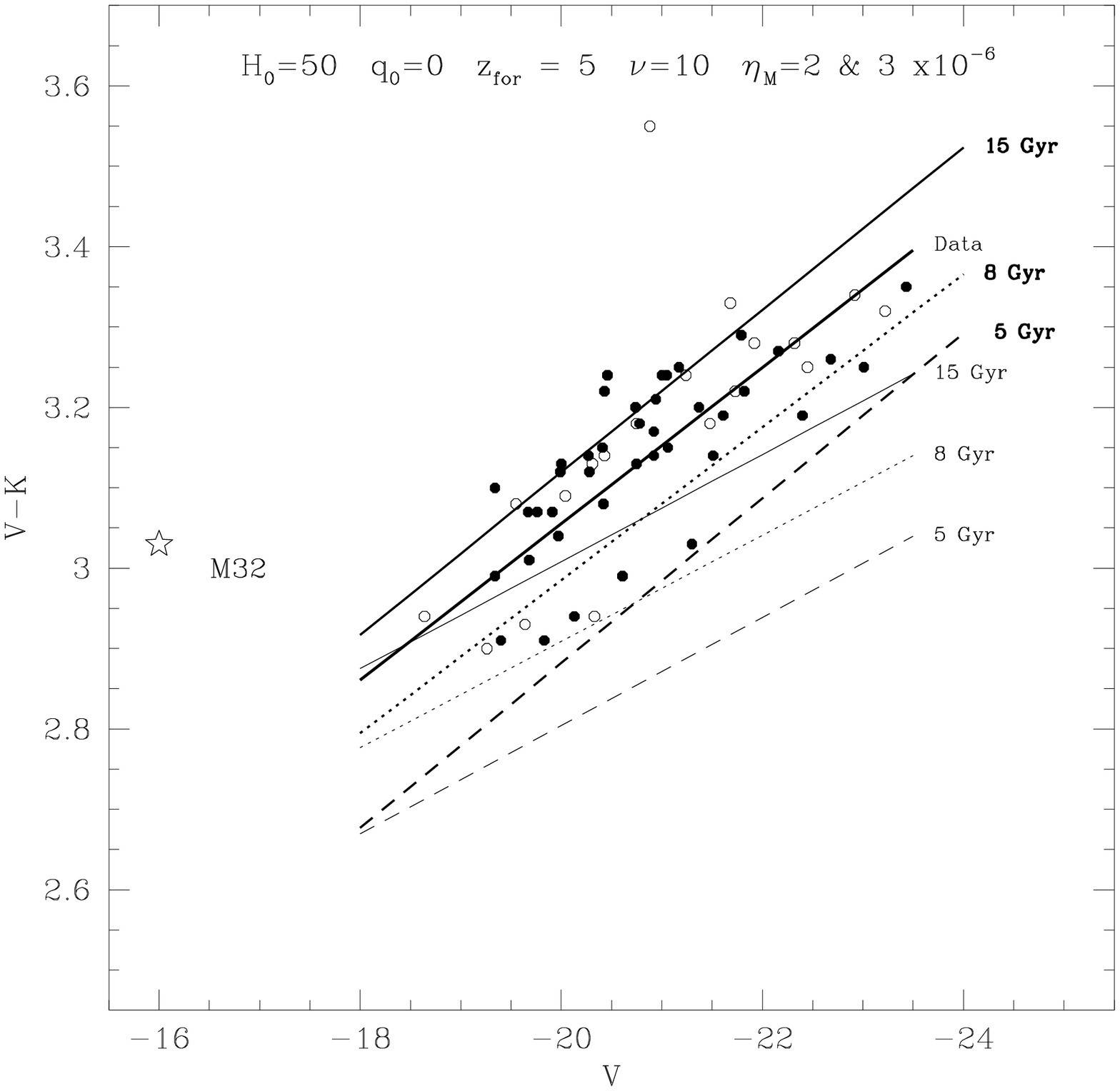,height=8.0truecm,width=8.5truecm}
\caption{ CMR. The filled and open circles are the data for the Virgo and Coma
galaxies of Bower et al. (1992). The distance modulus to Virgo is
$(m-M)_0=31.54$ (Branch \& Tammann 1992), whereas that to Coma is
$(m-M)_0=35.12$ (Bower et al. 1992). The thick solid line labelled {\it Data}
is the linear square fit of the data. The  theoretical relationships are
displayed by the linear square fit of the data of Tables~2 and 3: thick lines
for models with $\eta_M=3\times 10^{-6}$ and thin lines for models with
$\eta_M=2\times 10^{-6}$. Three values of the age are considered, namely  15,
8, and 5 Gyr,  as indicated  along each line.}
\label{cmr}
\end{figure}

The CMR predicted by our models is shown in  Fig.~\ref{cmr} for the three
values of the age  and it is compared to the data of Bower et al. (1992) for
the Virgo and Coma galaxies. The adopted distance modulus to Virgo is
$(m-M)_0=31.54$ (Branch \& Tammann 1992), whereas that to Coma is
$(m-M)_0=35.12$ (Bower et al. 1992). The thick line labelled {\it Data} is the
linear square fit of the observational data. The theoretical CMR are displayed
by the linear square fits of the data listed in Tables~2 and 3 separately for
models with $\eta_M=3\times 10^{-6}$ (thick lines) and  $\eta_M=2\times
10^{-6}$ (thin lines). Three values of the age are considered, i.e. 15, 8, and
5 Gyr. Finally, no correction $\Delta M_i$ is applied. Although the agreement with the data is not perfect, yet the slope of
the CMR is matched. Indeed the observational data is compatible with 
the 15 Gyr
relationship of the  $\eta_M=3 \times 10^{-6}$ models. 

It is
worth recalling that even in these models the CMR is a mass-metallicity
sequence, because despite the shorter duration of the star forming activity
and earlier winds in massive galaxies with respect to the low mass ones, the
mean metallicity of the former is higher than in the latter thanks to the
compensatory effect of the IMF  which skews toward the high mass end. {\it The
great advantage now is that the contradiction between CMR and } $\alpha$-{\it
enhancement as far as the overall duration of the star forming activity in
galaxies of different mass is concerned (cf. section 2.2 and 2.3) does no
longer occur}.

%%%%%%%%%%%%%%%%Figure 17   %%%%%%%%%%%%%%%%%%%%%%%%%%%%%%%%%%%%%%%%%%% 
%%%%%%%%%%%%%%%%%%%%%%%%%%%%%%%%%%%%%%%%%%%%%%%%%%%%%%%%%%%%%%%%%%%%%%

%%%%%%%%%%%%%%%%Figure 18   %%%%%%%%%%%%%%%%%%%%%%%%%%%%%%%%%%%%%%%%%%%%%%%% 
\begin{figure}
%\picplace{8.0cm}
%\psfig{file=blank.ps,height=8.0truecm,width=8.5truecm}
%\psfig{file=m_lb.ps,height=8.0truecm,width=8.5truecm}
\psfig{file=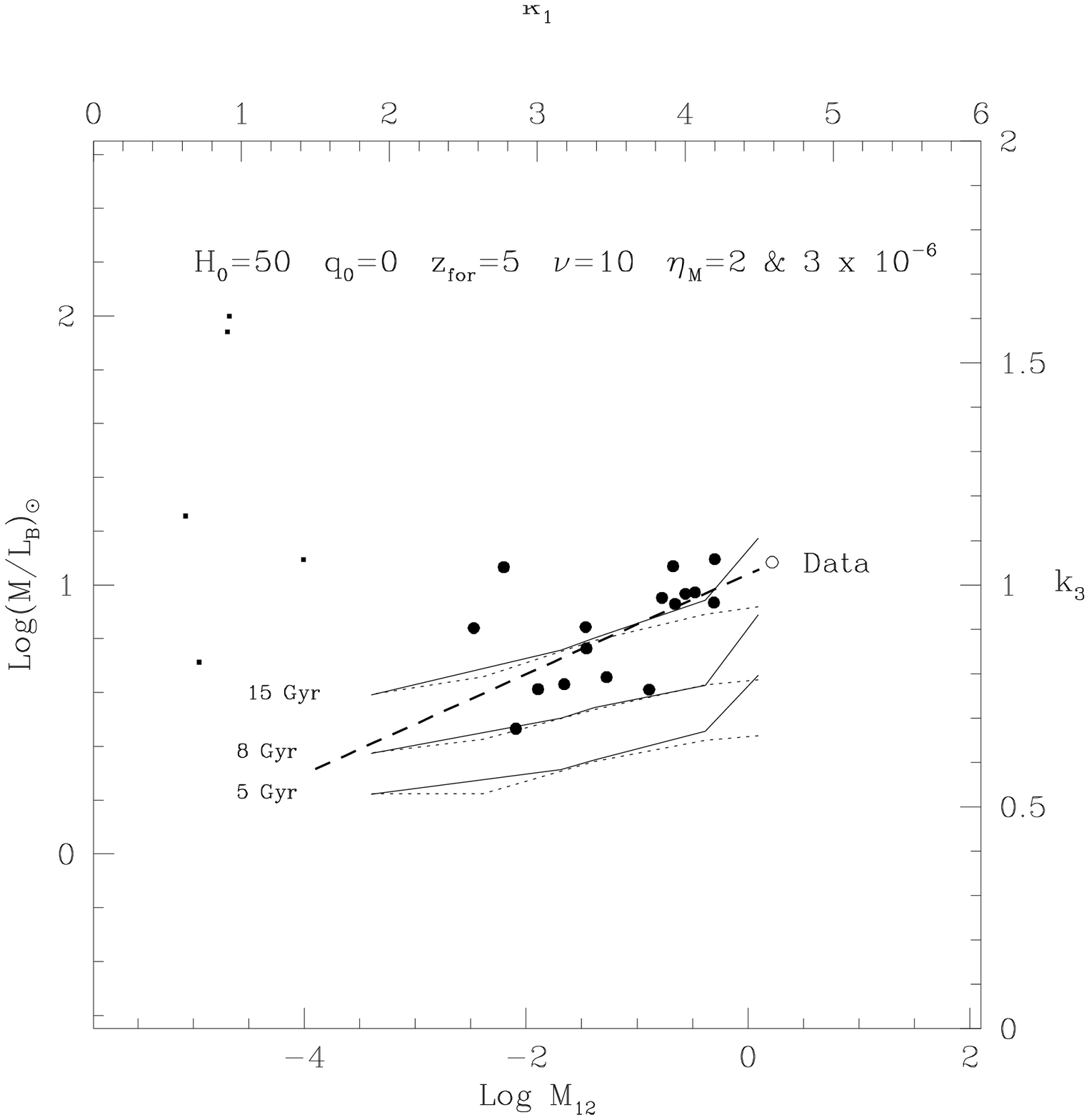,height=8.0truecm,width=8.5truecm}
\caption{FP. The theoretical  $log(M/L_B)_{\odot}$ ratios as a function of the
total mass $log M_{12}$ for the ages of 15, 8 and 5 Gyr as indicated. The
solid lines are for $\eta_M=3\times 10^{-6}$, whereas the dotted lines are for
$\eta_M=2\times 10^{-6}$. The thick dashed line labelled {\it Data} is the
relation for the Virgo galaxies,  whereas the filled  circles  are the data by
Bender et al. (1992) for the same objects. The small dots show a few dwarf
ellipticals for purposes of comparison.}
\label{m_lb}
\end{figure}
%%%%%%%%%%%%%%%%%%%%%%%%%%%%%%%%%%%%%%%%%%%%%%%%%%%%%%%%%%%%%%%%%%%%%%%

\subsection{Tilt of the fundamental plane}
The question to be addressed is:  can this IMF account for the slope of the FP
without invoking changes in the virial coefficients? To answer this question
we have calculated the mass-luminosity ratios  $(M/L_B)_{\odot}$ and
$(M/L_K)_{\odot}$ for Type A models with  $\eta_M=2\times 10^{-6}$ and
$\eta_M=3\times 10^{-6}$. and three values of the age (5, 8 and 15 Gyr). All
the data are listed in Table~4.
  
The comparison with the observational data for Virgo galaxies is  shown in
Fig.~\ref{m_lb}, which displays both the  $log(M/L_B)_{\odot}$ versus 
$logM_{12}$ plane and the $\kappa_3$ versus  $\kappa_1$ plane. The filled
circles are the data from Bender et al. (1992), the thick dashed line is the
relation $\kappa_3 = 0.15\kappa_1+0.36$ presented in section 2.8, the solid
lines are the model results for $\eta_M=3\times 10^{-6}$, whereas the dotted
lines are the same but for $\eta_M=2 \times 10^{-6}$. The FP for three values
of the age are displayed, i.e. 15, 8, and 5 Gyr. Remarkably, there is no
appreciable variation of the slope of the theoretical relation with the age.

Assuming that both the theoretical and the observational FP are represented by
linear relationships of the type

\begin{equation}
      log(M/L_B)_{\odot} = A \times logM_{12} + B
\end{equation}
\noindent
and
\begin{equation}
      \kappa_3 = \alpha \times \kappa_1 + \beta
\end{equation}

\noindent
and casting the theoretical relation in terms of the observational one (i.e.
$\alpha$ and $\beta$) we get

\begin{equation}
      log(M/L_B)_{\odot} = \alpha \sqrt{{ 3 \over 2}} \times logM_{12} + \gamma
\end{equation}
\noindent
with 
\begin{equation}
\gamma = -log( {c_1 \over c_2 }) + \alpha \sqrt{ {3 \over 2}} \times 
                     log ({1\over c_2}) + \sqrt {3} \times \beta . 
\end{equation}
\littleskip
\noindent
Therefore, the theoretical slope $A$ corresponds to the observational one
multiplied by the factor $\sqrt{3/2}$, whereas the theoretical zero point
contains the observational slope and zero point, and the virial coefficients 
$c_1$ and $c_2$ (see also Burstein et al. 1997).

To compare theory with data we calculate the linear-square fit of
$log(M/L_B)_{\odot}$ and $logM_{12}$ both for $\eta_M=2\times 10^{-6}$ and
$\eta_M=3\times 10^{-6}$ separately, and then for the two cases lumped
together to somehow take the uncertainty on this parameter into account. The
results are summarized in Table~5 which contains the theoretical slopes and
zero points together with their uncertainties, and the corresponding slopes in
the $\kappa_3 -\kappa_1$ space. Theoretical and observational slopes find good
agreement for the combined sample 2\&3$\times 10^{-6}$, which means that this
basic parameter of the models cannot be too different from the present values.
The scatter in the zero point is about $\pm 0.04$ which is fully compatible
with the observational $\sigma=0.05$.

What would the results be for $\eta_M$ either much lower or higher than $2\div
3 \times 10^{-6}$ ? We recall here that  $\eta_M$ determines how low in
temperature the cooling process can go in a collapsing cloud and  fixes $M_P$,
which ultimately drives the efficiency of all the other heating sources.
Therefore, the effects of $\eta_M$ are easy to foresee. If $\eta_M << 2\div
3\times 10^{-6}$, $M_P$ falls down to the lower limit of the IMF, which
becomes Salpeter-like, not many remnants are left over and the FP flattens
out. The effect is more pronounced in the high mass (low density) than low
mass (high density) galaxies. The opposite occurs if $\eta_M >> 2\div 3\times
10^{-6}$, in which case $log(M/L_B)_{\odot}$ for massive galaxies gets too
high and the FP too steep. The many numerical models computed to check this
points show that not only the FP goes wrong but also the CMR (because of the
effect of $\eta_M$ via $M_P$ on the mean metallicity of a galaxy).

In addition to $\eta_M$, the CBR limit concurs to set the minimum temperature
during the collapse of the star forming clouds in case $\eta_M$ allows for
even lower temperatures.

On the basis of the above considerations, the tilt of the FP and at some
extent the CMR as well reflect the thermal history of the star forming  gas
over the early evolutionary stages and its variation with the galaxy mass.
What really matters here is not the parameter $\eta_M$ {\it per se}, but  what
physical mechanism determines the temperature of the gas at the star forming
stage. We will came back to this below.

%%%%%%%%%%%%%%%Table 5 %%%%%%%%%%%%%%%%
\begin{table}
\begin{center}
\caption{Theoretical FP }
\begin{tabular}{lc cc } 
\hline
&&& \\
 $\lambda$    &A        & B      & $\alpha=A \times \sqrt{2 \over 3} $ \\
&&& \\
\hline
&&& \\
\multicolumn{3}{c}{B-Band~~~~Age: 15 Gyr}\\
&&& \\
\hline
&&& \\
 $2\times 10^{-6}$   & $0.171\pm 0.018$ & $1.026\pm 0.061$ & 0.139\\
 $3\times 10^{-6}$   & $0.209\pm 0.022$ & $1.136\pm 0.061$ & 0.170\\ 
2\&3$\times 10^{-6}$ & $0.188\pm 0.013$ & $1.080\pm 0.041$ & 0.153\\ 
&&& \\
 \hline
&&& \\
\multicolumn{3}{c}{K-Band~~~~Age: 15 Gyr}\\
&&& \\
\hline
&&& \\
 $2\times 10^{-6}$   & $0.004\pm 0.008$ & $0.054\pm 0.026$ & 0.003\\
 $3\times 10^{-6}$   & $0.007\pm 0.002$ & $0.095\pm 0.064$ & 0.005\\ 
2\&3$\times 10^{-6}$ & $0.006\pm 0.010$ & $0.074\pm 0.031$ & 0.005\\ 
&&&\\
\hline
\end{tabular}
\end{center}
\label{tab_fp}
\end{table}

Finally, we have also looked at the $log(M/L_K)_{\odot}$ ratio as a function
of $logM_{12}$: the slope  and zero point for the K-passband  are given in
Table~5. The slope is significantly flatter  than in the case of the 
B-passband.

%%%%%%%%%%%%%%%%%%%%%%%%%%%%%%%%%%%%%%%%%%
\section{Summary and conclusions }

In this paper we have presented the preliminary investigation of the effects
of Padoan's et al. (1997) IMF, the shape of which depends on the physical
conditions (temperature, density, and velocity dispersion) of the medium in
which stars are formed. The main goal of this study was to seek for a  picture
of galaxy evolution (both chemical and spectro-photometric) in which the
pattern of properties of elliptical galaxies can find a coherent explanation.
Our major concerns were the CMR, the tilt of the FP, and  the
enhancement of $\alpha$-elements in relation to the global duration of the
star formation activity, prior to the onset of galactic winds.

The greatest advantages of this IMF with respect to the classical Salpeter law
or similar laws in literature, e.g. Scalo (1986 and references therein),
Arimoto \& Yoshii (1987), Kroupa et al. (1993) are: (i) the slope, function of
the mass range; (ii) the natural mass cut-off below which the IMF tends to
vanish; (iii) the peak mass  $M_{P}$ that gets very small at decreasing
temperature and velocity dispersion, and increasing density of the gas
(remarkably, under typical values for these three physical quantities, e.g.
those holding for molecular clouds in the solar vicinity, the Salpeter law is
recovered); (iv) the departure of the new IMF from the standard ones under
different physical conditions, such as those likely met in elliptical galaxies
of different density and velocity dispersion. This brings a new
leverage to the problems under examination.

The new IMF has been applied to models of elliptical galaxies, in which the
radial dependence of mass density (both stars and gas) and star formation rate
are taken into account. The present analysis is limited to the region  inside
$R_e$. However, the results that one would expect in other (more external)
regions of the galaxies are easy to foresee.
  
The temperature, density, and velocity dispersion governing the IMF have been
derived by solving the energy equation in which various sources of heating and
cooling are  considered,  and by adopting a simple-minded model for the
thermo-dynamical description of the gas clouds prone to collapse and the star
formation process. 

Heating is mainly caused by the radiative cooling of supernova
remnants and stellar winds, whereas that by UV radiation from massive stars
is found to be marginal as  nearly all UV flux is re-processed by dust into the far infrared (Granato et al. 1997).
However, two more sources of mechanical nature are considered: the first one
[$H_{C}$] is active only in the very early stages of galaxy formation and
evolution and it is meant to take into account that part of the energy
liberated by the collapse of the primordial gas that will likely go into heat.
It provides a sort of initial condition to start with, or in other words the
initial flame of our models.
The second one  [$H_{M}$] is active at all times and somehow takes into
account that existing gas clouds will mechanically interact with one another
thus causing additional heating. Both sources of heat would require detailed studies
that go beyond the scope of this paper. Therefore, both heating rates are
affect by a certain degree of uncertainty and $H_{M}$, in particular, contains
the   parameter $\eta_M$ that has to be chosen  in advance and has to be
determined by comparing theory with observations.

The temporal evolution of the peak mass over the early evolutionary stages
drives the whole problem to which the minimum temperature attainable by
collapsing gas clouds, chemical enrichment, cooling, and under suitable
circumstances the CBR limit concur. It goes without saying that while the
quantitative results are expected to depend on the details of our models and on the their  parameters, in particular,  the
scenario emerging from our analysis does not. The following remarks are worth
making:
\smallskip

\noindent
(1) {\bf History of star formation}. 
The {\it overall duration of the star forming activity is inversely
proportional to the galaxy mass (or directly proportional to the  mean
density)}. Specifically, $\Delta t_{SF} \propto M_G^{-1} $ is shorter in
massive galaxies and longer in low mass ones. This in spite of the stronger
 gravitational potential in the former, and as a straight consequence of the
varying IMF. The same kind of reasoning applies to different regions of a
galaxy: star formation activity lasts longer in the center (high density) than
in the external (low density) regions.
\smallskip

\noindent
(2) {\bf Galactic winds}. 
The onset of galactic winds occur for the same reasons as in the classical
scheme, {\it however with the opposite trend as far as the galactic mass is
concerned}. Early on in massive galaxies, and much later in less massive ones.
Despite this, the mean and maximum metallicities increase with  the galaxy mass,
thus  providing the same bottom line for the interpretation of the CMR as in
the classical scenario (a mass metallicity sequence of nearly coeval objects).
\smallskip

\noindent
(3) {\bf G-Dwarf analog}. 
The excess of low metallicity stars is easily avoided independently of the
galaxy mass, because  the relatively top-heavy IMF during the early stages
secures prompt enrichment of the gas and a scarce population of
low-metallicity stars. This makes the infall scheme somehow superfluous.
\smallskip

\noindent
(4) {\bf Enhancement of $\alpha$-elements}.
This IMF naturally tends to produce different degrees of enhancement at
varying radial distance, age,  and galactic mass. At given age and position
within a galaxy, the enhancement in $\alpha$-elements increases with galactic
mass. This is the sort of trend expected from the line strength indices in
galaxies of different luminosity (mass), cf. section 2.2. At given age and 
galactic mass, the enhancement in $\alpha$-elements tends to be stronger going
from the center to more external regions, because of the decreasing density
favoring more massive stars and in turn shorter durations of the star forming
periods. Apparently this trend opposes to the gradients in the line strength
indices $Mg_2$ and $<Fe>$ observed in a number of elliptical galaxies (cf.
Carollo et al. 1993; Carollo \& Danziger 1994a,b). The reality is more
complicated than this straight conclusion because the line strength indices do
not simply correlate with the chemical abundances, but  depend also on the age
and partition function $N(Z)$, number of stars per metallicity bin (cf.
Tantalo  et al. 1997b). It turns out that a region with prolonged star
formation, higher metal content, and lower chemical enhancement (our central
regions) may have $Mg_2$ indices stronger than in the more external zones with
shorter star forming activity, less metal enrichment, and higher chemical
enhancement (cf. Tantalo  et al. 1997b for all details). Finally,  in any case
the enhancement in $\alpha$-elements decreases at increasing age (effect of
Type I supernovae).
\smallskip

\noindent
(5) {\bf Baryonic dark matter}. 
This IMF tends to favor the formation of remnants (black holes, neutron stars,
and white dwarfs, these latter in particular). The percentage of remnants may
easily outnumber that of  visible stars at least at the present epoch. This
holds in  massive galaxies in general and/or in low density regions of all
galaxies. These stars are obvious candidates to baryonic dark matter, with
strong physical and cosmological implications. Is there any observational hint
for a large percentage of white dwarfs in the stellar mix of elliptical
galaxies ? Bica et al. (1996) analyzing the UV excess in a sample of
elliptical galaxies noticed absorption features centered at about 1400 \AA\
and  1600 \AA\ that were common to both strong and weak sources. These
absorption features characterize the spectrum of moderately cool white dwarfs
(DA5). Easy arguments indicate that in order to reach detectability in the UV,
the relative mass fraction of such white dwarfs with respect to the luminous
component of the galaxy should be of the order of 10:1.
\smallskip

\noindent
(6) {\bf Tilt of the FP}.
The tilt and tightness of the FP follow from the systematic variation of the
IMF with galactic mass and the narrow range of temperatures $T_P$  that is
likely  established both  by internal and external conditions. The
observational  FP is best matched by old models (say 15 Gyr) with $\eta_M
\simeq 2\div 3 \times 10^{-6}$ and $T_P$ in the range 20 to 40 K. Other values
for $\eta_M$ and $T_P$ in turn would predict FPs either too flat or to steep
as compared to the observational data.

How a galaxy manages to get $\eta_M$ and $T_P$ in the above ranges is not
clear and perhaps beyond the scope of this paper. In the present models $T_P$
results from internal energy sources. External sources are not taken into
account but for the limit set by the CBR. This latter is mostly effective in
low mass galaxies (IMF skewed toward the low mass end) because in the initial
stages owing to the scarce energy input from supernovae and stellar wind from
massive stars, the temperature will easily fall below $T_{CBR}$. For the
purposes of the present study, we have adopted the Friedmann model of the
Universe with Hubble constant $H_0 = 50 {\rm ~km~s^{-1}~Mpc^{-1}}$, $q_0=0$
and red-shift of galaxy formation $z_{for} = 5$ (to which an initial value for
$T_{CBR} \simeq 15$ corresponds). Other values for the three parameters are
obviously possible.  They would anyhow lead to similar results. The effect of
increasing $z_{for}$ is straightforward, because at given metallicity (cooling)
during the early stages, it would be easier for the collapsing clouds to fall
below $T_{CBR}[z(t)]$. Needless to say that, increasing $z_{for}$, higher
values of $T_{CBR}[z(t)]$ in the very early stages are implied.

Another possibility is that  external conditions, for instance  the CBR
itself, set the initial gas temperature on the top of which  internal
processes should be added. This would lead to the interesting possibility that
$T_P$ correlates with $z_{for}$. In such a case large variations in $T_P$ 
over short periods of time are to be expected. However, a thorough exploration
of all  possible cases is beyond the aims of this paper.

Finally, our analysis of the tilt and tightness of the FP simply proves that a
solution is possible without invoking changes in the virial coefficients
$c_1$, and $c_2$ (see section 2.6) or other effects of dynamical nature (cf.
Ciotti et al. 1996). If  deviations from virial conditions and/or dynamical
effects do actually occur, they will be expected to constitute a minor
correction to the gross trend established by the varying IMF.
\smallskip

\noindent
(7) {\bf On the iron-discrepancy}.
Although it is beyond the scope of this paper to address the question of the
iron discrepancy raised by the ASCA data, we would like to briefly comment
that part of the disagreement could originate from the assignment of the iron
abundance to the stellar content of these galaxies. In fact, it stems from the
adoption of a particular $[Fe/H]$-$Mg_2$ calibration, in which no dependence on
age and possible enhancement of $\alpha$-elements is taken into account. We
start noticing that the general relationship linking the total metallicity $Z$
to the iron content $[Fe/H]$ is

\begin{displaymath}
[{Fe \over H}] = log[{Z \over Z_{\odot} }] -log[{X \over X_{\odot} }] + 
\end{displaymath}
\begin{equation}
~~~~~~~~~~~~~~~~ - 0.8 \times [{\alpha \over Fe}] - 
                 0.05 \times [{\alpha \over Fe}]^2
\label{Fe_Z}
\end{equation}

\noindent
in which $X_{\odot}=0.707$ and $Z_{\odot}=0.018$ are the adopted solar values,
and  the effect of  $\alpha$-elements enhancement  is also brought into
evidence. This relation reduces to the standard one given by Bertelli et al.
(1994) when $[{\alpha /Fe}]=0$ and $X_{\odot}=0.700$ and $Z_{\odot}=0.020$ are
adopted.

Considering that in the course of galactic evolution both helium and heavy
elements should have increased according to the popular enrichment law $\Delta
Y/\Delta Z=2\div 3 $ (cf. Pagel et al. 1992), the proper correlation between
$X$ and $Z$ of the theoretical SSP and the iron content $[Fe/H]$ can be
established. A recent calibration of the $Mg_2$ index as a function of the age
and metallicity $Z$ (and $X$) for SSPs is by BCT96 from whom we derive the
following analytical approximation

\begin{displaymath}
[{Fe \over H}] = [ 7.56 -0.340 \times t + 0.0148 \times t^2 ] \times Mg_2 + 
\end{displaymath}
\begin{equation}
~~~~~~~~~~~~~~~~~- [ 1.325 + 0.0031 \times t + 0.00074 \times t^2 ]
\label{Fe_Mg2}
\end{equation} 

\noindent
where $t$ is the age in Gyr, and no effect of $\alpha$-elements is yet
considered. The above relation holds  for $0.008 \leq Z \leq 0.05$, $0.20 \leq
Mg_2 \leq 0.35$, and ages older than about 5 Gyr, which is fully adequate to
our purposes.

For the age of 15 Gyr, this relation is similar to that of Buzzoni et al.
(1992)   used by Arimoto et al. (1997) in their analysis. Since elliptical
galaxies are likely to be rich in $\alpha$-elements (cf. section 2.2), before
 assigning $[Fe/H]$ to the stellar content on the basis of its  $Mg_2$
index, the proper correction for over-abundance by $\alpha$-elements should be
applied, i.e. the value $[Fe/H]$ read off from eq.(\ref{Fe_Mg2}) should be
decreased by the quantity $-0.8[\alpha /Fe]-0.05[\alpha /Fe]^2$. Considering
that $[\alpha/Fe]$ may vary from 0.5 to 1 according to current data (cf.
Matteucci 1997), the expected decrease in $[Fe/H]$ goes from 3 to 5, which would
reduce  if not eliminate the disagreement. Looking at the iron over-abundance
(with respect to the solar value) versus $log (L_X/L_B)$ diagram of Arimoto et
al. (1997), this is the case for NGC~4406, NGC~4636, and NGC~4472 (the
galaxies with the highest   $L_X/L_B$ ratio). Problems remain with NGC~4374
for which the stellar over-abundance of $Fe/Fe_{\odot}=0.85$ is derived as
compared to the value $Fe/Fe_{\odot}=0.11$ indicated by ASCA. Only very high
over-abundances of $\alpha$-elements would be able to remove the discrepancy.
That the $[\alpha/Fe]$ can vary from galaxy to galaxy is indicated by extant
data. That this applies to NGC 4374 (much higher value in particular) is not
granted. Passing we note that the fine structure parameter of this galaxy  is
2.3 (Schweizer \& Seitzer 1992). The galaxy has suffered from a certain degree
of dynamical youth or  internal rejuvenation (this is also suggested by its
location in \Hbeta\ versus  \MgFe\ plane, cf.  BCT96). How this would affect
the $Mg_2$ index is not easily assessable. The problems raised by Arimoto et
al. (1997) still persist. In any case, we like to point out that enhancement
of $\alpha$-elements is a natural  side product of the new IMF, in particular
in the low density regions of a galaxy.
\smallskip

\noindent
(8) {\bf On the intra-cluster gas}. 
The data presented in section 7.3 and Table~2 can perhaps alleviate the
difficulties encountered by the standard models (cf. Matteucci 1997 for a
recent review of the subject) in simultaneously account for amount of iron and
gas observed in the ICM.  The $R_e$-sphere of the  1 and 3 $M_{L,T}$
galaxies eject amounts of gas, oxigen, and iron going from 1.6 to 6.9$\times
10^{11} M_{\odot}$,  5.6 to 2.3 $\times 10^{10} M_{\odot}$ and 3.8 to 7.9
$\times 10^{9} M_{\odot}$, respectively.  These numbers can be easily doubled
considering the contribution from the remaining half of the galaxy not taken
into account here. As compared to the  estimates by Gibson \& Matteucci (1995)
for their standard model referred to the whole galaxy, for which they get
$\simeq 3.7\times 10^{11} M_{\odot}$ of gas, $\simeq 1.6\times 10^{9}
M_{\odot}$ of Fe, and $\simeq 1.7\times 10^{10} M_{\odot}$ of oxygen, there is
a mean increase by a factor of 1.2 to 2.3 in gas,  2.3 to 4.7 in O, and 3.6 to
7.3 in Fe. Whether this data can completely rule out the above difficulty
cannot be said without detailed model calculations convolved with the Schechter
(1976) luminosity functions of galaxies. Work is in progress to improve upon
the present models and cast light on this topic (Chiosi et al. 1997).
\smallskip

\noindent
(9) {\bf Going toward complete models}.
The models we have presented are limited to the region within the effective
radius, and therefore they   are not yet able to properly describe complete
galaxies nor to make detailed predictions for the amount and the chemical
composition of the gas ejected into the ICM. Furthermore, they cannot predict
properties specific to the very central regions of  galaxies, nor spatial
gradients in physical quantities. However, as the leading parameter is
density, the results are somehow independent from the particular region for
which they have been derived, and may be extended to other regions in other
galaxies provided they have the same density. What would the result be going
further out in radius, i.e. to regions of lower density ? The answer is
obvious and supported by preliminary numerical calculations. More external
regions would follow the same trend as passing from a low mass (mean high
density) galaxy to a high mass (low mean density) one. The lower density would
make it easier to form massive stars and to anticipate the onset of galactic
winds. The material expelled from the outer regions of a galaxy should be more
enhanced in $\alpha$-elements and in more generous relative proportions with 
respect to what is left over as stars.  If so, the construction of a galaxy
can be viewed as a sort of out-inward process with star formation lasting
progressively longer as we move inward.
\smallskip

\noindent
(10) {\bf Replacing galaxy masses with densities}.
All the discussion carried out in so far has been made using  the galactic
mass as the parameter ranking the various properties instead of the mean
density despite the fact that density was the underlying basic parameter.
Indeed, the whole discussion could be re-phrased in terms of this latter. This
makes the results of our analysis even more general in the sense that they be
could applied to sub-units of the same system (proto-galaxy) with different 
initial mean density of gas.  Let us imagine for the sake of argument, that a
number of such sub-units can exist within the total potential well of the
pro-galaxy (dark and baryonic matter), and start forming stars more or less at
the same time in the remote past. A low density cloud will have the peak mass
of its IMF skewed toward the high mass end ($M_P$ can easily be higher than
what found in our models with the lowest density) and, therefore, will soon
experience the conditions for local gas ejection and interruption of star
formation. If gas loss is too high, the newly born stellar system may even
disgregate dispersing its stars into the global potential well. If not, the
stellar aggregate will survive. In any case a scarce population of stars is
left over, the vast majority of which will evolve on time scales shorter than 
the Hubble time becoming compact remnants. All of the other stars (low mass
objects) will remain either as a diffuse medium or a low luminosity aggregate
within the potential well. In contrast, a  high density cloud that
preferentially will be located   in the centre of the potential well, will
have the peak mass of its IMF more skewed toward the low mass end, and
therefore will continue to form stars for longer periods of time before
meeting the conditions for local gas ejection and ceasing star formation.
Since high density  clouds  will lose much less gas than the low density ones,
they will survive as formed stellar aggregates eventually merging into a
single unit at the center of the galaxy. In the above picture of galaxy
formation, the bulk of star formation has taken place in the remote past. How
and whether the gas left over by the building process in each sub-unit is
thrown into the intergalactic medium cannot be assessed with this simple
scheme. Furthermore, the relative number of sub-units and how they eventually
coalesce into bigger aggregates is a typical problem to be addressed with the
aid of Tree-SPH simulations of galaxy formation and evolution in a given
cosmological context (cf. section 1) and goes beyond the aims of the present
study.

\smallskip

\vspace{0.5truecm} 
\noindent
C.C. wishes to thank the Pontificial Academy of Science for its invitation to
attend the Vatican Conference on "The Emergence of Structure in the Universe
at the Level of Galaxies" (November 25-29, 1996) where  a complete report of
the results described in this paper was delivered. This study has been financially
supported by the Italian Ministry of University, Scientific Research and
Technology (MURST), the Italian Space Agency (ASI), and the TMR grant
ERBFMRX-CT96-0086 from the European Community.

%*********************************************************

\end{document}